\documentclass[twocolumn,preprintnumbers,amsmath,amssymb]{revtex4}  

\usepackage{amsfonts}
\usepackage[T1]{fontenc}
\usepackage{amsmath,amsbsy,amssymb,graphicx}
\usepackage{times}
\usepackage{amsmath}
\usepackage{amssymb}
\usepackage{graphicx}
\usepackage{color}

\begin{document}

\title{Zigzag dice lattice ribbons: Distinct edge morphologies and structure-spectrum correspondences}

\author{Lei Hao}
 \address{School of Physics, Southeast University, Nanjing 211189, China}

\date{\today}

\begin{abstract}
Ribbons of two-dimensional lattices have properties depending sensitively on the morphology of the two edges. For regular ribbons with two parallel straight edges, the atomic chains terminating the two edges may have more than one choices for a general edge orientation. We enumerate the possible choices for zigzag dice lattice ribbons, which are regular ribbons of the dice lattice with edges parallel to a zigzag direction, and explore the relation between the edge morphologies and their electronic spectra. A formula is introduced to count the number of distinct edge termination morphologies for the regular ribbons, which gives 18 distinct edge termination morphologies for the zigzag dice lattice ribbons. For the pure dice model, because the equivalence of the two rim sublattices, the numerical spectra of the zigzag ribbons show qualitative degeneracies among the different edge termination morphologies. For the symmetrically biased dice model, we see a one-to-one correspondence between the 18 edge termination morphologies and their electronic spectra, when both the zero-energy flat bands and the dispersive or nonzero-energy in-gap states are considered. We analytically study several interesting features in the electronic spectra, including the number and wave functions of the zero-energy flat bands, and the analytical spectrum of novel in-gap states. The in-gap states of the zigzag dice lattice ribbons both exhibit interesting similarities and show salient differences when compared to the spectra of the zigzag ribbons of the honeycomb lattice.
\end{abstract}


\maketitle

\section{Introduction}

Real-world samples of two-dimensional (2D) and quasi-two-dimensional (q-2D) materials mostly have edges. The type of edge termination has a decisive influence on the energy spectrum, transport, thermal, and magnetic properties of the sample. Ribbons of 2D or q-2D materials, which have two edges running (roughly) parallel to each other, demonstrate this dependency clearly. For example, ribbons of the monolayer graphene with armchair-type, zigzag-type, and bearded-type edges all have distinct electronic and phononic spectra \cite{fujita96,nakada96,klein94,brey06,ryu02,mong11,delplace11}, which in turn lead to different transport properties \cite{wakabayashi01,yang09,evans10} or characteristic broken symmetry phases \cite{wakabayashi98,wakabayashi99,son06,son06nat,pisani07,wassmann08,palacios08}. The edge termination configuration also critically influences the properties of lateral heterostructures of 2D materials, such as the lateral heterostructures of graphene with hexagonal boron nitride \cite{zeng16,pruneda10,kim15,ci10,levendorf12,liu14}. With the technical progress for atomic-precision sample preparation and characterization \cite{chen07,li08,wang08,breybook,ci10,levendorf12,liu14,cano09,wang11,ma12,ma13,yan19}, it is highly promising to anticipate accurate control at the atomic level in the edge terminations of the ribbons.

A great variety of ribbon morphologies have been considered in previous works. The simplest kind of ribbons, which have attracted the most attention, have a pair of parallel straight edges and are free of any imperfections. They are translationally invariant along the edge direction. We call these ribbons the \emph{regular ribbons}. Interesting variations to the regular ribbons have also been explored. These include the influence of vacancies \cite{palacios08,topsakal08,song11,deng15,li18}, impurities \cite{filho07,fang08,biel09,friedrich20,vannucci20,touski21}, and edge roughness \cite{evans10,moors19}. Another class of works consider ribbons with periodically patterned edges \cite{nakada96,akhmerov08,wakabayashi10,delplace11}, which are in some sense in between regular ribbons and imperfect ribbons. More significant modifications to regular ribbons include samples with shapes different from a strip \cite{yang09,bellucci20}, which may more properly be considered as flakes of the 2D lattice. The richness in the ribbon morphologies studied on one hand signifies the importance of the ribbons and on the other hand suggests the necessity of a reasonable classification for the various kinds of ribbons. In particular, it is desirable to know the number of distinct edge terminations for certain interesting or experimentally accessible edge orientations for the ribbons. This information, together with the energy spectra of the distinct ribbons, serve as the starting point for studying other properties of the ribbons. In this work, taking the zigzag dice lattice ribbons as examples, we enumerate the edge termination morphologies of the regular ribbons and explore their relations with the corresponding electronic spectra.

The dice lattice (or, $\mathcal{T}_{3}$ lattice) is an interesting physical system well known for a completely flat band in its band structures and is under intensive studies \cite{sutherland86,vidal98,abilio99,naud01,rizzi06,bercioux09,wang11b,moller12,illes15,illes16,biswas16,xu17,betancur17,
oriekhov18,bugaiko19,chen19,alam19,soni20,tan21,wang21,gorbar21,wang21b,zhou21,tamang21,hao21,hao21b,cunha21,iurov21,soni21}. It may be considered as a variant of the honeycomb lattice by adding a new atom to the center of each hexagon of the honeycomb lattice. The additional sublattice in the center of the hexagons form bonds only with a single sublattice of the original honeycomb lattice. Compared to ribbons of the honeycomb lattice (i.e., the graphene lattice) with zigzag-type and bearded-type edges that have been studied in great details and are well known to support novel in-gap states, properties of the dice lattice ribbons with edges along a zigzag direction are much less known. With the experimental progress in fabricating novel lattices in terms of arrays of quantum dots \cite{sikdar21}, photonic lattices \cite{mukherjee15}, and optical lattices of cold atoms \cite{ozawa17}, the study of dice lattice ribbons is becoming an experimentally relevant problem. It is therefore highly desirable to make an in-depth analysis over the structures and electronic spectra of the zigzag dice lattice ribbons.

We firstly derive a formula that determines the number of qualitatively distinct regular ribbons as regards the combinations of the two edges, for a general 2D or q-2D lattice. We illustrate the formula in terms of the regular ribbons of three representative 2D lattices, including the Lieb lattice \cite{lieb89,shen10}, the honeycomb lattice \cite{fujita96,nakada96}, and the dice lattice \cite{sutherland86,rizzi06}. The zigzag ribbons of the honeycomb lattice and the dice lattice, with two parallel straight edges along a zigzag direction of the lattices, constitute the simplest nontrivial realizations of the formula. According to the formula, there are 18 distinct edge termination morphologies for the zigzag dice lattice ribbons.

Then, for the 18 edge termination morphologies of the zigzag dice lattice ribbons, we study their structure-spectrum correspondences, for both the pure dice model and the symmetrically biased dice model. In terms of the dispersive or nonzero-energy edge states and the zero-energy flat bands, we find a one-to-one correspondence between the electronic spectra and the edge termination configurations, for zigzag ribbons of the symmetrically biased dice model. On the other hand, there are qualitative degeneracies in the energy spectra for zigzag ribbons of the pure dice model. Based on the full tight-binding model defined on the ribbons, we study analytically several interesting features in the numerical electronic spectra, including the zero-energy flat bands and the in-gap states. The zigzag dice lattice ribbons have segmental flat band edge states, similar to the zigzag ribbons of the honeycomb lattice. There are also additional in-gap states unique to the zigzag dice lattice ribbons. Both of them are studied in details.

The rest of the paper is organized as follows.
In Sec.II, we introduce and explain a general formula which gives the number of distinct regular ribbons, for a specific combination of a bulk 2D lattice and cutting conditions for the ribbons. Then we illustrate the formula by applying it to several ribbons of typical 2D lattices, including the honeycomb lattice, the dice lattice, and the Lieb lattice. In Sec.III, we introduce the models for the electrons in the zigzag dice lattice ribbons and calculate numerically their electronic spectra. We consider both the pure dice model and the symmetrically biased dice model. We establish a one-to-one correspondence between the energy spectra and the edge termination morphologies, for the 18 zigzag ribbons of the symmetrically biased dice model. In Sec.IV, we analyze in details several interesting features in the numerical spectra, including the zero-energy flat bands and the in-gap states, and their relations with the corresponding edge termination morphologies. The results are summarized in Section V.

\section{counting distinct edge termination morphologies of regular ribbons}

\subsection{General formula}

Cutting an ideal 2D or q-2D lattice apart along a pair of parallel straight lines in the lattice plane, the regular ribbon that we consider is left in between the two cutting lines. We assume the two straight cutting lines break all the bonds intersecting them and do not cross any sites, so that all the atoms in the ribbon are intact. For a general cutting direction, the above assumption is equivalent to assuming atoms of negligible size on the lattice sites. Thin sheets of laser beam with uniform irradiance across the beam profile might qualify as the blade to cut a regular ribbon defined in the above sense, in addition to other (mechanical or chemical) approaches \cite{chen07,li08,wang08,breybook,ci10,levendorf12,liu14,cano09,wang11,ma12,ma13,yan19}.

The properties of very narrow ribbons depend sensitively on their widths, owing to prominent quantum confinement effect. As the ribbon becomes wide enough, the finite-size effect diminishes and the qualitative properties of the ribbons are determined mostly by the original bulk lattice and the morphologies of the two edges. We are interested in these relatively wide ribbons. For these regular and wide ribbons, the lattice structures are distinguished according to the direction along the two edges and the two outermost chains of lattice sites. We therefore enumerate the different types of wide and regular ribbons in terms of the direction vector along the two edges and the combination of the two outermost chains of lattice sites.

A 2D or q-2D lattice is defined by specifying a primitive unit cell and the lattice sites included within it. For a given 2D or q-2D lattice, we consider a primitive unit cell subtended by two primitive lattice vectors $\mathbf{v}_{1}$ and $\mathbf{v}_{2}$. The area of the primitive unit cell is
\begin{equation}
\Omega=\hat{\mathbf{z}}\cdot(\mathbf{v}_{1}\times\mathbf{v}_{2}).
\end{equation}
$\mathbf{v}_{1}$ and $\mathbf{v}_{2}$ are chosen so that $\Omega>0$.
This primitive unit cell is assumed to be \emph{minimal}, with the shortest possible perimeter. Even though the minimal primitive unit cell is not unique, any viable choice is equally valid. A general lattice vector is uniquely represented as a linear combination of $\mathbf{v}_{1}$ and $\mathbf{v}_{2}$ with integer coefficients.

Among the infinite ways of choosing the cutting directions for the regular ribbons, we focus on the cases where the translational invariance along the edge direction is preserved. We denote the minimal lattice vector along the edge direction as
\begin{equation}
\mathbf{v}_{mn}=m\mathbf{v}_{1}+n\mathbf{v}_{2},
\end{equation}
where $m$ and $n$ are coprime integers. Once the atomic species and the positions of the lattice sites in the unit cell are given, the number of distinct edge terminations is determined uniquely by the vector $\mathbf{v}_{mn}$. Specifically, as we show in what follows, this number is
\begin{equation}
n_{\perp}^{2}|\tilde{s}|,
\end{equation}
where both $n_{\perp}$ and $|\tilde{s}|$ are positive integers.
Let us imagine a process of moving along the direction perpendicular to the edge of the ribbon. As we move we come across different atomic chains all parallel to the edge. We will see that the atomic chain always repeat after every $n_{\perp}$ chains. The consecutive $n_{\perp}$ chains are distinct from each other either by the atomic species in the chain or by their alignments along the chain. However, the $n$th atomic chain and the $(n+n_{\perp})$th atomic chain are usually not aligned along the direction we move (i.e., perpendicular to the edge). In other words, those two chains are connected by a lattice vector NOT perpendicular to the edge. Instead, the $n$th atomic chain and the $(n+n_{\perp}|\tilde{s}|)$th atomic chain are aligned along the direction perpendicular to the edge.

To corroborate the above result and to determine $n_{\perp}$ and $|\tilde{s}|$ from $\mathbf{v}_{mn}$, we define two further lattice vectors, $\mathbf{v}_{pq}$ and $\mathbf{v}_{st}$, which satisfy
\begin{equation}
\begin{cases}
\hat{\mathbf{z}}\cdot(\mathbf{v}_{pq}\times\mathbf{v}_{mn})=\Omega,  \\
\mathbf{v}_{st}\cdot\mathbf{v}_{mn}=0.
\end{cases}
\end{equation}
$\{p,q\}$ and $\{s,t\}$ are two pairs of coprime integers, in terms of which the above equations become
\begin{equation}
\begin{cases}
pn-qm=1,  \\
s(m|\mathbf{v}_{1}|^{2}+n\mathbf{v}_{1}\cdot\mathbf{v}_{2})=-t(n|\mathbf{v}_{2}|^{2}+m\mathbf{v}_{1}\cdot\mathbf{v}_{2}),
\end{cases}
\end{equation}
where $|\mathbf{v}_{i}|^{2}=\mathbf{v}_{i}\cdot\mathbf{v}_{i}$ ($i=1,2$). $\gcd(m,n)=\gcd(p,q)=\gcd(s,t)=1$ by definition, where $\gcd(m,n)$ means the greatest common divisor of two integers $m$ and $n$.

According to a theorem by Bezout in number theory \cite{everestbook}, the first equation of Eq.(5) has an infinite series of coprime integer solutions for $\{p,q\}$. For a particular solution $\{p_{0},q_{0}\}$ to the first equation of Eq.(5), $\{p_{0}+lm,q_{0}+ln\}$ are also solutions for arbitrary integer $l$. This ambiguity is the consequence of the fact that the volume subtended by $\mathbf{v}_{p_{0}q_{0}}$ and $\mathbf{v}_{mn}$ does not change when the $\mathbf{v}_{p_{0}q_{0}}$ vector is supplemented by an integral multiple of $\mathbf{v}_{mn}$, namely
\begin{equation}
\hat{\mathbf{z}}\cdot(\mathbf{v}_{p_{0}q_{0}}\times\mathbf{v}_{mn}) =\hat{\mathbf{z}}\cdot[(\mathbf{v}_{p_{0}q_{0}}+l\mathbf{v}_{mn})\times\mathbf{v}_{mn}]=\Omega.
\end{equation}
Among this series of solutions, we take $\mathbf{v}_{pq}$ as the particular solution that has the minimal length. A particular set of solution, $\{p_{0},q_{0}\}$, to the first equation of Eq.(5) may usually be obtained by inspection and trial-and-error. For larger $\{m,n\}$, a more systematic method of obtaining $\{p_{0},q_{0}\}$ is by applying the Euclidean Algorithm of calculating $\gcd(m,n)$ in the reverse order \cite{everestbook}.

From the first of Eq.(4), $\{\mathbf{v}_{pq},\mathbf{v}_{mn}\}$ form a basis set of a primitive unit cell, which is usually non-minimal for a general $\mathbf{v}_{mn}$ vector. Suppose there are $n_{uc}$ atoms (sites) within each primitive unit cell, subtended by $\mathbf{v}_{mn}$ and $\mathbf{v}_{pq}$. According to their distances from one edge, which is parallel to $\mathbf{v}_{mn}$, these $n_{uc}$ sites may separate into $n_{\perp}$ sets. The sites within the same set have the same distance from the edge and constitute a chain parallel to the edge. Depending on the arrangements of the atoms within the unit cell, $n_{\perp}$ may vary from $1$ to $n_{uc}$.

Next we consider the solution $\{s,t\}$ to the second equation of Eq.(5).
Among the four basic lattice types for 2D and q-2D crystals \cite{dresselhausbook}, both the hexagonal and the square lattices allow integer solutions to the second equation of Eq.(5), for arbitrary $\{m,n\}$. For the rectangular and the oblique lattices, however, there may not always be integer solutions to $\{s,t\}$. In this work, we focus on the cases for which integer solutions to $\{s,t\}$ do exist. As examples, we will consider the honeycomb lattice, the dice lattice, and the Lieb lattice. Both the honeycomb lattice and the dice lattice are hexagonal lattices. The Lieb lattice, on the other hand, is a square lattice. Among the possible integer solutions to the second equation of Eq.(5), we choose a minimal solution for which the lattice vector $\mathbf{v}_{st}$ has the smallest length.

As a lattice vector, $\mathbf{v}_{st}$ may be uniquely represented as a linear combination of the primitive lattice vectors $\{\mathbf{v}_{pq},\mathbf{v}_{mn}\}$ with integer coefficients
\begin{equation}
\mathbf{v}_{st}=\tilde{s}\mathbf{v}_{pq}+\tilde{t}\mathbf{v}_{mn}.
\end{equation}
In terms of Eqs. (1) and (4), the two coefficients are
\begin{equation}
\begin{cases}
\tilde{s}=\frac{\hat{\mathbf{z}}\cdot(\mathbf{v}_{st}\times\mathbf{v}_{mn})}{\hat{\mathbf{z}}\cdot(\mathbf{v}_{pq}\times\mathbf{v}_{mn})} =\left|\begin{array}{cc} s & t  \\ m & n \end{array}\right|=sn-tm,  \\
\tilde{t}=\frac{\hat{\mathbf{z}}\cdot(\mathbf{v}_{st}\times\mathbf{v}_{pq})}{\hat{\mathbf{z}}\cdot(\mathbf{v}_{mn}\times\mathbf{v}_{pq})} =-\left|\begin{array}{cc} s & t  \\ p & q \end{array}\right|=tp-sq.  \\
\end{cases}
\end{equation}
Because
\begin{equation}
\hat{\mathbf{z}}\cdot(\mathbf{v}_{st}\times\mathbf{v}_{mn}) =\tilde{s}\hat{\mathbf{z}}\cdot(\mathbf{v}_{pq}\times\mathbf{v}_{mn})=\tilde{s}\Omega,
\end{equation}
$|\tilde{s}|$ denotes the number of copies of the primitive unit cells contained in the supercell subtended by $\mathbf{v}_{st}$ and $\mathbf{v}_{mn}$.

Taking advantage of the periodicity of the underlying lattice, all the $|\tilde{s}|$ repetitions of atoms in the supercell subtended by $\mathbf{v}_{st}$ and $\mathbf{v}_{mn}$ may be related to those atoms in the supercell subtended by $\mathbf{v}_{mn}$ and $\tilde{s}\mathbf{v}_{pq}$. Once an atom in the latter supercell lies outside the supercell subtended by $\mathbf{v}_{mn}$ and $\mathbf{v}_{st}$, we move it back by shifting a certain integer multiple of $\mathbf{v}_{mn}$. The largest lattice vector required to implement the shifting is simply $\tilde{t}\mathbf{v}_{mn}$. These displacements of the atoms do not change their distances to the edge. Therefore, the $|\tilde{s}|$ copies of atoms arrange into $|\tilde{s}|$ sets according to their distances to the edge. Each set has exactly one copy of the atoms in a single primitive unit cell, which separate into $n_{\perp}$ subsets according to their distances to the edge. There are $n_{\perp}|\tilde{s}|$ distinct subsets of atoms in the supercell subtended by $\mathbf{v}_{st}$ and $\mathbf{v}_{mn}$. As we move along the direction perpendicular to the edge, these $n_{\perp}|\tilde{s}|$ subsets repeat until broken by the edge.

Now we define the unit cell of the regular ribbons. The ribbons as quasi-1D periodic structures have the rungs bordered along the two edges by the lattice vector $\mathbf{v}_{mn}$ as the unit cells.  Among the possible choices for the unit cells, the rectangular rung whose second pair of edges are parallel to $\mathbf{v}_{st}$ is special for two reasons. Firstly, this rectangular rung is minimal among all rungs bordered along the two edges by $\mathbf{v}_{mn}$, because it has the smallest perimeter. Secondly, $\mathbf{v}_{st}$ and $\mathbf{v}_{mn}$ define a rectangular supercell of the bulk lattice. For this rectangular supercell, the reduced Brillouin zone (rBZ) of the bulk lattice is also a rectangle. The 1D BZ of the ribbon corresponds directly to one reciprocal lattice vector of the rectangular rBZ. All other rungs, which have the shape of parallelograms, have neither of these two merits. In this work, we take this rectangular rung as the unit cell of the regular ribbon. The supercell of the bulk lattice subtended by $\mathbf{v}_{st}$ and $\mathbf{v}_{mn}$ is a minimal repetitive unit of the unit cell.

For a regular ribbon with two edges parallel to $\mathbf{v}_{mn}$, each edge may terminate at any of the $n_{\perp}|\tilde{s}|$ subsets. However, because of the translational invariance of the ribbon along $\mathbf{v}_{mn}$, and the corresponding freedom in choosing the two borders of the unit cell parallel to $\mathbf{v}_{st}$, not all the $(n_{\perp}|\tilde{s}|)^{2}$ combinations of the edge termination morphologies are distinct from each other. To be concrete, let us assume a regular ribbon lying in the $xy$ plane with two edges along $x$, and define the direction parallel to the edges and pointing from left to right as the positive $x$ axis. So $\mathbf{v}_{mn}$$\parallel$$\hat{\mathbf{x}}$ and $\mathbf{v}_{st}$$\parallel$$\hat{\mathbf{y}}$. Among the $n_{\perp}$ subsets of sites in a primitive unit cell defined above, we assume the upper edge terminates at the $\alpha$-th subset of sites ($\alpha=1,2,\cdots,n_{\perp}$). As we have pointed out below Eq.(3), this $\alpha$-th subset of atoms repeats $|\tilde{s}|$ times along $\mathbf{v}_{st}$$\parallel$$\hat{\mathbf{y}}$, which are not aligned with each other along $\mathbf{v}_{st}$. It appears that we should specify which of the $|\tilde{s}|$ copies of the $\alpha$-th set of atoms actually terminates the upper edge of the ribbon. However, by shifting the two borders along $\mathbf{v}_{st}$ of the rectangular unit cell by $\mathbf{v}_{mn}/|\tilde{s}|$ successively, we may make the unit cell terminate with any one of the $|\tilde{s}|$ copies of the $\alpha$-th subset of atoms. Therefore, taking into account this flexibility in choosing the unit cell, there are only
\begin{equation}
n_{\perp}|\tilde{s}|/|\tilde{s}|=n_{\perp}
\end{equation}
qualitatively distinct edge terminations on the upper edge.
Since the upper and lower edges are bound together by the ribbon lattice, the translational invariance may be applied only once. Therefore, the total number of qualitatively distinct edge termination configurations is
\begin{equation}
(n_{\perp}|\tilde{s}|)^{2}/|\tilde{s}|=n_{\perp}^{2}|\tilde{s}|.
\end{equation}
This reproduces Eq.(3).

\subsection{Examples for selected 2D lattices}

To make the discussions concrete, we illustrate the above formula in terms of typical regular ribbons of the honeycomb lattice, the dice lattice, and the Lieb lattice. In the ideal undeformed cases, the honeycomb and the dice lattices are hexagonal, the Lieb lattice is square. For all the three lattices, according to the previous discussions, any given $\mathbf{v}_{mn}$ leads to a minimal $\mathbf{v}_{pq}$. If we fix the direction of $\mathbf{v}_{st}$ by requiring $\hat{\mathbf{z}}\cdot(\mathbf{v}_{st}\times\mathbf{v}_{mn})=\tilde{s}\Omega>0$, then $\mathbf{v}_{st}$ is also uniquely determined by $\mathbf{v}_{mn}$.

For large coprime $m$ and $n$, all three vectors (i.e., $\mathbf{v}_{mn}$, $\mathbf{v}_{pq}$, and $\mathbf{v}_{st}$) are generally large, and the minimal repetitive unit of the unit cell contains a large number of atoms. For example, let us consider $\mathbf{v}_{mn}=\mathbf{v}_{5,7}$ in the honeycomb or dice lattice, and take $\mathbf{v}_{1}\cdot\mathbf{v}_{1}=\mathbf{v}_{2}\cdot\mathbf{v}_{2}=2\mathbf{v}_{1}\cdot\mathbf{v}_{2}$. Then, according to the previous section, we get $\mathbf{v}_{pq}=\mathbf{v}_{-2,-3}$ and $\mathbf{v}_{st}=\mathbf{v}_{19,-17}$. The minimal repetitive unit subtended by $\mathbf{v}_{mn}$ and $\mathbf{v}_{st}$ contains
\begin{equation}
\tilde{s}=\left|\begin{array}{cc} s & t \\ m & n \end{array}\right|
=\left|\begin{array}{cc} 19 & -17 \\ 5 & 7 \end{array}\right|=218      \notag
\end{equation}
copies of the atoms (sites) within each primitive unit cell of the lattice.

To visualize a minimal repetitive unit for a general regular ribbon, we consider a regular ribbon of the Lieb lattice with $\mathbf{v}_{mn}=\mathbf{v}_{2,3}$. As shown in Fig. 1, the two primitive lattice vectors are taken to satisfy $\mathbf{v}_{1}\cdot\mathbf{v}_{1}=\mathbf{v}_{2}\cdot\mathbf{v}_{2}=\hat{z}\cdot(\mathbf{v}_{1}\times\mathbf{v}_{2})$ and $\mathbf{v}_{1}\cdot\mathbf{v}_{2}=0$. For this ribbon, we find $\mathbf{v}_{pq}=\mathbf{v}_{1,1}$ and $\mathbf{v}_{st}=\mathbf{v}_{3,-2}$. The minimal repetitive unit of the unit cell subtended by $\mathbf{v}_{mn}$ and $\mathbf{v}_{st}$ contains
\begin{equation}
\tilde{s}=\left|\begin{array}{cc} s & t \\ m & n \end{array}\right|
=\left|\begin{array}{cc} 3 & -2 \\ 2 & 3 \end{array}\right|=13      \notag
\end{equation}
copies of the atoms (sites) of a single primitive unit cell of the Lieb lattice. The minimal repetitive unit of the unit cell for this ribbon is bordered by the dashed lines in Fig. 1, from which we have $n_{uc}=3$ and $n_{\perp}=2$. According to Eq.(11), there are $n_{\perp}^{2}|\tilde{s}|=2^{2}\times13=52$ distinct edge terminations. More explicitly, from Fig. 1, the two $n_{\perp}=2$ factors count the types of atomic chains that terminate the two edges. For each of the $n_{\perp}^{2}=4$ combinations of such edge terminations, the relative positions of the two outermost atomic chains can occur in $|\tilde{s}|=13$ ways according to their alignment along the direction parallel to $\mathbf{v}_{st}$.

\begin{figure}[!htb]\label{fig1} \centering
\includegraphics[width=6.6cm,height=6.52cm]{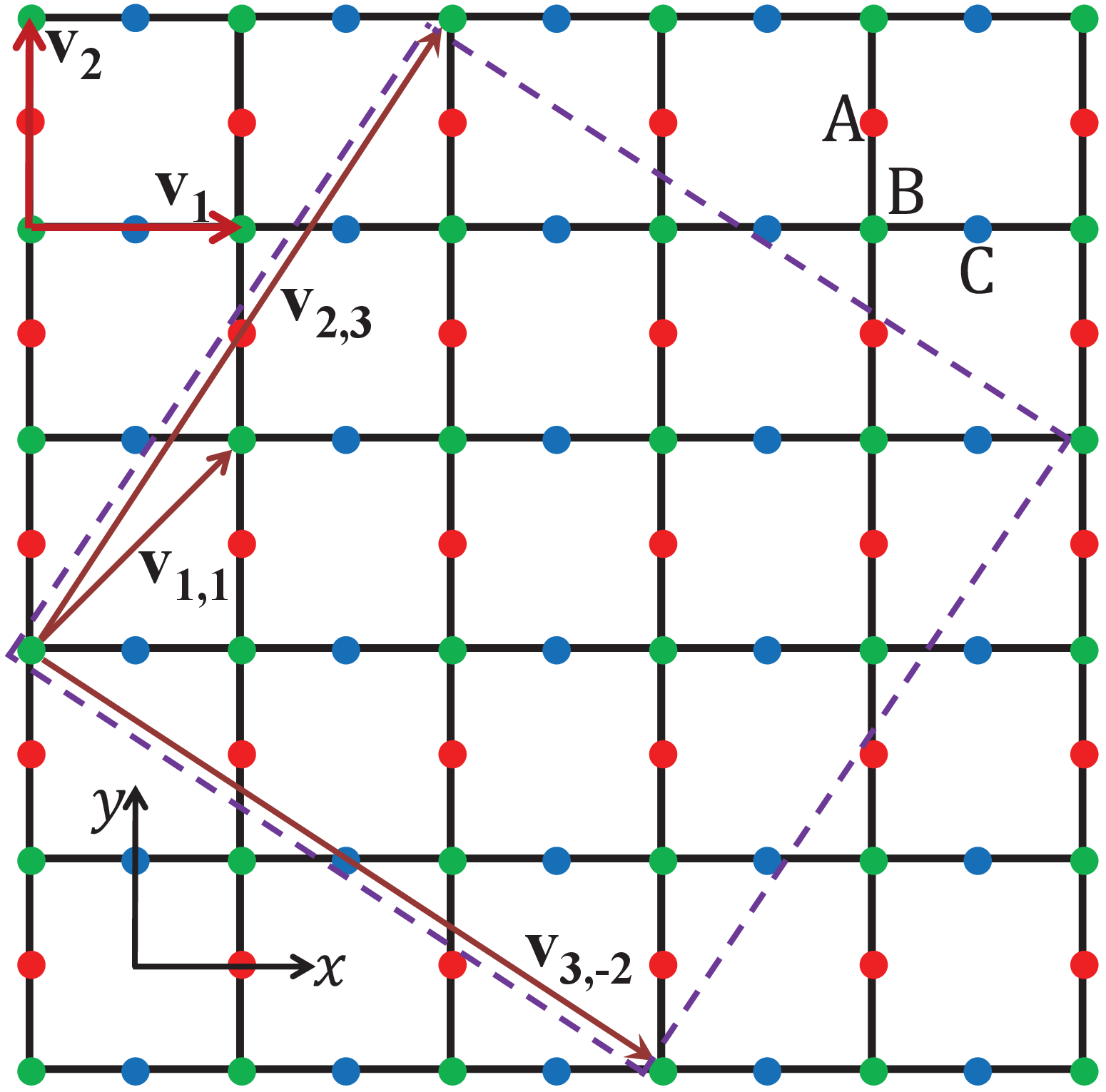}
\caption{Cutting of the Lieb lattice, with the two edges running parallel to $\mathbf{v}_{mn}=\mathbf{v}_{2,3}$, for which $\mathbf{v}_{pq}=\mathbf{v}_{1,1}$ and $\mathbf{v}_{st}=\mathbf{v}_{3,-2}$. The minimal repetitive unit of the rectangular unit cell is encircled by the purple dashed square.}
\end{figure}

52 is still too large for realistic discussions for the relationship between the edge terminations and other physical properties. In most theoretical discussions, focus are put on ribbons with the two edges parallel to a certain minimal primitive lattice vector. That is $\mathbf{v}_{mn}=\mathbf{v}_{1}$, $\mathbf{v}_{2}$, or another minimal primitive lattice vector related to them through symmetry operations. For the Lieb lattice, this gives ribbons with edges parallel to $\mathbf{v}_{1}$ or $\mathbf{v}_{2}$. Clearly we have $n_{\perp}=2$, $|\tilde{s}|=1$, and so the number of distinct edge terminations is $2^{2}\times1=4$.

\begin{figure}[!htb]\label{fig2} \centering
\hspace{-2.95cm} {\textbf{(a)}} \hspace{3.8cm}{\textbf{(b)}}\\
\hspace{0cm}\includegraphics[width=4.2cm,height=4.31cm]{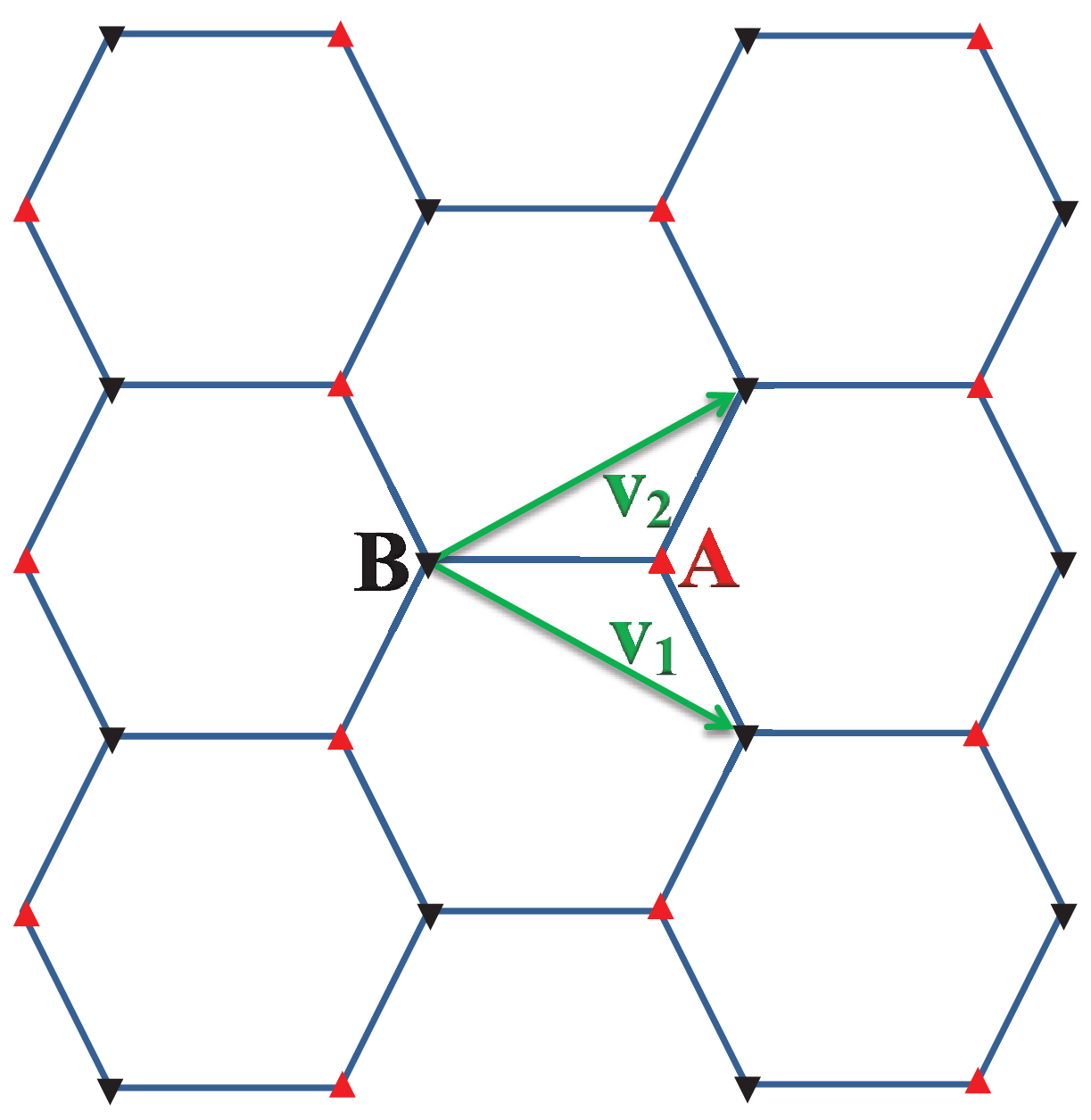}
\includegraphics[width=4.2cm,height=4.31cm]{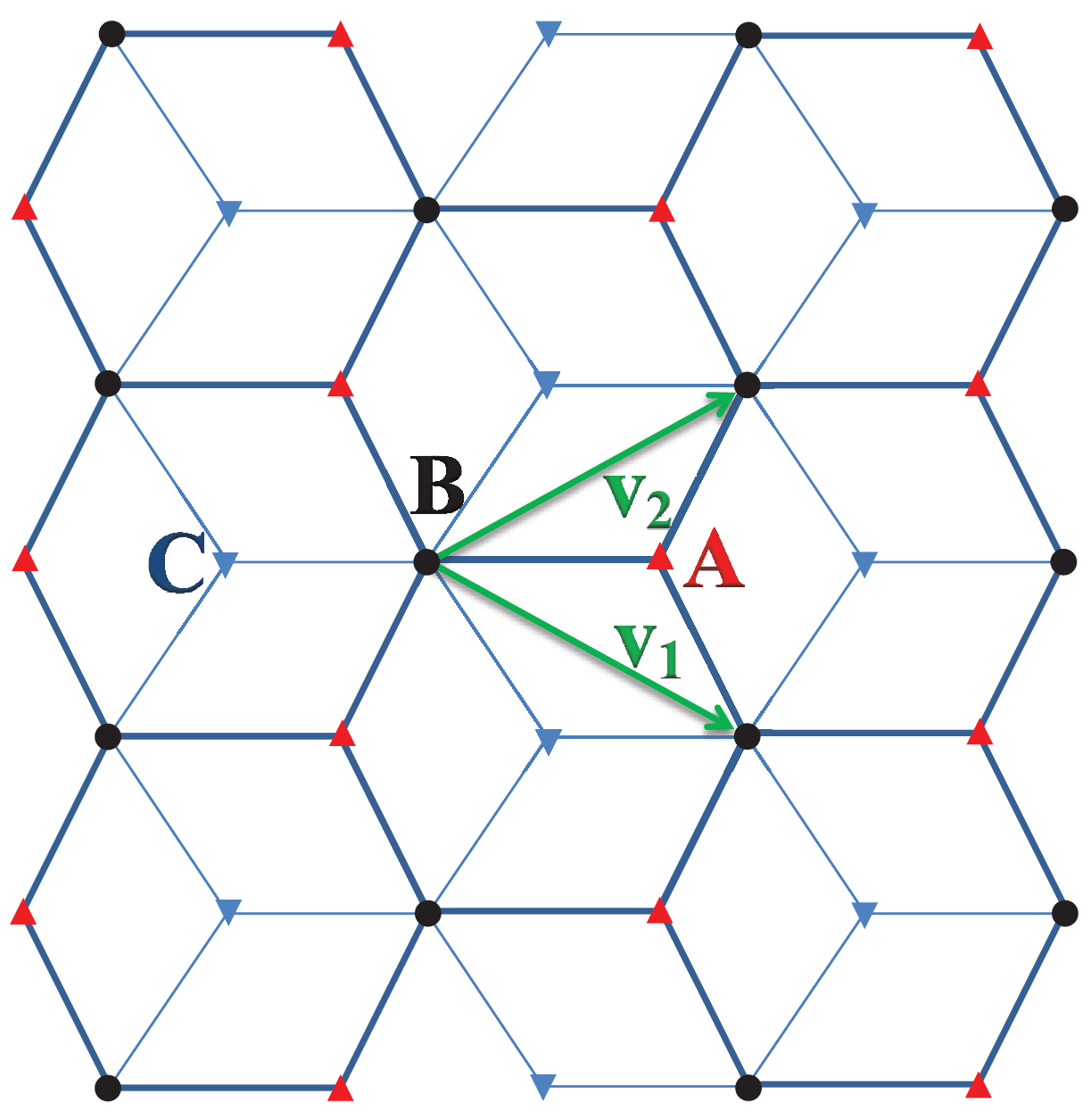}
\caption{The lattice and the minimal primitive lattice vectors for (a) the honeycomb lattice, and (b) the dice lattice. $\mathbf{v}_{1}=(\frac{\sqrt{3}}{2},-\frac{1}{2})a$ and $\mathbf{v}_{2}=(\frac{\sqrt{3}}{2},\frac{1}{2})a$. The honeycomb lattice has two sublattices A and B. The dice lattice has three sublattices, including the hub sublattice B, and two rim sublattices A and C.}
\end{figure}

For the honeycomb lattice and the dice lattice, which are both hexagonal lattices according to Fig. 2, ribbons with zigzag and armchair edges have attracted most of the attention. For zigzag ribbons of the honeycomb lattice, $\mathbf{v}_{mn}=\mathbf{v}_{-1,1}$, $\mathbf{v}_{pq}=\mathbf{v}_{1,0}$ or $\mathbf{v}_{0,1}$, and $\mathbf{v}_{st}=\mathbf{v}_{1,1}$. $\tilde{s}=sn-tm=2$ and $n_{\perp}=2$. According to Eq.(11), there are $2^{2}\times2=8$ distinct types of regular zigzag ribbons. Among these 8 configurations, two have a pair of closed zigzag-type edges, two have a pair of open bearded-type edges, the remaining 4 have one zigzag-type edge and one bearded-type edge.

For the regular ribbons of the dice lattice with two edges along a zigzag direction, hereafter called zigzag ribbons for simplicity, the three characteristic lattice vectors, $\mathbf{v}_{mn}$, $\mathbf{v}_{pq}$, and $\mathbf{v}_{st}$, may be taken the same as those for the honeycomb lattice. It is clear from Fig. 2(b) that $n_{\perp}=3$. The number of distinct regular zigzag ribbons of the dice lattice is thus $3^{2}\times2=18$.

For the armchair ribbons of the honeycomb lattice and the dice lattice, we may take $\mathbf{v}_{mn}=\mathbf{v}_{1,1}$, $\mathbf{v}_{pq}=\mathbf{v}_{1,0}$ or $\mathbf{v}_{0,-1}$, and $\mathbf{v}_{st}=\mathbf{v}_{1,-1}$. We have $\tilde{s}=2$ and $n_{\perp}=1$ for both the honeycomb lattice and the dice lattice. Therefore, there are $1^{2}\times2=2$ distinct regular armchair ribbons for the honeycomb lattice and the dice lattice, according to the edge termination morphologies.

Overall, the regular zigzag ribbons of the honeycomb lattice and the dice lattice, whose edges are parallel to a zigzag direction of the lattices, constitute the simplest nontrivial examples of Eq.(11) in the sense that on one hand both $n_{\perp}>1$ and $|\tilde{s}|>1$, and on the other hand they are among the smallest possible realizations of the inequality.

\begin{figure}[!htb]\label{fig3}
\centering
\hspace{-6.6cm} {\textbf{(a)}}\\
\includegraphics[width=6.5cm,height=5.67cm]{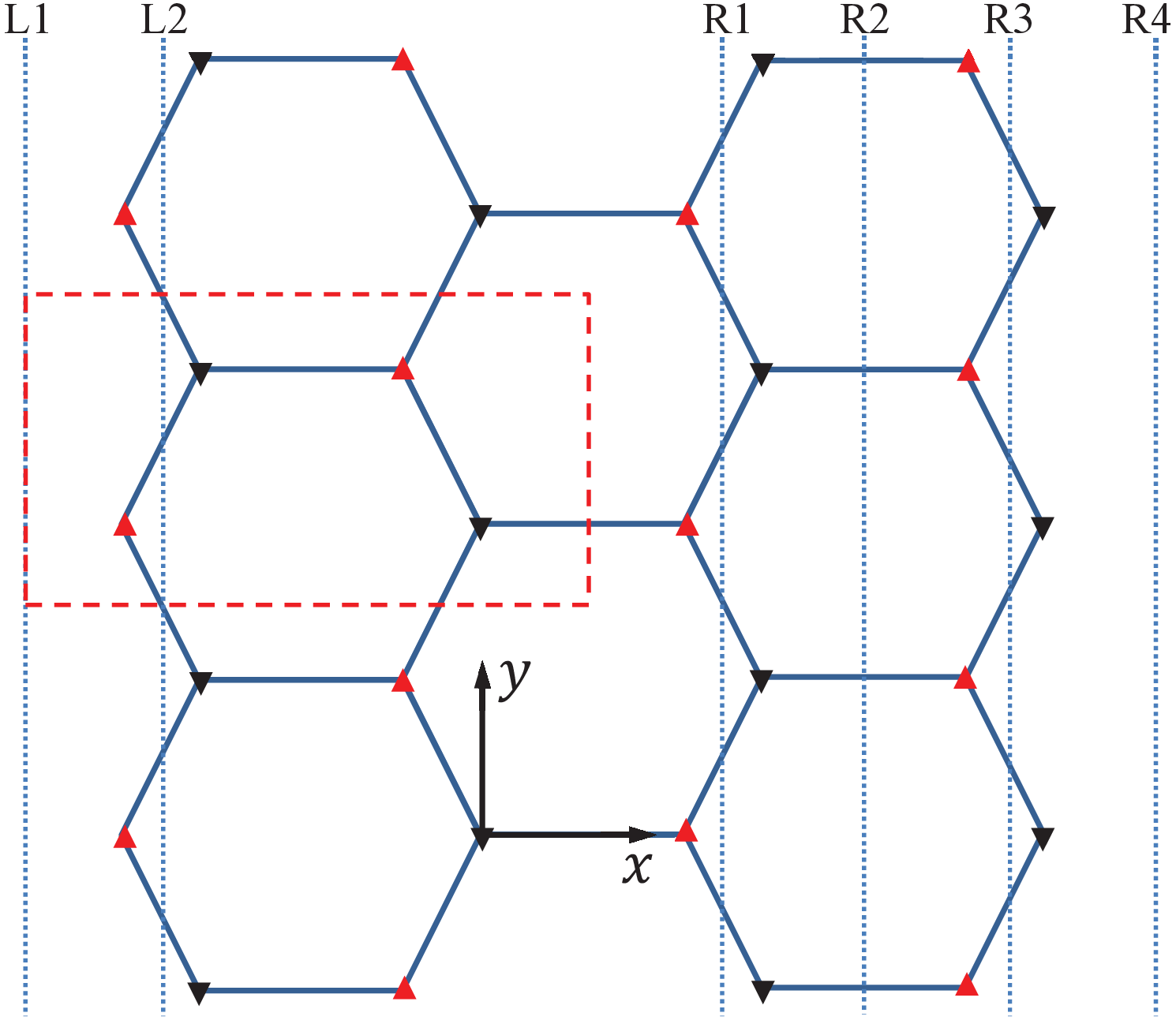} \\ \vspace{-0.05cm}
\hspace{-6.6cm} {\textbf{(b)}}\\
\includegraphics[width=7.0cm,height=6.12cm]{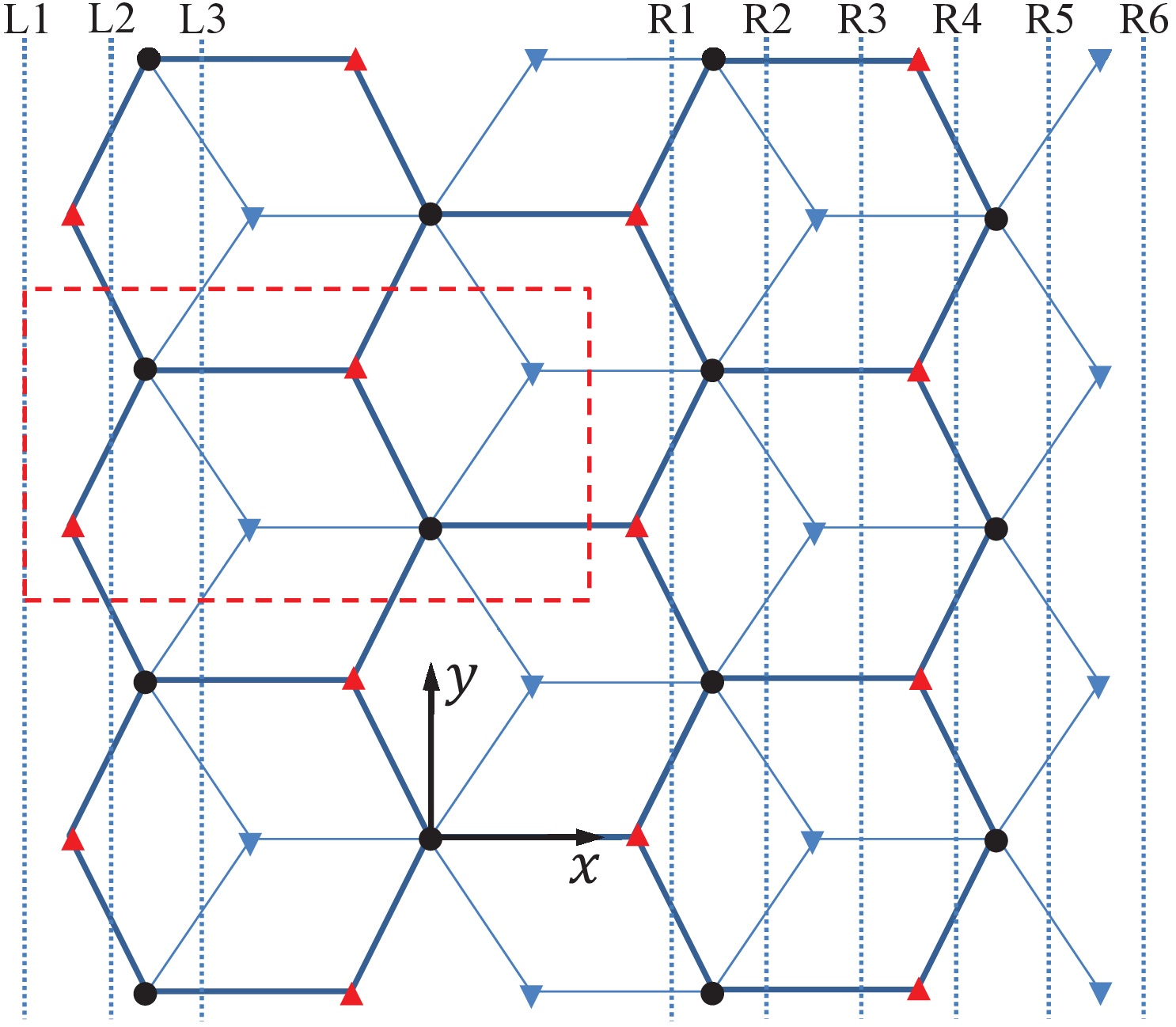}  \\
\caption{The distinct edge terminations of zigzag regular ribbons for (a) the honeycomb lattice, and (b) the dice lattice. The vertical dotted lines mark the optional positions of the two edges. A ribbon between the two edges L$_{\alpha}$ and R$_{\beta}$ is denoted as L$_{\alpha}$R$_{\beta}$. The region surrounded by the red dashed rectangle is one of the two (i.e., $|\tilde{s}|=2$) choices for the minimal repetitive unit of the unit cell. The section of the ribbon covered by translating the minimal repetitive unit of the unit cell along $+y$ and $-y$ directions is the minimal repetitive unit of the ribbon along its width. For the ribbon with a specific kind of edge termination, its width may be increased (decreased) by adding (cutting off) an integral multiple of the minimal repetitive unit of the ribbon to (from) the left.}
\end{figure}

As shown in Fig.3 are the distinct edge termination combinations for the regular ribbons of the honeycomb lattice and the dice lattice, with the two edges parallel to the zigzag chains of the lattices along the $y$ axis. Among the two edges of the ribbons, we apply the translational invariance of the ribbon along $y$ to the left edge, to reduce the number of distinct edge termination configurations of the left edge from $n_{\perp}|\tilde{s}|$ to $n_{\perp}$. For the right edge, there are $n_{\perp}|\tilde{s}|$ distinct edge termination configurations. The $n_{\perp}$ ($n_{\perp}|\tilde{s}|$) different terminations on the left (right) edge of the ribbons are marked by $n_{\perp}$ ($n_{\perp}|\tilde{s}|$) dotted straight lines that do not cross any site and are parallel to the $y$ axis. The lines at the left and right edges of the ribbon are separately denoted as L$_{\alpha}$ ($\alpha=1,...,n_{\perp}$) and R$_{\beta}$ ($\beta=1,...,n_{\perp}|\tilde{s}|$). Each distinct edge termination morphology is thus represented as L$_{\alpha}$R$_{\beta}$ ($\alpha=1,...,n_{\perp}$; $\beta=1,...,n_{\perp}|\tilde{s}|$). For ribbons of the honeycomb lattice, L$_{1}$R$_{2}$ and L$_{1}$R$_{4}$ are the two kinds of zigzag ribbons with odd and even numbers of zigzag chains \cite{fujita96}, L$_{2}$R$_{1}$ and L$_{2}$R$_{3}$ are the ribbons with two bearded-type edges \cite{klein94}, the remaining four edge termination morphologies give zigzag ribbons with one zigzag-type edge and one bearded-type edge. For the 18 zigzag ribbons of the dice lattice shown in Fig.3(b), we may also distinguish between the zigzag-type edges and the bearded-type edges. For example, L$_{3}$ gives bearded-type edge (the C sublattice) on the left edge, R$_{1}$ and R$_{4}$ give bearded-type edges (the A sublattice) on the right edge.

Let us consider the regular ribbons defined in Figs. 3(a) and 3(b).
Suppose we move an extremely thin cutting blade that is parallel to $\mathbf{v}_{mn}$ at a constant speed perpendicular to $\mathbf{v}_{mn}$ (i.e., moving parallel to $\mathbf{v}_{st}$), and let us consider pointlike atoms on the lattice sites. For ribbons of Fig. 3(a), the chance of getting a zigzag-type edge is twice as large as that of getting a bearded-type edge. This justifies the popularity of the L$_{1}$R$_{2}$ and L$_{1}$R$_{4}$ ribbons of the honeycomb lattice over the other ribbons with one or two bearded-type edges. In contrast, for the dice lattice, we have equal chances of creating any of the six types of terminations. In this sense, all the $18$ types of zigzag ribbons defined in Fig.3(b) for the dice lattice are equally probable and should be considered on an equal footing.

\section{Zigzag ribbons of the dice lattice: Correspondences between edge morphologies and electronic spectra}

Electronic spectrum of a ribbon underlies many other properties such as transport and phase transitions. As regards the above results for the distinct edge termination morphologies, a natural question is how many qualitatively distinct electronic spectra exist among the $n_{\perp}^{2}|\tilde{s}|$ groups of regular ribbons sharing the same edge orientation. In this section, we study this correspondence for the zigzag ribbons of the dice lattice. We numerically calculate the electronic spectra for the various zigzag ribbons and leave the analytical analysis to the next section.

From Sec.III, we have $n_{\perp}^{2}|\tilde{s}|$=$3^2$$\times$2=18 distinct kinds of zigzag dice lattice ribbons, which separate into nine pairs (corresponding to $|\tilde{s}|$=2). In each pair of ribbon types whose left (right) edges are both terminated with sites of the $\alpha$-th ($\beta$-th) sublattice, the atoms on the two edges are aligned in the direction perpendicular to the edge in one case but misaligned in the other case. We label these two cases as $\alpha\beta$-in and $\alpha\beta$-off, respectively. $\alpha$ and $\beta$ take values among A, B, and C. For examples, referring to Fig. 3(b), the L$_1$R$_1$ and L$_1$R$_4$ ribbons are separately AA-in and AA-off ribbons, the L$_1$R$_5$ and L$_1$R$_2$ ribbons are separately AB-in and AB-off ribbons, and so on.

According to Fig.3(b), we measure the width of a regular zigzag ribbon in terms of the number of atomic chains parallel to the $y$ axis, hereafter called $y$-chains. Each atomic $y$-chain has a width of $\frac{1}{2\sqrt{3}}a$, where $a$ is the length of the primitive lattice vectors defined in Fig. 2(b). Along the width of the ribbon (i.e., along $\mathbf{v}_{st}$), the atomic chains repeat by every six ($n_{\perp}|\tilde{s}|=6$) $y$-chains. After an integer number $N_{x}$ of repetitions of the units with six $y$-chains, there may be an incomplete unit with $n_{x}$ atomic $y$-chains. $n_{x}$ ranges from 0 to 5 and is determined uniquely by the type of the edge termination. The values of $n_{x}$ for $\alpha\beta$-in and $\alpha\beta$-off ribbons differ by 3, corresponding to one $y$-chain of A sublattice sites, one $y$-chain of B sublattice sites, and one $y$-chain of C sublattice sites. The total width of the ribbon is
\begin{equation}
(6N_{x}+n_{x})\frac{a}{2\sqrt{3}}=N_{s}\frac{a}{2\sqrt{3}},
\end{equation}
where we have defined $N_{s}$ to represent the total number of sites within a unit cell of the ribbon. From Fig. 3(b), $n_{x}=1$ for $\alpha\alpha$-in ($\alpha$$=$A, B, C). $n_{x}=3$ for AC-in, CB-in, and BA-in. $n_{x}=5$ for AB-in, BC-in, and CA-in. The value of $n_{x}$ for $\alpha\beta$-off is obtained from the $n_{x}$ for $\alpha\beta$-in by adding or subtracting 3, so that the integer $n_{x}$ ranges from 0 to 5.

Several members of the $18$ kinds of zigzag dice lattice ribbons have been studied in previous works \cite{xu17,oriekhov18,bugaiko19,chen19,alam19,soni20,tan21}. Xu et al studied the spectrum of the AC-off zigzag ribbons, for both the pure dice model and the dice model with an on-site energy term that amounts to a symmetric bias \cite{xu17}. Chen et al studied the influence of magnetic field on the spectrum and optical conductance of the zigzag ribbons, and separated the zigzag ribbons into two classes \cite{chen19}. Alam et al studied six types of zigzag ribbons (BA-in, BA-off, BC-in, BC-off, BB-in and BB-off) of the pure dice model and found two types of spectra according to whether the low-energy spectrum is gapped or gapless \cite{alam19}. Soni et al studied the AC-in and AC-off ribbons of the dice lattice, for the dice model supplemented by the Rashba spin-orbit coupling, the Zemann term, and an on-site energy term of the B sublattice sites \cite{soni20}. Oriekhov et al \cite{oriekhov18,bugaiko19} studied the low-energy spectrum of $9=n_{\perp}^{2}$ configurations of the zigzag dice lattice ribbons, which is up to now the most complete study of the zigzag dice lattice ribbons. Oriekhov et al did not differentiate the $\alpha\beta$-in and $\alpha\beta$-off configurations. By focusing on the low-energy spectrum except the flat bands for the pure dice model, they also identified two types of spectra, gapped or gapless. On the other hand, their numerical results for several kinds of zigzag ribbons (CA, AC, and BB) exhibit clear differences in the in-gap states close to the boundary of the Brillouin zone (BZ) \cite{oriekhov18}. It is unclear presently how many distinct types of electronic spectra exist among the 18 types of zigzag dice lattice ribbons, if both the low-energy spectrum and the higher energy in-gap states are taken into account.

Since on one hand previous studies have shown in-gap states in the electronic spectra of some zigzag ribbons of the dice model \cite{xu17,oriekhov18,chen19,alam19}, and on the other hand the flat bands are distinctive features of the electronic spectrum of the present system \cite{sutherland86,xu17,oriekhov18,bugaiko19,chen19,alam19}, we will take these two features as the criteria to differentiate the quasi-1D band structures of the 18 kinds of zigzag dice lattice ribbons. We consider wide regular ribbons in this work, so that the finite-size effect to the qualitative aspects of the band structures of the 18 kinds of ribbons has diminished to be negligible.

\begin{figure*}[!htb]\label{fig4} \centering
\hspace{-4.0cm} {\textbf{(a)}} \hspace{5.10cm}{\textbf{(b)}}  \hspace{5.10cm}{\textbf{(c)}}\\
\hspace{0cm}\includegraphics[width=5.5cm,height=4.24cm]{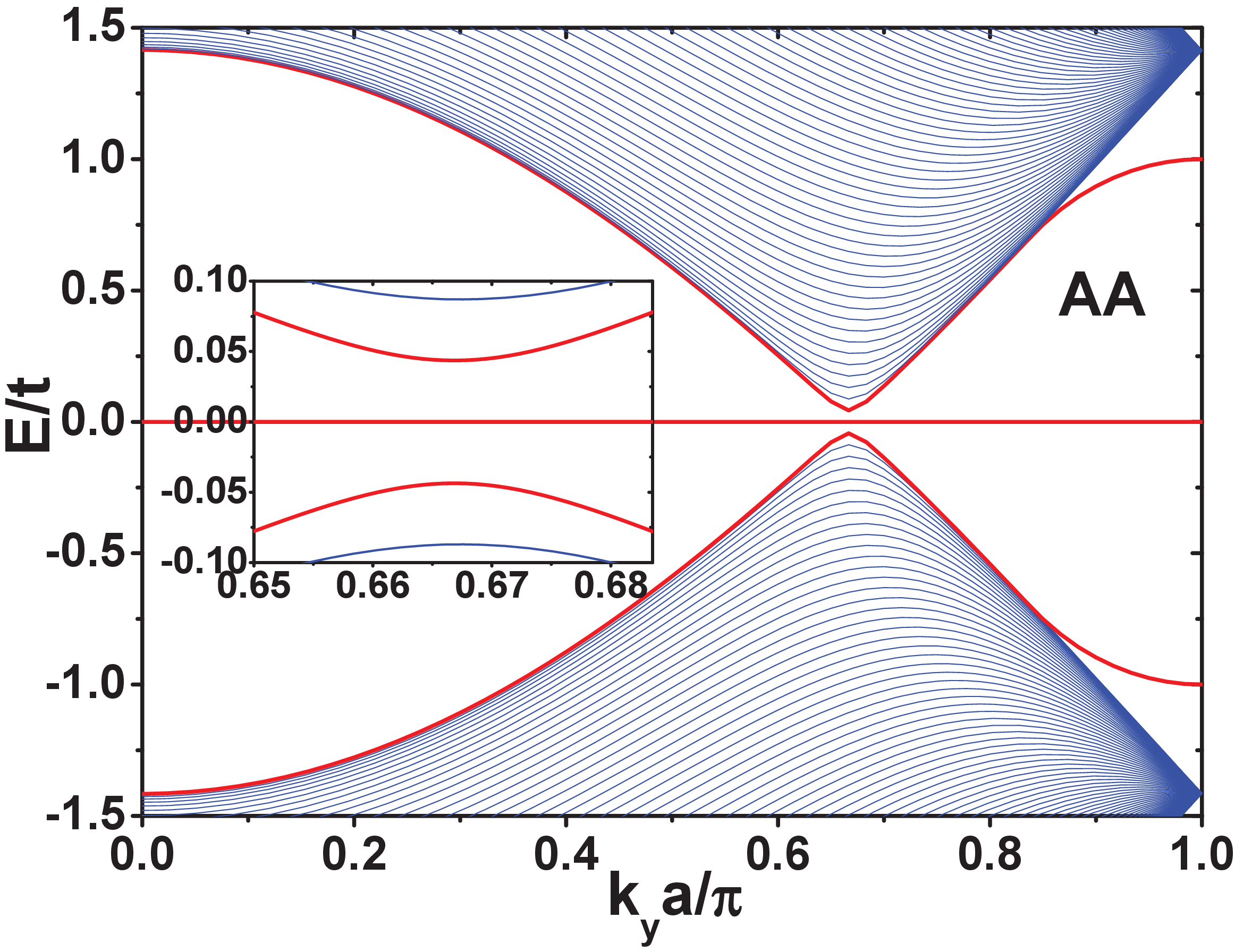}
\includegraphics[width=5.5cm,height=4.24cm]{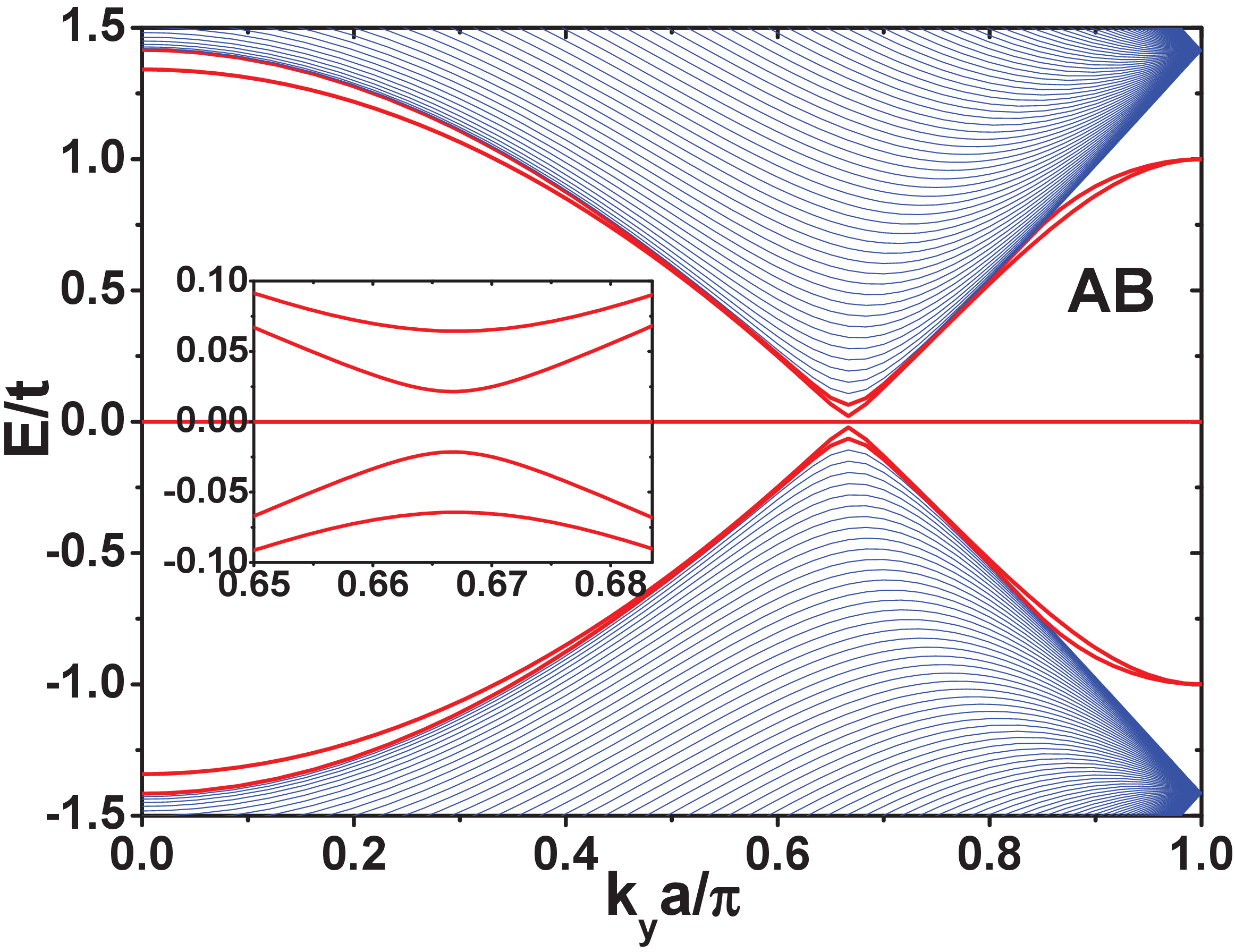}
\includegraphics[width=5.5cm,height=4.24cm]{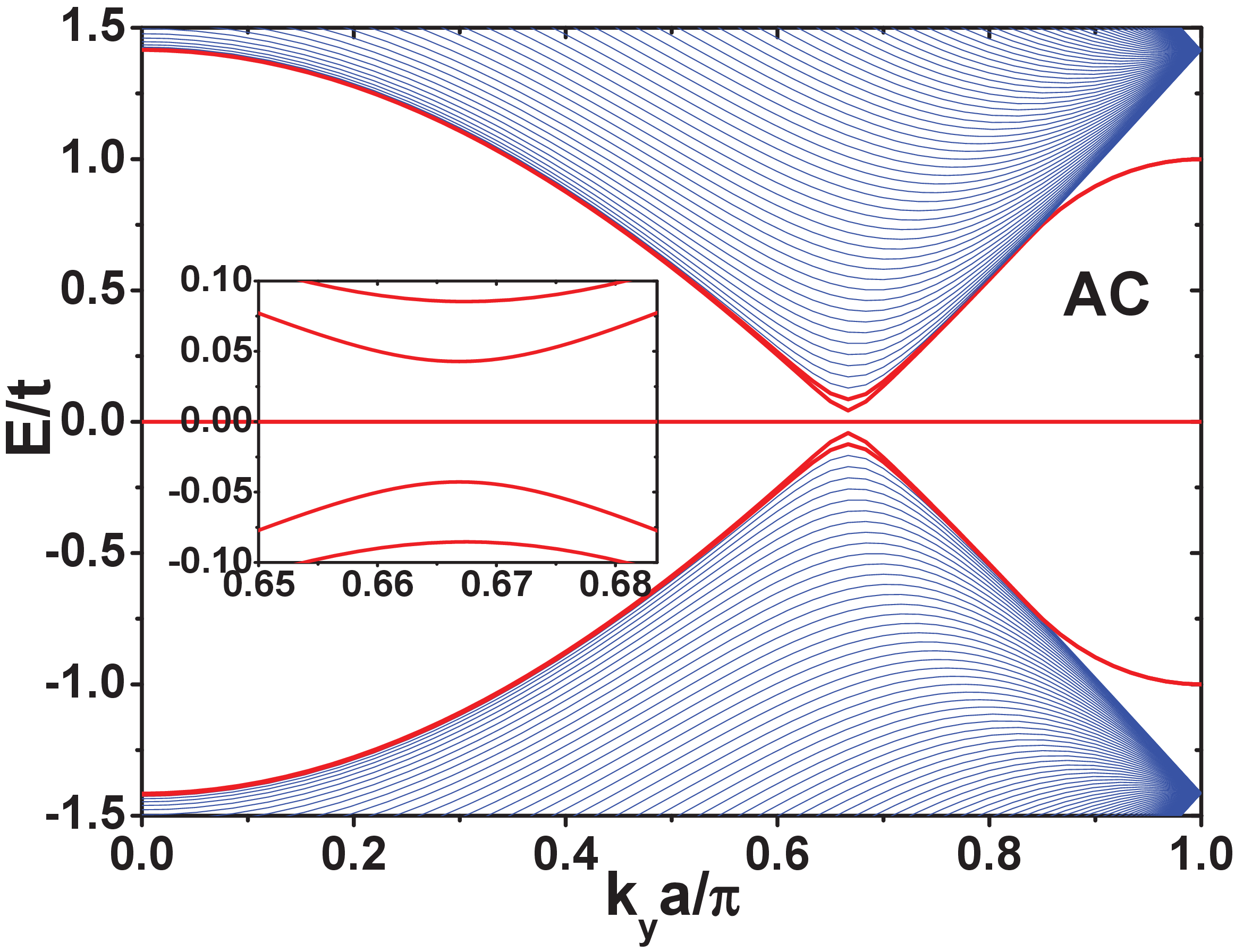}\\
\vspace{-0.10cm}
\hspace{-4.0cm} {\textbf{(d)}} \hspace{5.1cm}{\textbf{(e)}} \hspace{5.1cm}{\textbf{(f)}}\\
\hspace{0cm}\includegraphics[width=5.5cm,height=4.24cm]{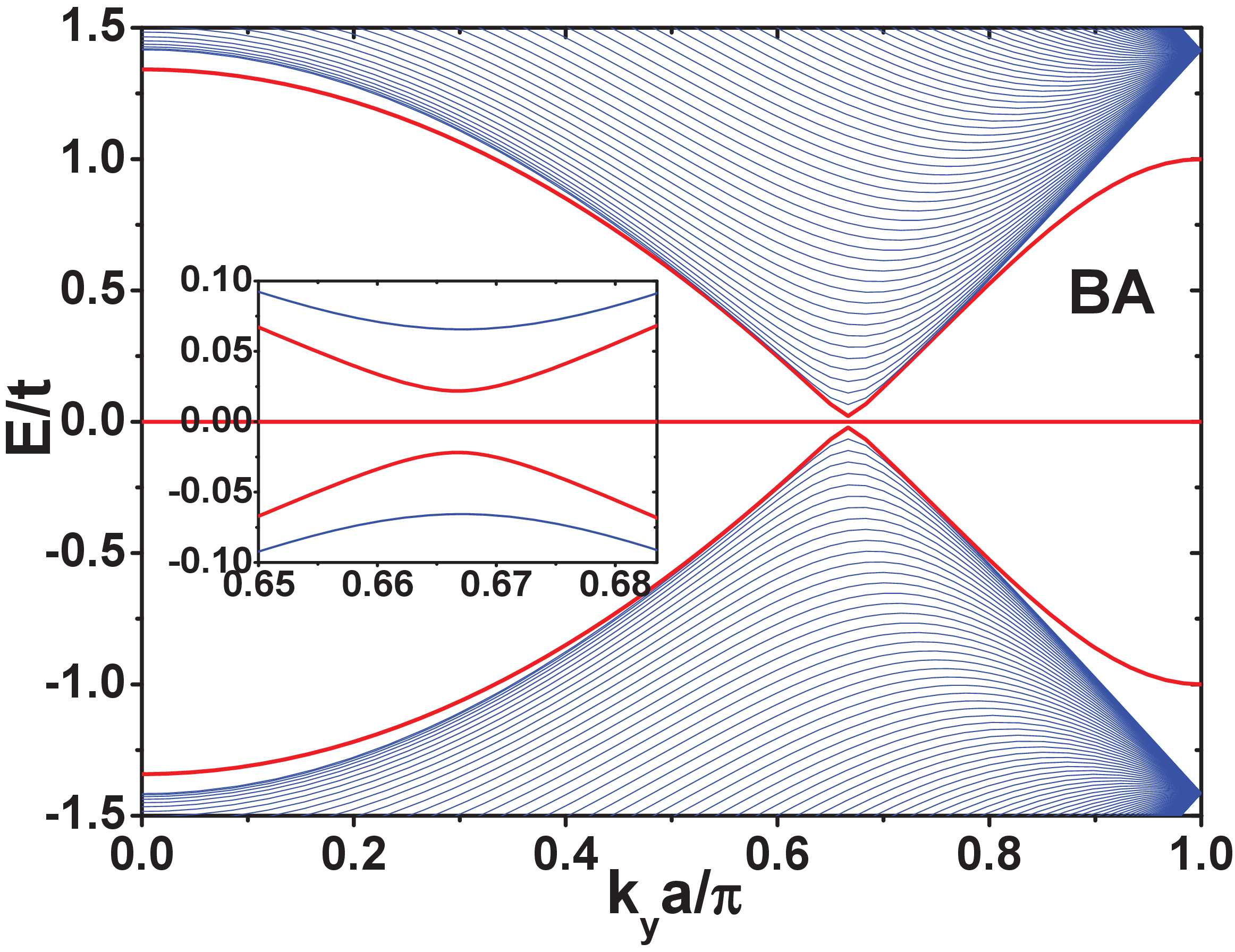}
\includegraphics[width=5.5cm,height=4.24cm]{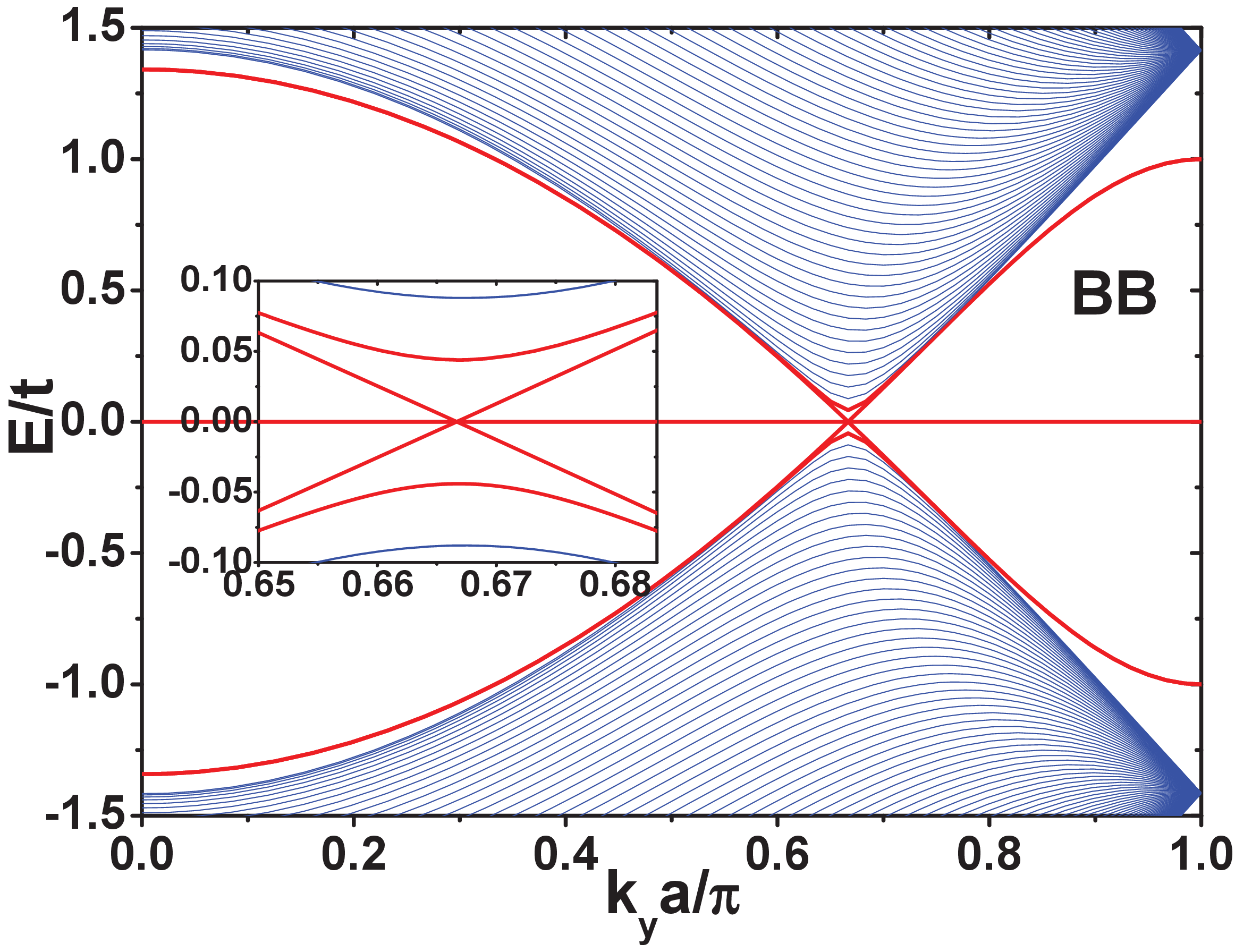}
\includegraphics[width=5.5cm,height=4.24cm]{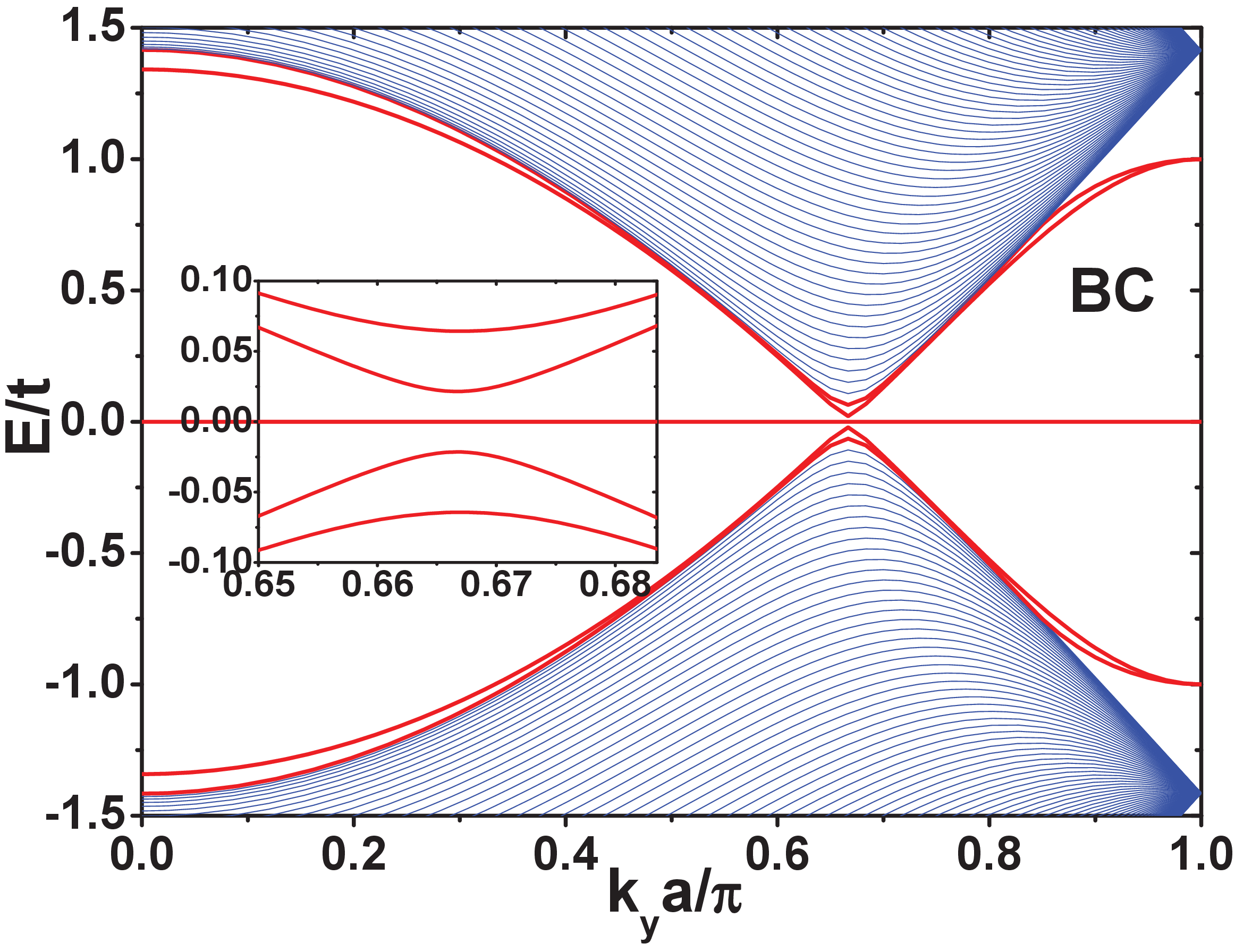}\\
\vspace{-0.10cm}
\hspace{-4.0cm} {\textbf{(g)}} \hspace{5.1cm}{\textbf{(h)}} \hspace{5.1cm}{\textbf{(i)}}\\
\hspace{0cm}\includegraphics[width=5.5cm,height=4.24cm]{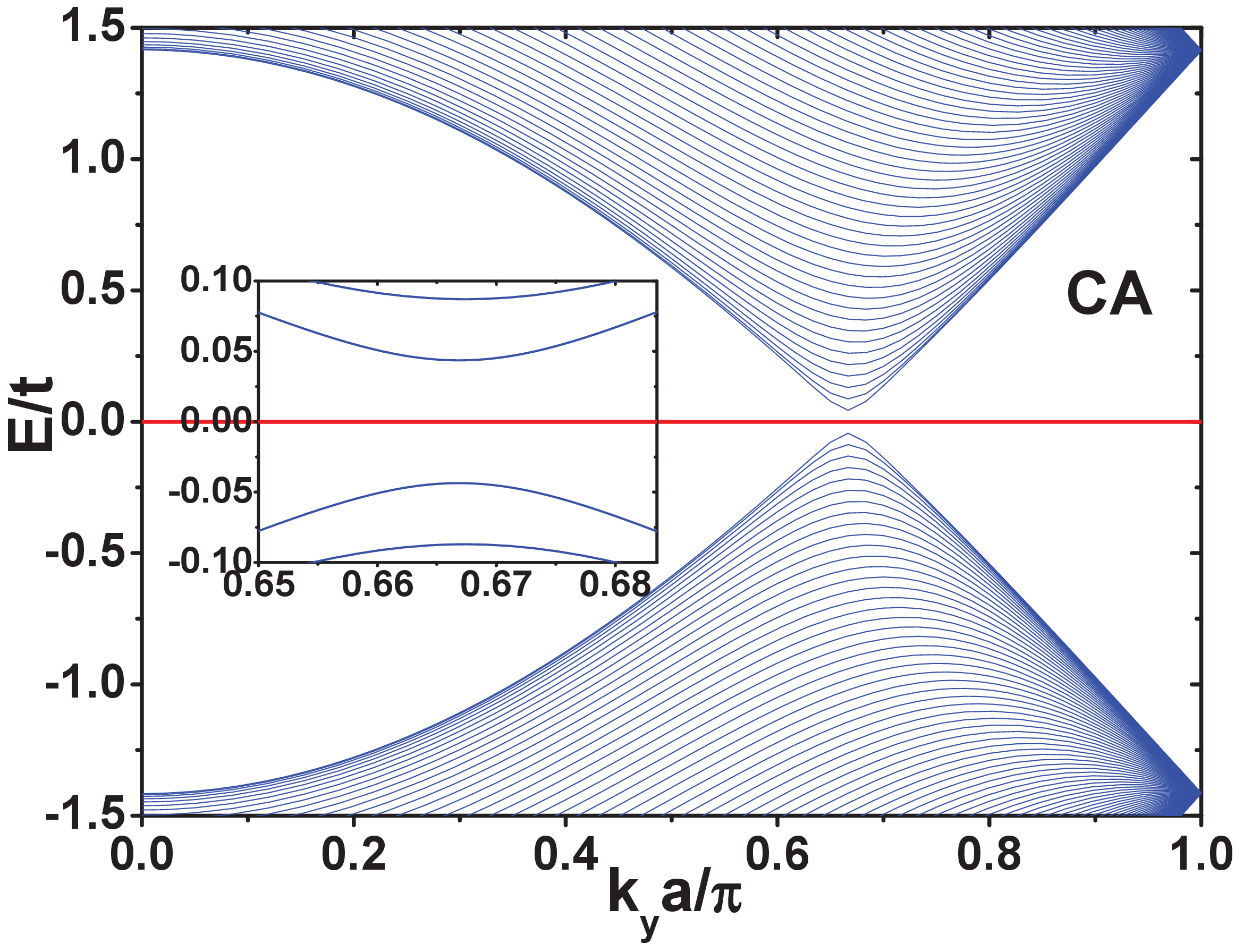}
\includegraphics[width=5.5cm,height=4.24cm]{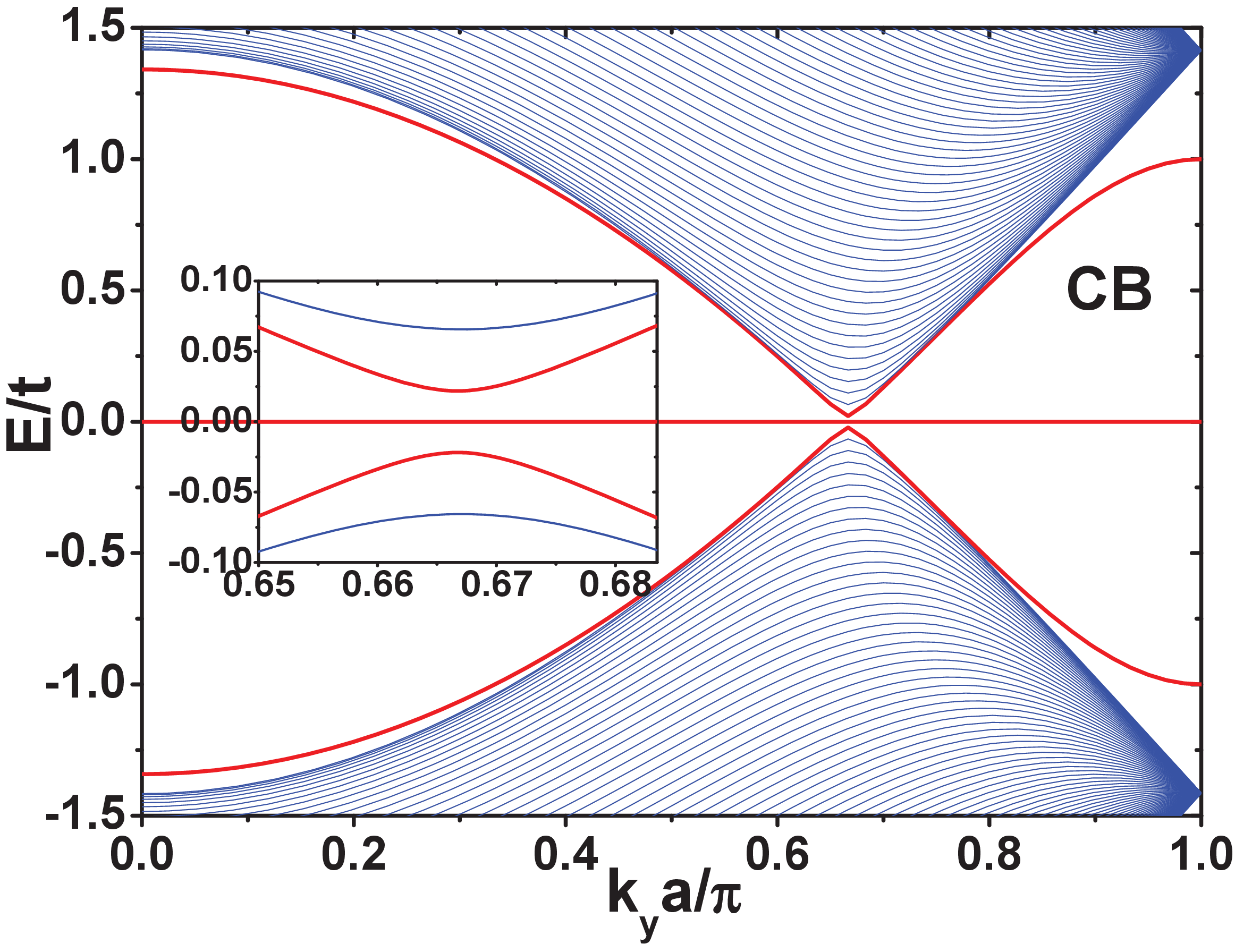}
\includegraphics[width=5.5cm,height=4.24cm]{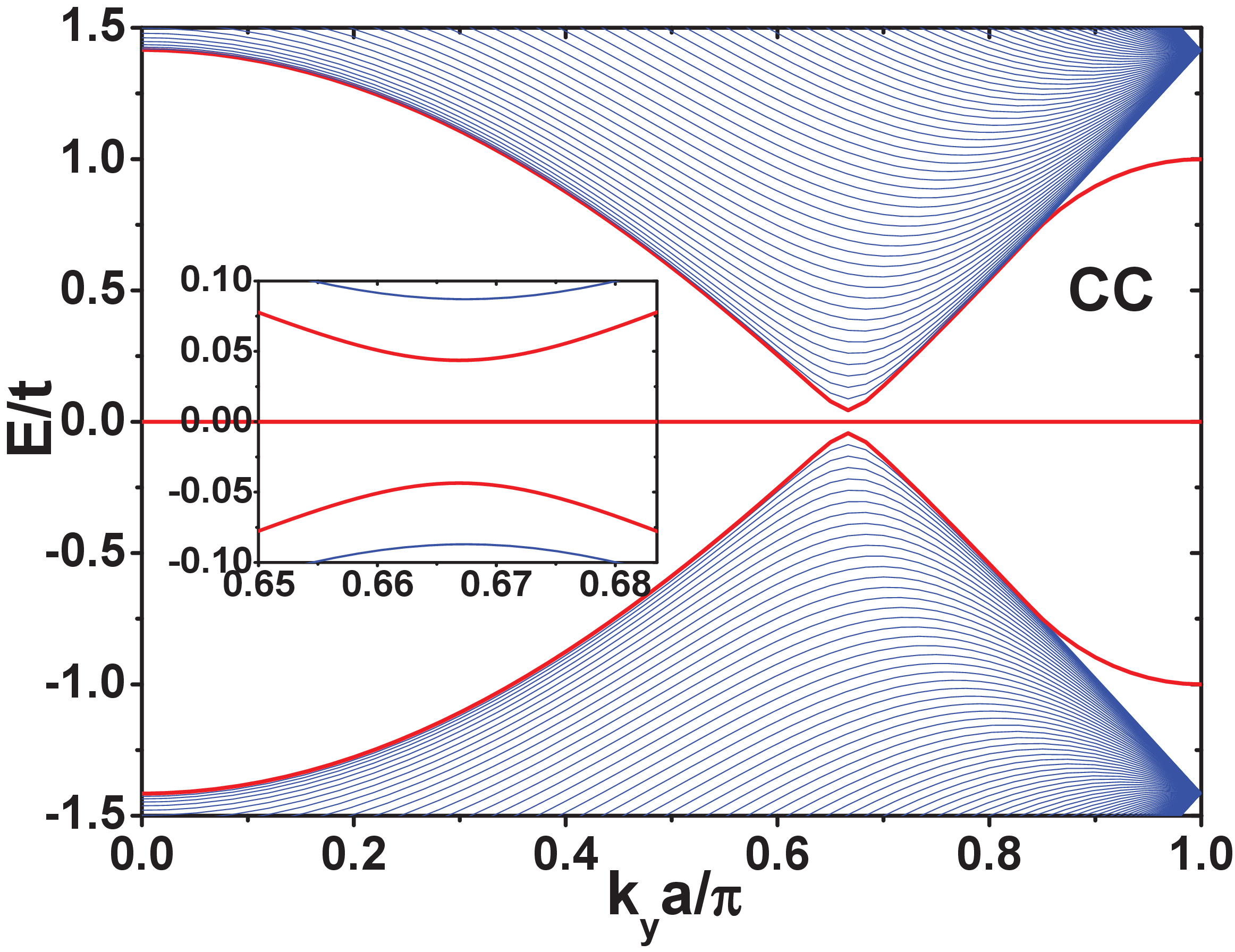}
\caption{The electronic spectra for the $\alpha\beta$-in zigzag ribbons ($\alpha$, $\beta$= A, B, C) of the pure dice model, for $t=1$, $\Delta=0$, and $N_{x}=50$. Only the results for half of the 1D BZ are shown. The zero-energy flat bands and the bands involving the in-gap states are thickened. The inset of each figure is the magnified display of the low-energy portion of the energy spectra close to $k_{y}a=2\pi/3$.}
\end{figure*}

We consider the following tight-binding model for the electrons in the dice lattice
\begin{eqnarray}
\hat{H}&=&t\sum\limits_{\langle i,j\rangle}(b^{\dagger}_{i}a_{j}
+b^{\dagger}_{i}c_{j}+\text{H.c.})   \notag \\
&&+\Delta\sum\limits_{i}(a^{\dagger}_{i}a_{i}
-c^{\dagger}_{i}c_{i}).
\end{eqnarray}
$a_{i}$, $b_{i}$, and $c_{i}$ separately annihilates an electron on the A, B, and C sublattices of the $i$-th unit cell of the bulk lattice. H.c. means the Hermitian conjugate of the terms explicitly written out. Since the model is spin-independent, we have neglected the dummy spin index in the model. As a result, we also neglect the two-fold spin degeneracy of the bands. The first term is the nearest-neighboring (NN) hopping term. The summation $\langle i,j\rangle$ runs over NN intersublattice bonds, which are connected by the NN vectors $\boldsymbol{\delta}_{1}=(-1,0)a_{0}$, $\boldsymbol{\delta}_{2}=(\frac{1}{2},-\frac{\sqrt{3}}{2})a_{0}$, or $\boldsymbol{\delta}_{3}=(\frac{1}{2},\frac{\sqrt{3}}{2})a_{0}$. $a_{0}$ is the length of an NN bond. The second term is the on-site energy term representing a symmetric bias of the two rim sublattices A and C relative to the hub sublattice B. The symmetric bias may be introduced in several ways. In dice model realized in a trilayer material \cite{wang11b}, the bias may be introduced by applying an electric field perpendicular to the layer plane. In dice model realized in an array of quantum dots \cite{sikdar21}, the bias term is implemented by biasing the A and C sublattices oppositely with reference to the B sublattice. In addition, the symmetric bias may be realized naturally in a tunable displaced optical dice lattice of cold atoms \cite{hao21}.

For the zigzag dice lattice ribbons defined in Fig.3(b), the primitive lattice vector is $\mathbf{v}_{mn}=\mathbf{v}_{-1,1}=(0,1)a$, where $a=\sqrt{3}a_{0}$. The 1D BZ of the zigzag ribbons is $k_{y}\in(-\frac{\pi}{a},\frac{\pi}{a}]$. The unit cell of the zigzag ribbons contain $N_{s}$ sites according to Eq.(12). The low-energy part of the bulk band structures of the pure dice model consists of a zero-energy flat band and two valleys, with each valley having a 2D Dirac cone \cite{sutherland86}. The two Dirac points of the bulk band structures project to $k_{y}=\pm\frac{2\pi}{3a}$ of the 1D BZ of the zigzag ribbons. For each $k_{y}$ of the 1D BZ, the spectrum of the zigzag ribbon is obtained by solving the eigenproblem of an $N_{s}\times N_{s}$ matrix $h(k_{y})$. This matrix consists of three kinds of terms: The diagonal on-site energies of each site in the unit cell, a matrix element of $t$ for each horizontal NN bond, and a matrix element of $2t\cos(\frac{k_{y}a}{2})\equiv t'$ for each NN bond that is not horizontal. $h(k_{y})=h(-k_{y})$ is an even function of $k_{y}$.

\begin{figure*}[!htb]\label{fig5} \centering
\hspace{-4.0cm} {\textbf{(a)}} \hspace{5.10cm}{\textbf{(b)}}  \hspace{5.10cm}{\textbf{(c)}}\\
\hspace{0cm}\includegraphics[width=5.5cm,height=4.24cm]{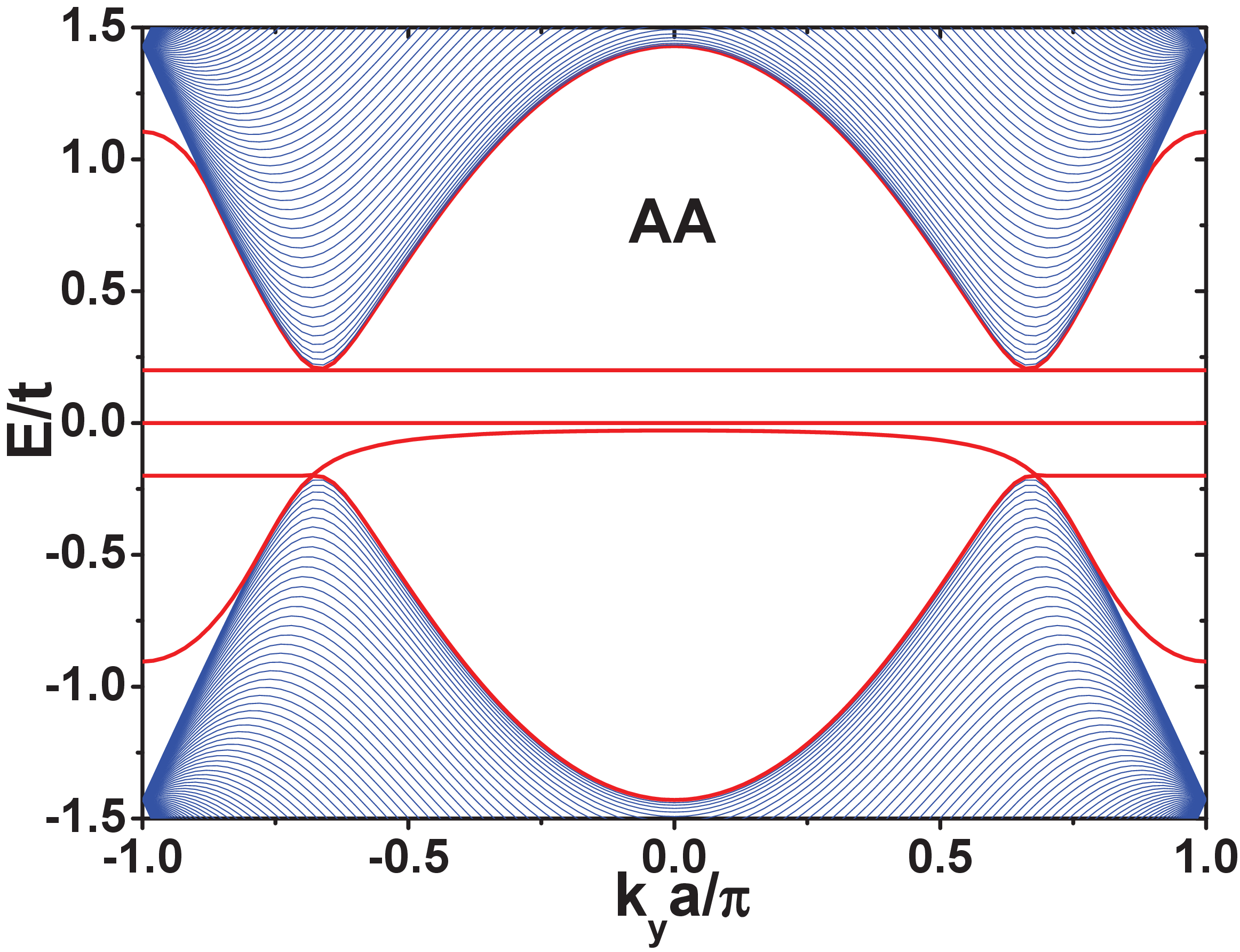}
\includegraphics[width=5.5cm,height=4.24cm]{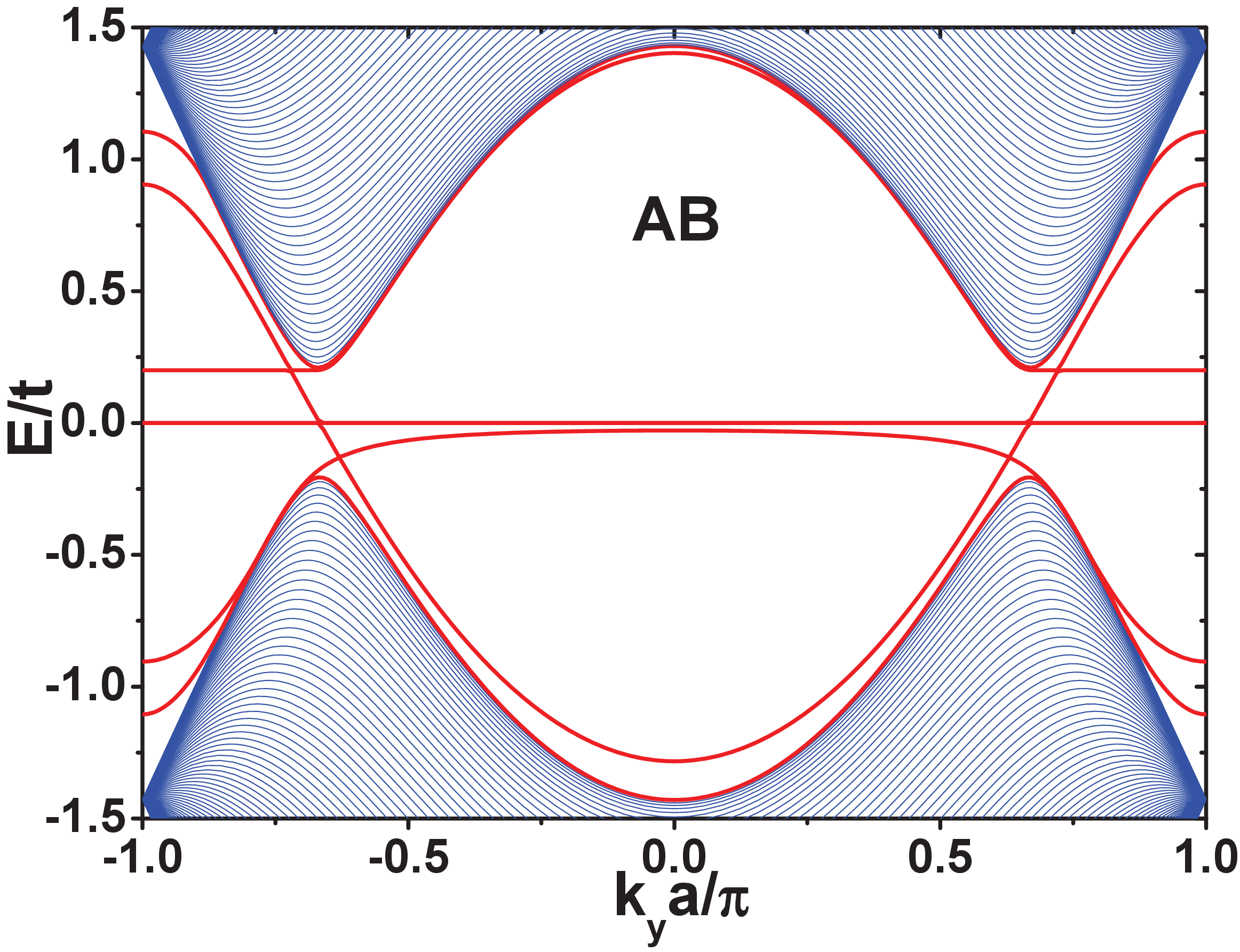}
\includegraphics[width=5.5cm,height=4.24cm]{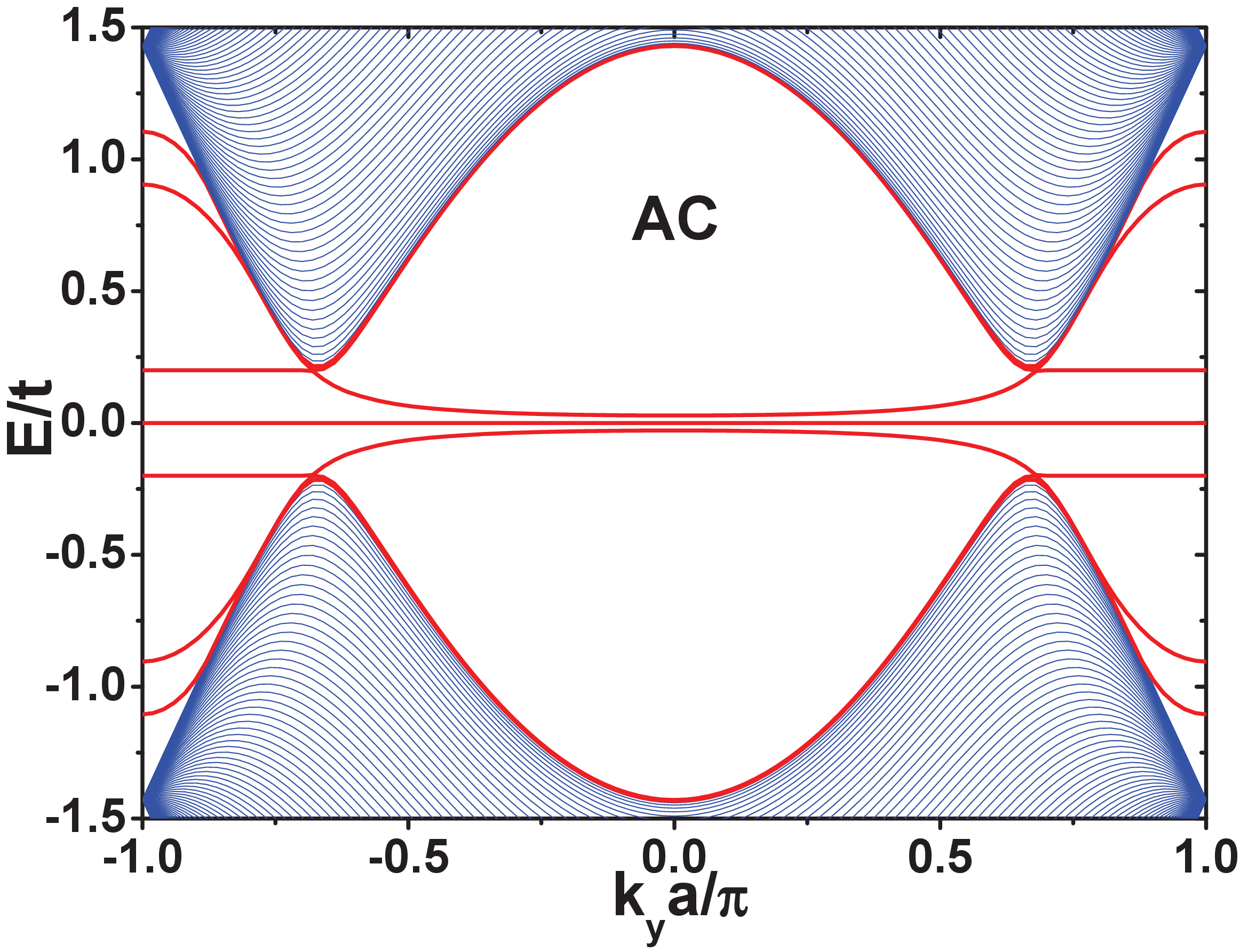}\\
\vspace{-0.10cm}
\hspace{-4.0cm} {\textbf{(d)}} \hspace{5.1cm}{\textbf{(e)}} \hspace{5.1cm}{\textbf{(f)}}\\
\hspace{0cm}\includegraphics[width=5.5cm,height=4.24cm]{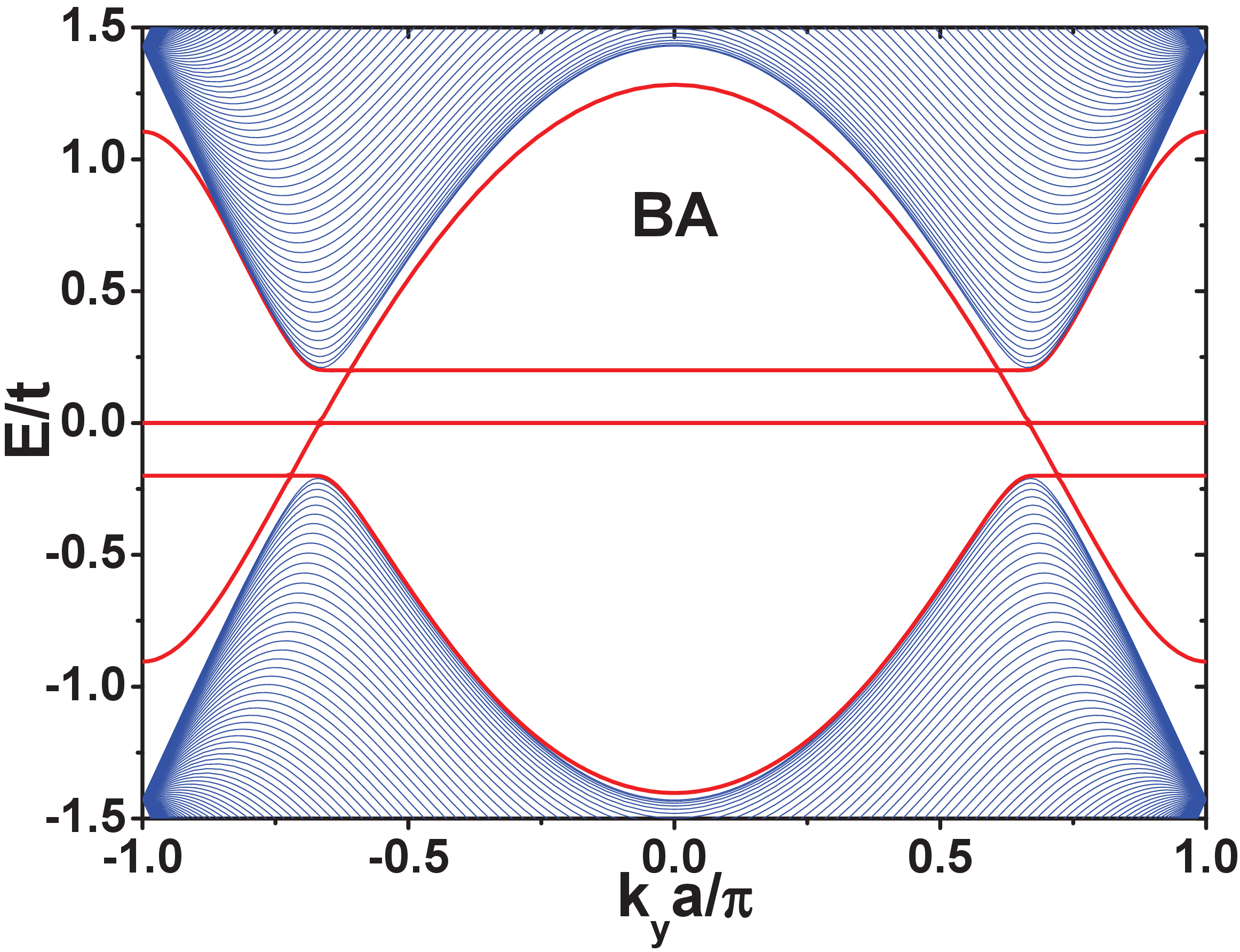}
\includegraphics[width=5.5cm,height=4.24cm]{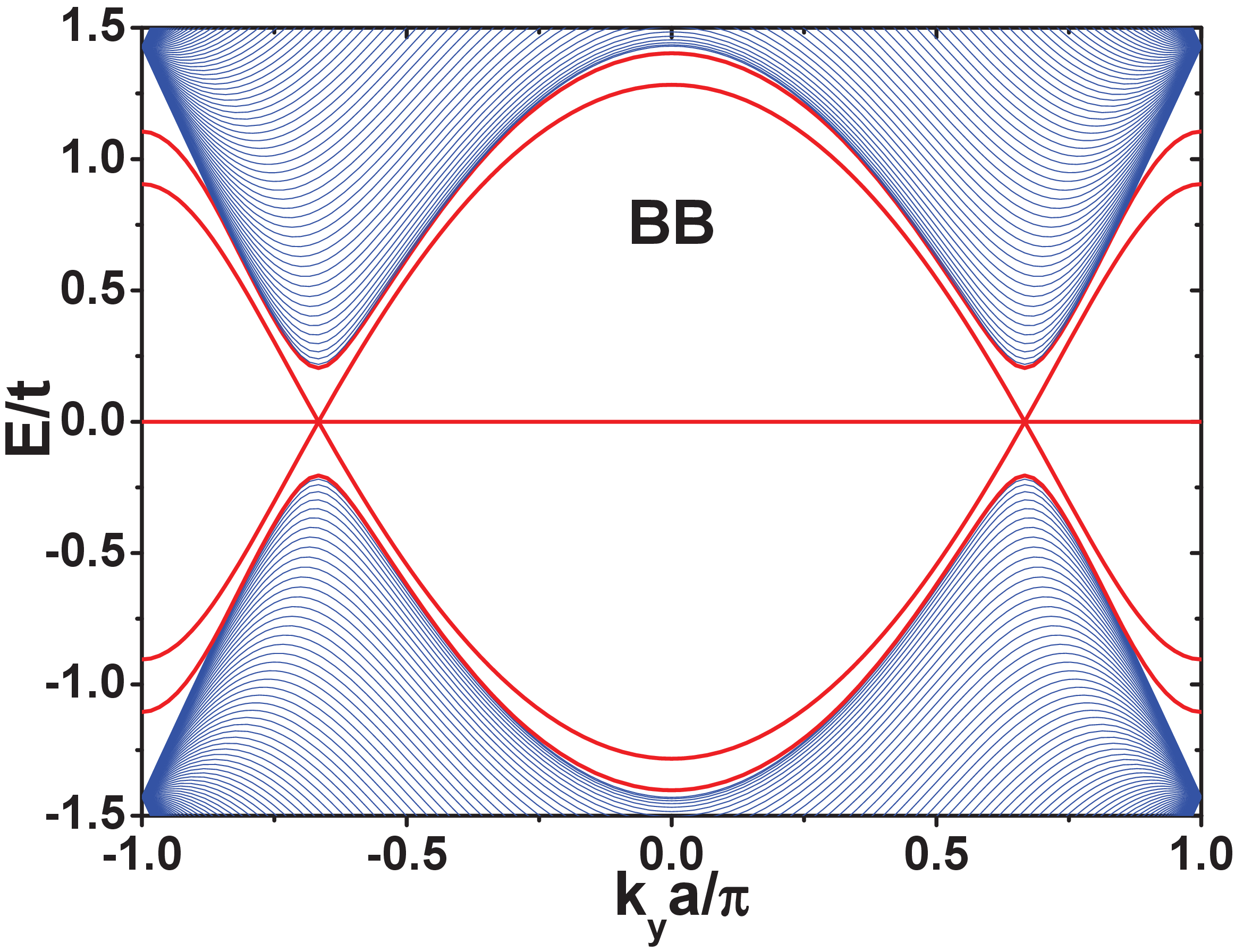}
\includegraphics[width=5.5cm,height=4.24cm]{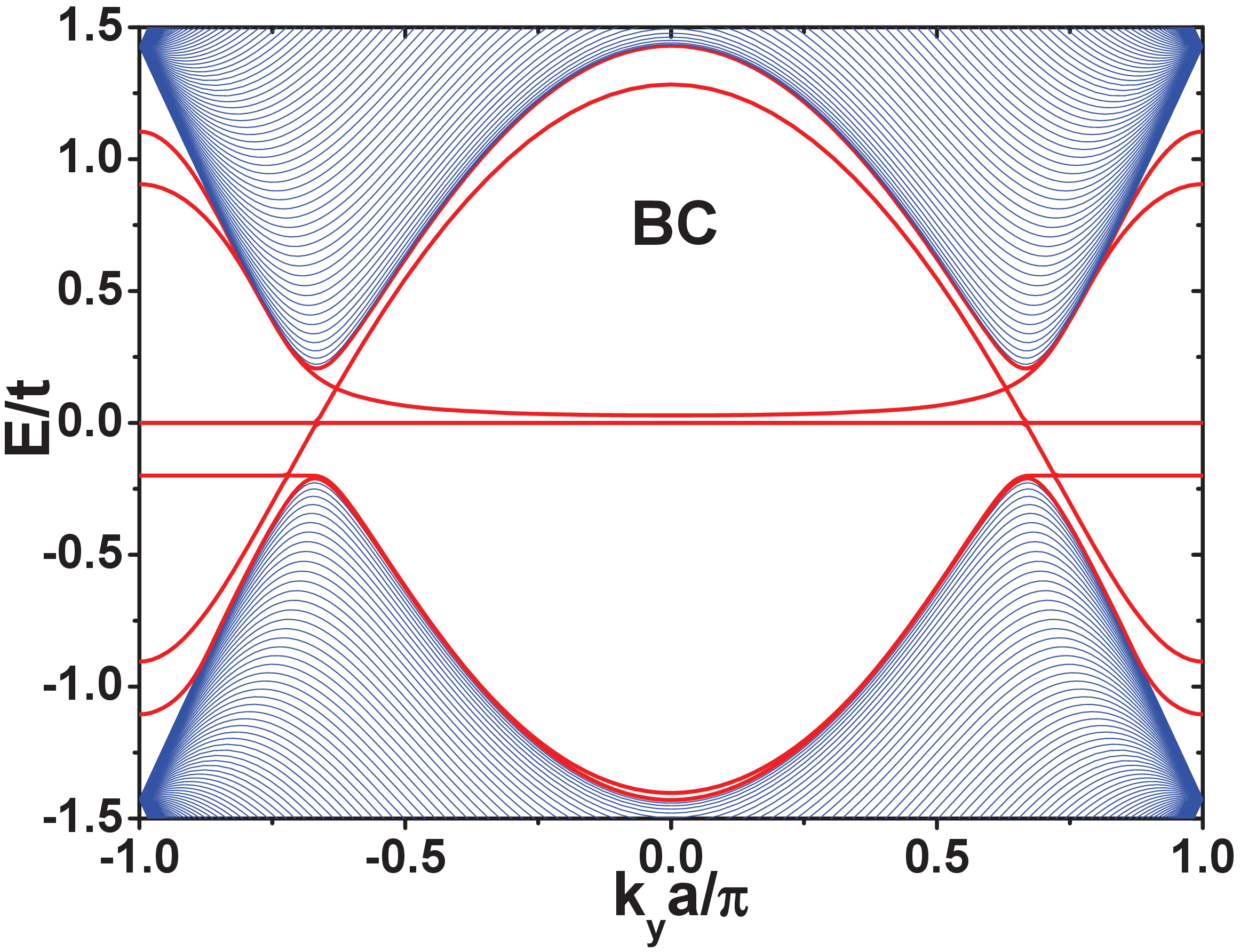}\\
\vspace{-0.10cm}
\hspace{-4.0cm} {\textbf{(g)}} \hspace{5.1cm}{\textbf{(h)}} \hspace{5.1cm}{\textbf{(i)}}\\
\hspace{0cm}\includegraphics[width=5.5cm,height=4.24cm]{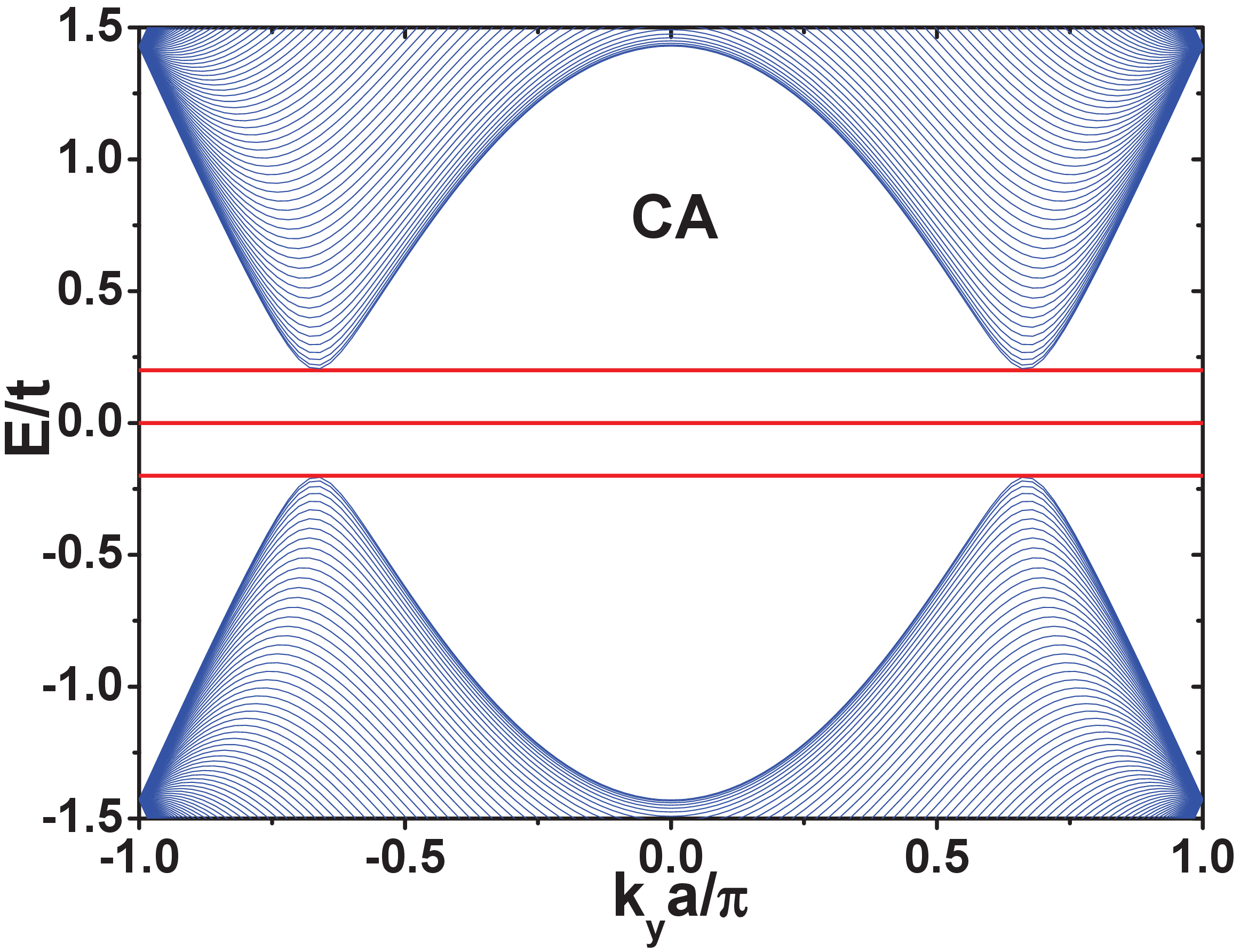}
\includegraphics[width=5.5cm,height=4.24cm]{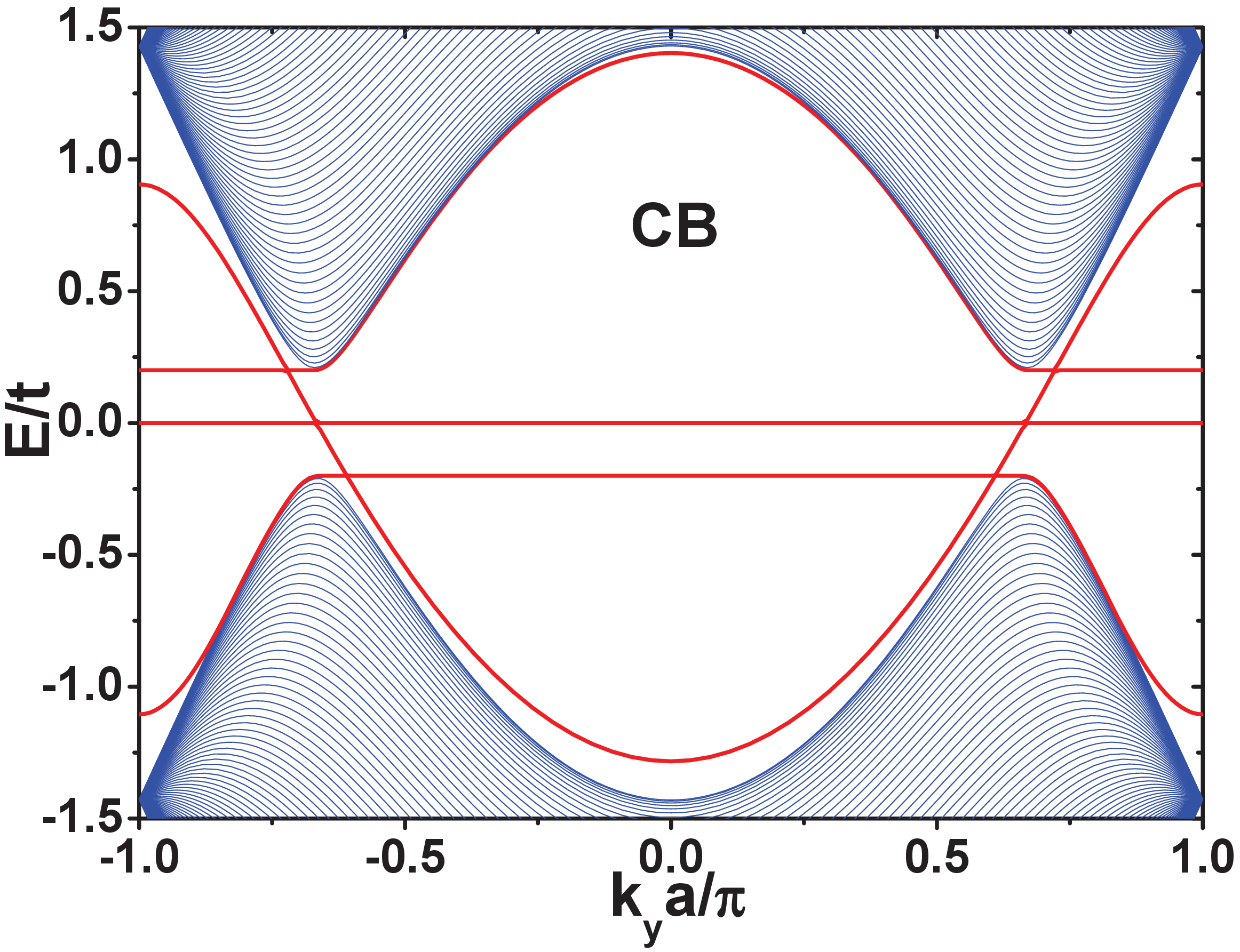}
\includegraphics[width=5.5cm,height=4.24cm]{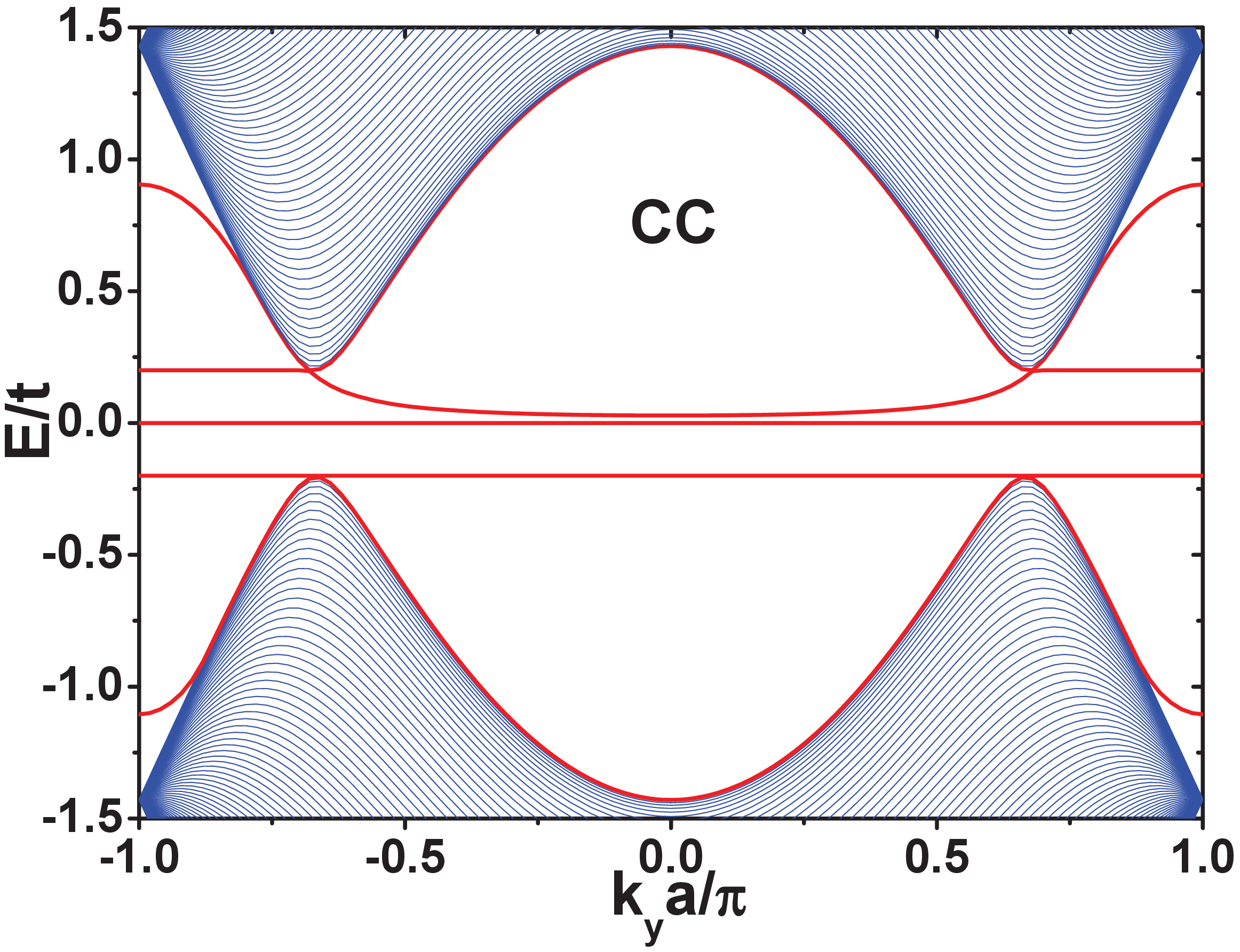}
\caption{The electronic spectra for the $\alpha\beta$-in zigzag ribbons ($\alpha$, $\beta$= A, B, C) of the symmetrically biased dice model, for $t=1$, $\Delta=0.2t$, and $N_{x}=50$. The zero-energy flat bands and the bands involving the in-gap states are thickened.}
\end{figure*}

We firstly consider zigzag ribbons of the pure dice (pure $\mathcal{T}_{3}$) model, for $t=1$ and $\Delta=0$.
In two previous works, one by Oriekhov et al \cite{oriekhov18} and another by Alam et al \cite{alam19}, a partial classification of the zigzag ribbons of this model were carried out in terms of the low-energy electronic spectrum close to the zero-energy flat band. Here, we make a more detailed and exhaustive classification based on all delicate features in the in-gap states and the 1D flat bands of the full band structures obtained by solving numerically the tight-binding model on the ribbons.

As shown in Fig. 4 are the numerical results of the electronic spectra for the nine $\alpha\beta$-in ($\alpha$, $\beta$ are A, B, or C) zigzag ribbons of the pure dice model, for $N_{x}$$=$$50$. Only the results for $k_{y}a$$\in$$[0,\pi]$ are plotted. The spectra for $k_{y}a$$\in$$[-\pi,0)$ are the same as those for $-k_{y}a$$\in$$(0,\pi]$. The result for the $\alpha\beta$-off ribbon ($\alpha$, $\beta$= A, B, C) with the same $N_{x}$$=$$50$ is indistinguishable by eye from the subfigure in Fig.4 for the corresponding $\alpha\beta$-in ribbon, and is therefore not shown in Fig.4. The low-energy spectra close to $k_{y}a$$=$$2\pi/3$ are shown as insets on each subfigure. Numerical calculations show that the low-energy gaps in the electronic spectra of the ribbons (apart from BB, which is always gapless) decrease monotonically as $1/N_{x}$, in agreement with the analytical result of Oriekhov et al \cite{oriekhov18}.

The first thing to note is the clear symmetry among the nine spectra of Fig. 4, with respect to the three spectra on the off-diagonal line for AC, BB, and CA. The two spectra (i.e., AB versus BC, AA versus CC, and BA versus CB) at symmetric positions with respect to the off-diagonal line are indistinguishable by eye, from both the low-energy and the higher-energy parts of the spectra. This symmetry roots in the equivalence of the A and C sublattices in the pure dice model. That is, rotating by $\pi$ about a line joining two NN B-sublattice sites leaves the bulk dice lattice and the pure dice model of electrons defined on it unchanged. For the zigzag ribbons with the same $N_{x}$, rotating the lattice by $\pi$ about the $y$ axis turns AB to BC, AA to CC, and BA to CB, and vice versa.
This symmetry therefore removes 3 possibly distinct electronic spectra, BC, CB, and CC. All the remaining 6 spectra are qualitatively different. The spectrum of Fig.4(c) for AC ribbons may seem identical with the spectra of Figs. 4(a) for AA ribbons and 4(i) for CC ribbons. However, the number of dispersive in-gap states close to the boundary of the 1D BZ are 4 for AC ribbons but 2 for AA ribbons and CC ribbons. For AC ribbons, there is a pair of dispersive edge modes on both the left and the right edges, one above $E=0$ and the other below $E=0$. For AA (CC) ribbons, there is a pair of dispersive edge modes only on the left (right) edge, one above $E=0$ and the other below $E=0$. Similarly, the number of dispersive edge modes close to the boundary of the 1D BZ are 4 for BB ribbons but 2 for BA ribbons and CB ribbons. Therefore, the spectra for the BB ribbons are qualitatively distinct from the spectra of the BA and CB ribbons, both from the low-energy spectrum close to $k_{y}=\pm\frac{2\pi}{3a}$ and from the in-gap states.

Because the spectra and degeneracies of the in-gap states for the $\alpha\beta$-off ribbon are indistinguishable from those of the $\alpha\beta$-in ribbon for the same $N_{x}$ that is sufficiently large, there are 6 distinct types of in-gap spectra among the 18 kinds of zigzag ribbons of the pure dice model. The two-fold degeneracy between $\alpha\beta$-in and $\alpha\beta$-off is broken when we count the number of zero-energy flat bands. Namely, explicit number counting for the same $N_{x}$ shows that the $\alpha\beta$-in and $\alpha\beta$-off ribbons separately have an odd and an even number of zero-energy flat bands. Overall, in terms of the in-gap states and the parity of the number of zero-energy flat bands, there are $12$ distinct electronic spectra for the pure dice model defined on the 18 distinct zigzag dice lattice ribbons.

Now we consider the symmetrically biased dice model with $\Delta\ne0$. The bulk band structures of this model have an isolated zero-energy flat band in the middle of two dispersive bands \cite{xu17,betancur17}. The energy gaps between the flat band and the two dispersive bands are both $|\Delta|$. In the biased model, the symmetry between the A and C sublattices in the pure dice model is broken explicitly. As shown in Fig. 5 are the spectra of the nine $\alpha\beta$-in ($\alpha$, $\beta$ are A, B, or C) zigzag ribbons of the symmetrically biased dice model with $t=1$, $\Delta=0.2t$, and $N_{x}=50$. The spectra of the $\alpha\beta$-off ribbon for the same $N_{x}=50$ is indistinguishable by eye from that of the $\alpha\beta$-in ribbon. All the nine spectra of Fig.5 are clearly different from each other, in terms of the nonzero energy in-gap states. The two-fold degeneracy in the in-gap states of the AB, AC, BB, and BC ribbons are broken explicitly. On the other hand, the parities of the numbers of zero-energy flat bands are still opposite for the $\alpha\beta$-in and $\alpha\beta$-off ribbons, which provide a qualitative distinction between the $\alpha\beta$-in and $\alpha\beta$-off ribbons. Overall, the 18 distinct zigzag dice lattice ribbons all have qualitatively distinct electronic spectra. Therefore, the distinct edge termination morphologies, whose total number is determined by Eq.(11), not only differ in their lattice structures but may lead to qualitative differences in other physical properties including their electronic spectra and associated properties.


\section{structure-spectrum relations for distinctive features of the electronic spectra}

In this section, we study several interesting features in the numerical spectra of Figs. 4 and 5 that are used to differentiate the electronic spectra of different zigzag ribbons. These include the zero-energy flat bands, the in-gap states at and close to the boundary of the 1D BZ, the new in-gap states emerging in Fig.5 for $\Delta\neq0$, and the Dirac cones for the BB ribbons. The flat bands and in-gap states are both novel features that make ribbons fascinating subjects of research.

\subsection{States at $k_{y}a=\pm\pi$}

There are in-gap states at the boundary of the 1D BZ, $k_{y}a=\pm\pi$, for all the 18 edge termination morphologies of the symmetrically biased dice model. There are also in-gap states for the zigzag ribbons of the pure dice model, except for the CA-in and CA-off ribbons. The Bloch wave functions for the states at $k_{y}a=\pm\pi$ are the multiplication of the phase factor $\exp[ik_{y}n_{y}a]=(-1)^{n_{y}}$ and a vector independent of $n_{y}$, where $n_{y}$ labels the consecutive unit cells of the ribbon along $y$. The Bloch phase factor therefore gives a sign-alternating modulation to the periodic state vector of any eigenstate, indicative of the destructive interference playing a decisive role. On the other hand, because
\begin{equation}
t'=2t\cos(\frac{k_{y}a}{2})=2t\cos(\frac{\pm\pi}{2})=0,
\end{equation}
the $N_{s}\times N_{s}$ Hamiltonian matrix of the ribbon at $k_{y}a=\pm\pi$ has only the diagonal on-site energy terms and the NN hopping terms on the horizontal bonds. In terms of Fig.3(b), this means that the $N_{s}\times N_{s}$ Hamiltonian matrix at $k_{y}a=\pm\pi$ describes a set of decoupled horizontal trimers supplemented by dimers and (or) monomers at the two ends of the unit cell. Each trimer consists of a B sublattice site in the center, a C sublattice site on the left, and an A sublattice site on the right. There is one BA dimer at the left end of the unit cell for the left edge terminated with a $y$-chain of A sublattice sites or a $y$-chain of B sublattice sites. There is one BC dimer at the right end of the unit cell for the right edge terminated with a $y$-chain of C sublattice sites or a $y$-chain of B sublattice sites. There is one A-site (C-site) monomer at the left (right) end of the unit cell for the left (right) edge terminated with a $y$-chain of A sublattice sites or a $y$-chain of C sublattice sites. The states at $k_{y}a=\pm\pi$ are therefore completely localized at these isolated trimers, dimers, and monomers.

The $3\times3$ matrix for a trimer, in the basis of $[a^{\dagger},b^{\dagger},c^{\dagger}]$ is
\begin{equation}
\begin{pmatrix} \Delta & t & 0 \\ t & 0 & t \\ 0 & t & -\Delta \\ \end{pmatrix}.
\end{equation}
The three eigenenergies of the trimer are
\begin{equation}
E_{0}=0, \hspace{0.5cm} E_{\pm}=\pm\sqrt{2t^{2}+\Delta^{2}}.
\end{equation}
The eigenvector of the zero-energy mode is
\begin{equation}
\frac{1}{\sqrt{2t^{2}+\Delta^{2}}}\begin{pmatrix} -t \\ \Delta \\ t \\ \end{pmatrix}.
\end{equation}
The eigenvector for the $E_{\alpha}$ ($\alpha=\pm$) mode is
\begin{equation}
\frac{\text{sgn}(t)}{2|E_{\alpha}|}\begin{pmatrix} E_{\alpha}+\Delta \\ 2t \\ E_{\alpha}-\Delta \\ \end{pmatrix}.
\end{equation}
$\text{sgn}(t)=t/|t|$ is the sign function.

In the basis of $[b^{\dagger},a^{\dagger}]$ for the BA dimer or $[b^{\dagger},c^{\dagger}]$ for the BC dimer, the $2\times2$ matrix for a dimer is
\begin{equation}
\begin{pmatrix} 0 & t \\ t & \alpha\Delta \\ \end{pmatrix},
\end{equation}
where $\alpha=1$ for the BA dimer and $\alpha=-1$ for the BC dimer. For a specific $\alpha$, the above model for the dimer has two eigenstates with energies
\begin{equation}
E_{\alpha\beta}=\frac{\alpha\Delta}{2}+\beta\sqrt{(\frac{\Delta}{2})^{2}+t^{2}},
\end{equation}
where $\beta=\pm$. The two eigenvectors are
\begin{equation}
\begin{pmatrix} u_{\alpha\beta}  \\  v_{\alpha\beta}  \end{pmatrix}=\frac{1}{\sqrt{2E_{\alpha\beta}(E_{\alpha\beta}-\frac{\alpha\Delta}{2})}}\begin{pmatrix} t  \\  E_{\alpha\beta} \end{pmatrix}.
\end{equation}

The two monomers, if existing, both have a single eigenvalue equal to the corresponding on-site energy: An A-site monomer at the left edge gives an eigenstate with eigenenergy $\Delta$, while a C-site monomer at the right edge gives an eigenstate with eigenenergy $-\Delta$. The corresponding eigenvectors reside completely on the A-site monomer at the left edge or the C-site monomer at the right edge.

The above eigenstates of the trimers, dimers, and monomers fully account for all the eigenmodes of the zigzag dice lattice ribbons at $k_{y}a=\pm\pi$. Firstly, each trimer corresponds to a fully coordinated $y$-chain of B-sublattice sites. The three eigenstates of the trimer correspond directly to the three bulk band states at $k_{y}a=\pm\pi$, including one state of the zero-energy flat band and the two states of the dispersive bands. The two dispersive bulk 2D bands have the dispersions
\begin{equation}
\begin{cases}
E_{\pm}(\mathbf{k})=\pm\sqrt{2|\xi(\mathbf{k})|^{2}+\Delta^{2}},  \\
\xi(\mathbf{k})=t(e^{i\mathbf{k}\cdot\boldsymbol{\delta}_{1}}+e^{i\mathbf{k}\cdot\boldsymbol{\delta}_{2}} +e^{i\mathbf{k}\cdot\boldsymbol{\delta}_{3}}).
\end{cases}
\end{equation}
$\mathbf{k}=(k_{x},k_{y})$ is a 2D wave vector. Clearly, for all $k_{x}$, $E_{\pm}(k_{x},k_{y}=\pm\frac{\pi}{a})=\pm\sqrt{2t^{2}+\Delta^{2}}$, which coincide with the two nonzero eigenenergies of the trimer in Eq.(16).
The isolated dimers account for the in-gap states at (for $\Delta=0$) or close to (for small but nonzero $\Delta$) $t$ and $-t$. The variations in the numerical values of these in-gap states with $\Delta$ agree very well with the formula of Eq.(20). For example, for $t=1$ and $\Delta=0.2t$, the two eigenvalues are approximately $1.105t$ and -$0.905t$ for the BA dimer, and are approximately $0.905t$ and -$1.105t$ for the BC dimer. These agree exactly with the in-gap states in the spectra of Fig.5 at $k_{y}a=\pm\pi$.
The monomers, on the other hand, account for the new in-gap states at $\Delta$ (the A-site monomers) or $-\Delta$ (the C-site monomers). In addition, if we continuously reduce the magnitude of $\Delta$ from a nonzero value, these eigenenergies of the monomers also continuously approach zero. However, even if $\Delta=0$, these eigenmodes of the monomers are always localized on the monomers at the edges. Therefore, the zero-energy flat bands of the zigzag dice lattice ribbons, excluding BB-in and BB-off ribbons, of the pure dice model include bands localized at the edges. These states become the new in-gap states at energies equal to or smaller in magnitude than $|\Delta|$.

Away from the boundary of the 1D BZ, the spectrum of each band changes continuously. We consider three kinds of interesting states for general $k_{y}$ in what follows.

\subsection{Zero-energy flat bands robust to nonzero $\Delta$}

Here, we consider the zero-energy flat bands that are robust to a nonzero $\Delta$. As we noted above, each horizontal CBA trimer of the unit cell gives a zero-energy state at the BZ boundary, which is independent of the bias $\Delta$. A horizontal CBA trimer in the unit cell of the ribbon corresponds to a fully coordinated $y$-chain of B-sublattice sites. These zero-energy flat bands are thus intimately related to the bulk zero-energy flat band, which also survives a nonzero $\Delta$.

In the bulk lattice and in the basis of $[a^{\dagger}_{\mathbf{k}},b^{\dagger}_{\mathbf{k}},c^{\dagger}_{\mathbf{k}}]$, the eigenvector for the zero-energy flat band at the 2D wave vector $\mathbf{k}$ is
\begin{equation}
\frac{1}{\sqrt{2|\xi(\mathbf{k})|^{2}+\Delta^{2}}}\begin{pmatrix} -\xi(\mathbf{k}) \\ \Delta \\ \xi^{\ast}(\mathbf{k}) \\  \end{pmatrix},
\end{equation}
where $\xi(\mathbf{k})$ is defined by Eq.(22). The opposite signs in front of the first and third elements of the eigenvector correspond to the destructive interference of electrons hopping between the A and C sublattice sites and the B sublattice sites.

In the zigzag ribbons of the dice lattice, each fully coordinated $y$-chain of B sublattice sites involves five sites in the unit cell of the ribbon, consisting of one horizontal CBA trimer, an A site and a C site connected to the central B site by the NN bond vectors $-\boldsymbol{\delta}_{2}$ and $\boldsymbol{\delta}_{3}$. In the light of Eq.(23) for the bulk eigenvector, and Eq.(17) for the eigenvector of an isolated trimer, we write the eigenvector for the state of the zero-energy flat band associated to this fully coordinated $y$-chain of B sublattice sites at $k_{y}$ as
\begin{equation}
\frac{1}{\sqrt{2t^{2}+2t'^{2}+\Delta^{2}}}
\begin{pmatrix} -t \\ t \\ \Delta \\ -t' \\ t' \\ \end{pmatrix},
\end{equation}
in the basis of $[a^{\dagger}_{1k_{y}},c^{\dagger}_{1k_{y}},b^{\dagger}_{k_{y}}, a^{\dagger}_{2k_{y}},c^{\dagger}_{2k_{y}}]$. $t'=2t\cos(\frac{k_{y}a}{2})$. $a^{\dagger}_{1k_{y}}$, $c^{\dagger}_{1k_{y}}$, and $b^{\dagger}_{k_{y}}$ create an electron of wave vector $k_{y}$ on the three sites of the horizontal trimer centering at the considered B site. $a^{\dagger}_{2k_{y}}$ and $c^{\dagger}_{2k_{y}}$ create an electron on the A site and C site connected to the considered B site by the vectors $-\boldsymbol{\delta}_{2}$ and $\boldsymbol{\delta}_{3}$. The weights of this state on the other sites of the unit cell are completely zero. It is easy to check that Eq.(24) is indeed a zero-energy eigenstate of the $N_{s}\times N_{s}$ model $h(k_{y})$. For $t'=0$ at the BZ boundary, Eq.(24) reduces to Eq.(17).

For each fully coordinated $y$-chain of B sublattice sites, we may construct one band of zero-energy states defined by Eq.(24). It is easy to see that the eigenvectors of two zero-energy flat bands corresponding to two NN $y$-chains of fully coordinated B sublattice sites are nonorthogonal to each other. But these states are linearly independent of each other, and therefore it is appropriate to take them as the complete basis set for the 1D zero-energy flat bands. The number of zero-energy flat bands that are independent of the bias $\Delta$ coincides exactly with the number of zero-energy flat bands constructed in the above manner, for each of the 18 types of zigzag dice lattice ribbons defined in Fig.3(b). Since the $\alpha\beta$-in and $\alpha\beta$-off ribbons ($\alpha$, $\beta$=A, B, C) with the same $N_{x}$ differ by one fully coordinated $y$-chain of B sublattice sites, the present analysis explains why the parities of the numbers of their zero-energy flat bands are opposite.

\subsection{In-gap states at $\pm\Delta$ and smaller: Comparison with the zigzag graphene ribbons}

The segmental flat bands at $\Delta$ and $-\Delta$ are present for all the zigzag ribbons of the dice lattice, except for the BB-in and BB-off ribbons. These states are present explicitly in Fig.5 and hidden inside the zero-energy bands in Fig.4. They are reminiscent of the well-known segmental flat-band edge states of the honeycomb lattice ribbons (i.e., graphene ribbons) with zigzag-type and bearded-type edges \cite{fujita96,nakada96,klein94,brey06,ryu02,mong11,delplace11,breybook}. For regular zigzag ribbons of the honeycomb lattice, as shown in Fig.3(a), a zigzag-type edge [the L$_{1}$, R$_{2}$, and R$_{4}$ edges of Fig.3(a)] supports dispersionless edge states between $k_{y}=\alpha\frac{2\pi}{3a}$ and $k_{y}=\alpha\frac{\pi}{a}$ ($\alpha=\pm$) of the 1D BZ \cite{fujita96,nakada96,brey06}, and a bearded-type edge [the L$_{2}$, R$_{1}$, and R$_{3}$ edges of Fig.3(a)] supports dispersionless edge states between $k_{y}=-\frac{2\pi}{3a}$ and $k_{y}=\frac{2\pi}{3a}$ \cite{klein94}.

\begin{figure}[!htb]\label{fig6}
\centering
\hspace{-8.0cm} {\textbf{(a)}}\\
\includegraphics[width=7.5cm,height=1.826cm]{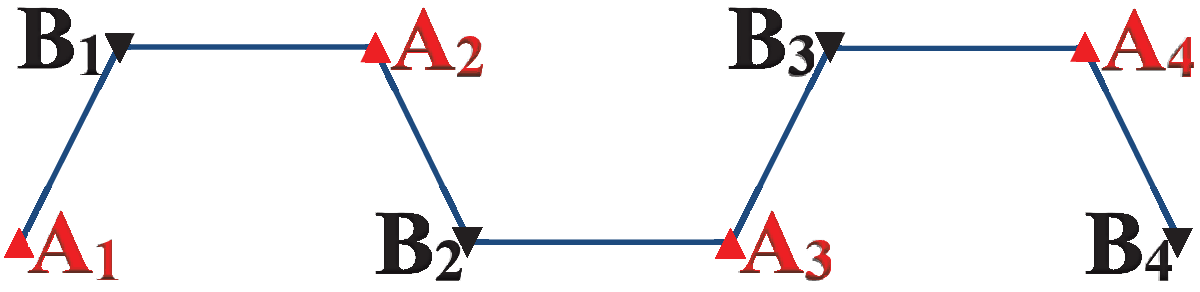} \\ \vspace{0.05cm}
\hspace{-8.0cm} {\textbf{(b)}}\\
\includegraphics[width=8.2cm,height=1.555cm]{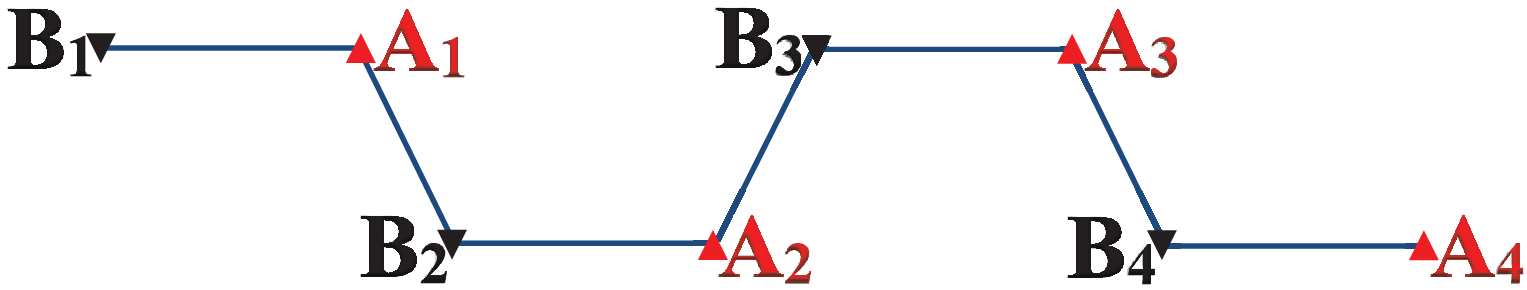}  \\
\caption{A part of the unit cell of a semiinfinite honeycomb lattice terminated with (a) a zigzag-type and (b) a bearded-type left edge. The sites in the unit cell are labeled by a positive integer index increasing with the distance away from the edge. Only the eight leftmost sites in the unit cell are shown, which contain two repetitions of the minimal repetitive unit of the unit cell. The same pattern repeats and extends rightward indefinitely.}
\end{figure}

The analytical solutions to the segmental flat-band states of the zigzag ribbons of the honeycomb lattice are well known from previous works \cite{fujita96,nakada96,klein94}. To compare with the results of the dice lattice ribbons, we consider a tight-binding model on the honeycomb lattice with NN hopping amplitudes to be $t$, and the on-site energies of electrons on the A and B sublattices to be $\Delta$ and $-\Delta$. We consider the edge states localized on the left edge of a semiinfinite sample of the honeycomb lattice. A part of the unit cells for the samples with zigzag-type (A-type) and bearded-type (B-type) left edges are separately shown in Figs.6(a) and 6(b). As shown by Fujita et al \cite{fujita96}, for a zigzag-type left edge terminated with the A sublattice sites, the edge states have weights only on the A sublattice sites and the amplitude of the wave function decays proportional to $-t'/t=-2\cos(\frac{k_{y}a}{2})$. The edge state exists for $|t'/t|<1$ which leads to $k_{y}a\in[-\pi,-\frac{2\pi}{3})\cup(\frac{2\pi}{3},\pi]$. The energies of these edge states are exactly $\Delta$. As shown by Klein \cite{klein94}, for a bearded-type left edge terminated with the B sublattice sites, the edge states have weights only on the B sublattice sites and the amplitude of the wave function decays proportional to $-t/t'=1/[-2\cos(\frac{k_{y}a}{2})]$. The edge state exists for $|t/t'|<1$ which leads to $k_{y}a\in(-\frac{2\pi}{3},\frac{2\pi}{3})$. The energies of these edge states are exactly $-\Delta$. The edge states for a semiinfinite honeycomb lattice terminated with a right edge has a simple correspondence with the above results: A zigzag-type left edge terminated with the A sublattice sites corresponds to a zigzag-type right edge terminated with the B sublattice sites and a sign reversal of $\Delta$, a bearded-type left edge terminated with the B sublattice sites corresponds to a bearded-type right edge terminated with the A sublattice sites and a sign reversal of $\Delta$. Therefore, a zigzag-type right edge terminated with the B sublattice sites has a segmental flat band edge states for $k_{y}a\in[-\pi,-\frac{2\pi}{3})\cup(\frac{2\pi}{3},\pi]$ with energy $-\Delta$, and a bearded-type right edge terminated with the A sublattice sites has a segmental flat band edge states for $k_{y}a\in(-\frac{2\pi}{3},\frac{2\pi}{3})$ with energy $\Delta$. Combination of these results accounts for all the edge states of the 8 types of edge termination morphologies of Fig.3(a) for the honeycomb lattice.

The segmental flat bands at $\pm\Delta$ on the subfigures of Fig.5 for zigzag ribbons of the dice lattice are described by exactly the same kind of wave functions as those for the flat band edge states of the zigzag ribbons of the honeycomb lattice. This is possible because the dice lattice may be seen as two intersecting honeycomb lattices, one consisting of the A and B sublattices versus the other consisting of the C and B sublattices, which merge together at the common B sublattice sites. The additional sublattice (A or C) do not influence the above segmental flat bands because these states have weights only on the A or C sublattices, and the direct hopping between the A and C sublattice sites are zero for the tight-binding model up to the NN hopping integrals.

As a result of the above coexistence and decoupling of the A and C sublattices, the left edge of the dice lattice terminated with a $y$-chain of C sublattice sites is a bearded-type edge for the honeycomb lattice consisting of the C and B sublattices (hereafter, the CB honeycomb lattice) and is a zigzag-type edge for the honeycomb lattice consisting of the A and B sublattices (hereafter, the AB honeycomb lattice). From the above results for the honeycomb lattice, a C-terminated left edge has edge states on the C sublattice of energy $-\Delta$ for $k_{y}a\in(-\frac{2\pi}{3},\frac{2\pi}{3})$ and edge states on the A sublattice of energy $\Delta$ for $k_{y}a\in[-\pi,-\frac{2\pi}{3})\cup(\frac{2\pi}{3},\pi]$. An A-terminated right edge has edge states on the A sublattice of energy $\Delta$ for $k_{y}a\in(-\frac{2\pi}{3},\frac{2\pi}{3})$ and edge states on the C sublattice of energy $-\Delta$ for $k_{y}a\in[-\pi,-\frac{2\pi}{3})\cup(\frac{2\pi}{3},\pi]$. An A-terminated left edge and a C-terminated right edge give separately zigzag-type edge of the AB honeycomb lattice and the CB honeycomb lattice. They separately give left edge states on the A sublattice of energy $\Delta$ and right edge states on the C sublattice of energy $-\Delta$, for $k_{y}a\in[-\pi,-\frac{2\pi}{3})\cup(\frac{2\pi}{3},\pi]$. These states account for most of the additional nonzero energy in-gap states for $\Delta\ne0$. B-terminated left and right edges do not give these edge states corresponding to the honeycomb lattice, because a B-terminated edge is the mixture of the zigzag-type edge for one honeycomb lattice and the bearded-type edge for the other honeycomb lattice. Neither of the two typical kinds of edge states survive the mixing.

\begin{figure}[!htb]\label{fig7}
\centering
\hspace{-8.0cm} {\textbf{(a)}}\\
\includegraphics[width=8.5cm,height=1.66cm]{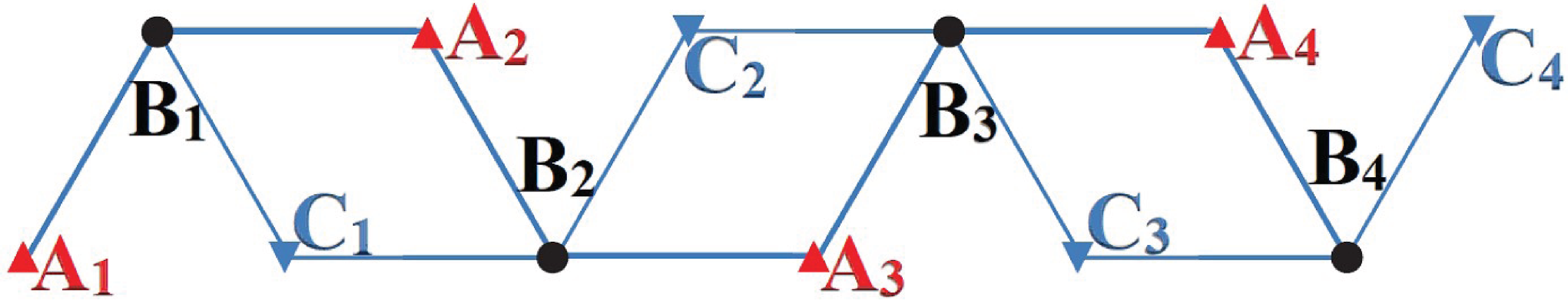} \\ \vspace{0.05cm}
\hspace{-8.0cm} {\textbf{(b)}}\\
\includegraphics[width=8.0cm,height=5.992cm]{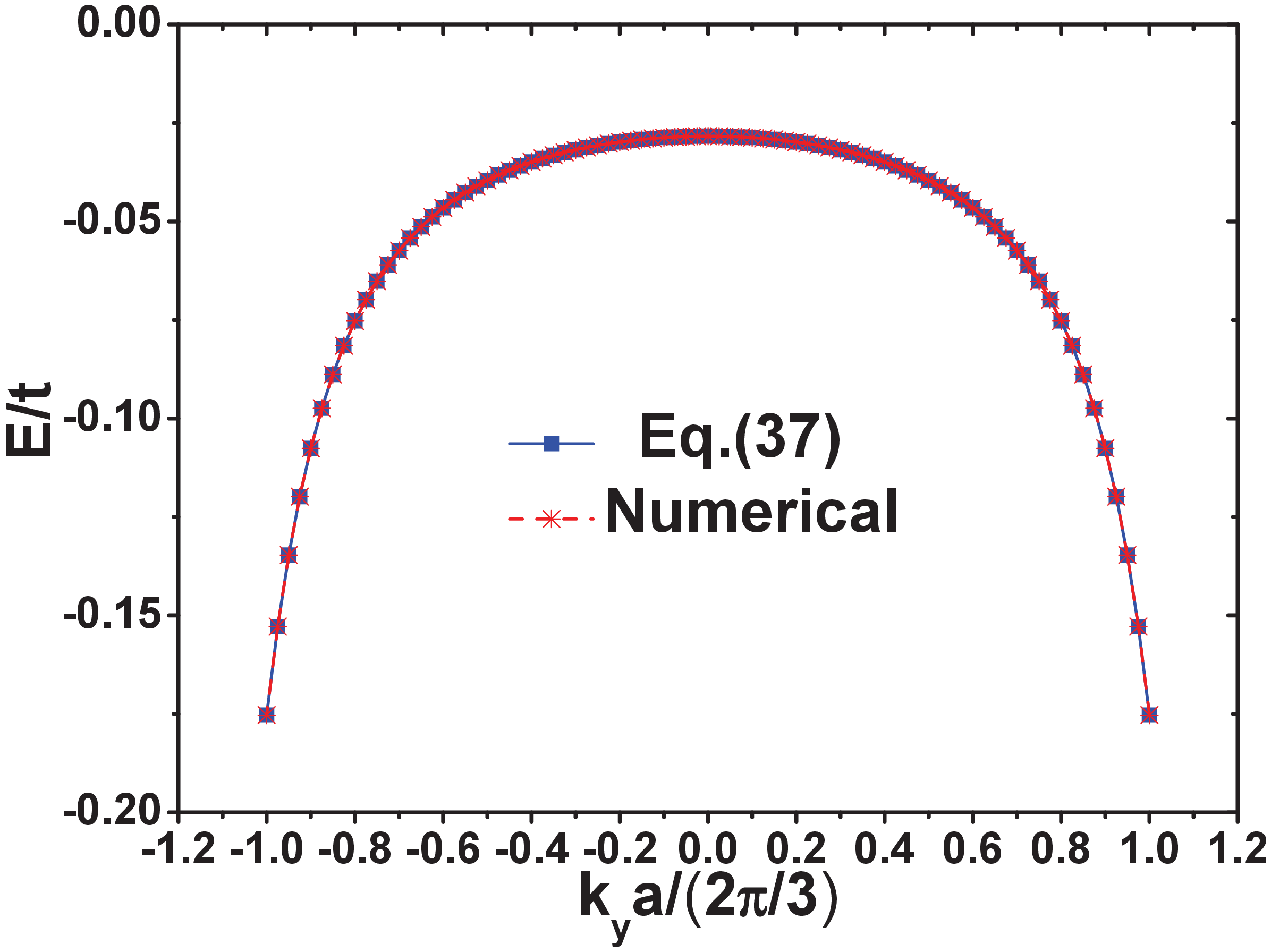}  \\
\hspace{-8.0cm} {\textbf{(c)}}\\
\includegraphics[width=8.0cm,height=6.053cm]{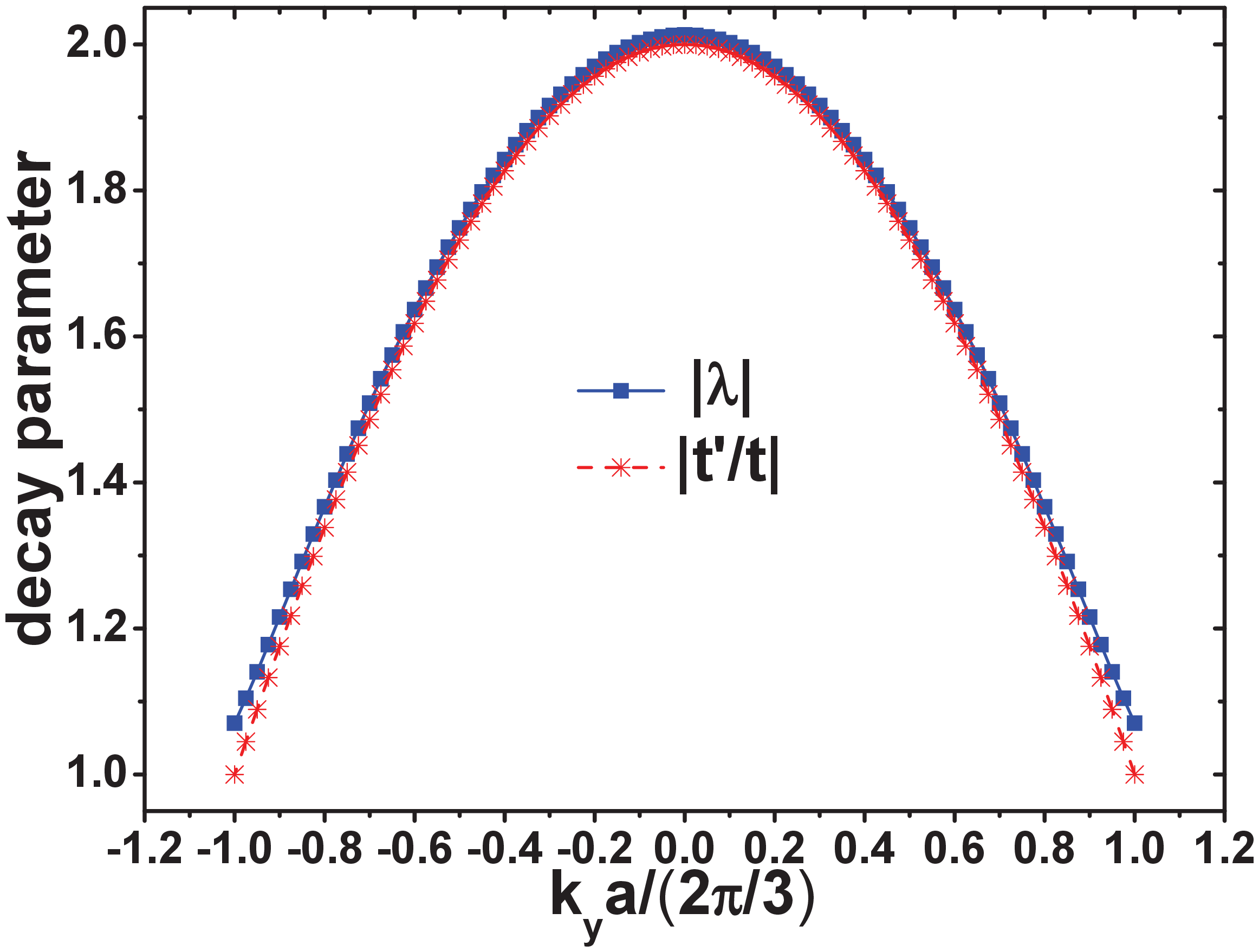}  \\
\caption{(a) A part of the unit cell of a semiinfinite dice lattice terminated with a $y$-chain of A sublattice sites on the left. The sites in the unit cell are labeled by an index increasing with the distance away from the edge. Only the twelve leftmost sites in the unit cell are shown, which contain two repetitions of the minimal repetitive unit of the unit cell. The same pattern repeats and extends rightward indefinitely. (b) A comparison of the spectrum of Eq.(37) and the numerical results for the AB-in ribbon with $N_{x}=50$. $t=1$ and $\Delta=0.2t$ are used in the calculations. (c) A comparison of the decay parameter $|\lambda|$ determined by Eq.(33) for the edge states of Eq.(37) with the decay parameter $|t'/t|$ for the edge states corresponding to an ideal bearded-type edge of the honeycomb lattice.}
\end{figure}

An interesting feature missing in the above picture is the in-gap states smaller in magnitude than $|\Delta|$ in the range $k_{y}\in(-\frac{2\pi}{3a}, \frac{2\pi}{3a})$, which are present in Figs.5(a,b,c,f,i). These states are present for A-type edge on the left edge (these states are localized also on the left edge) and for C-type edge on the right edge (these states are localized also on the right edge). The energy scales, the wave vector ranges, and the signs of the energies for these states strongly suggest that they have something to do with the edge states of a bearded-type edge of the honeycomb lattice. As shown in Fig.7(a) are the leftmost sites of the unit cell for a zigzag dice lattice ribbon with a left edge terminated with a $y$-chain of A sublattice sites. There is certainly a zigzag-type edge of the AB honeycomb lattice, which accounts for the edge states of energy $\Delta$ for $k_{y}\in[-\pi,-\frac{2\pi}{3a})\cup(\frac{2\pi}{3a},\pi]$. The bearded-type left edge for the CB honeycomb lattice is disrupted by the A$_1$ and B$_1$ sites at the left end of the unit cell. Locally, the three leftmost sites (A$_1$, B$_1$, and C$_1$) of the unit cell constitute the unit cell of a diamond chain lattice \cite{hyrkas13,tovmasyan18,cartwright18}. The destructive interference of the A$_1$ and C$_1$ sublattices with the B$_1$ sublattice results in a zero-energy flat band of the diamond chain lattice \cite{hyrkas13,tovmasyan18,cartwright18}. The wave function components on the A$_1$ and C$_1$ sites has equal amplitudes but opposite signs for states in this flat band. Combining this local mode related to the A$_1$B$_1$C$_1$ triplet and the edge states associated with the bearded-type left edge of the CB honeycomb lattice beginning from C$_1$, we may construct an edge state mostly associated with the bearded-type left edge of the CB honeycomb lattice. This simple combination is however disrupted by the B$_{1}$A$_{2}$ bond of the unit cell and the finite weight of the local mode on B$_{1}$ due to $\Delta\ne0$, which propagate the wave function weights along the channel B$_{1}$A$_{2}$B$_{2}$A$_{3}$$\cdots$ to A$_{i}$ and B$_{i}$ for $i\ge2$.

According to the above physical picture, we try the following trial solution for an edge state on the left edge terminated with a $y$-chain of A sublattice sites:
\begin{equation}
\begin{cases}
\psi_{A1}=-\psi_{C1};  \\
\psi_{Bi}=\psi_{B1}\lambda_{B}^{-(i-1)},  (i\ge2);  \\
\psi_{Ci}=\psi_{C1}\lambda_{C}^{-(i-1)},  (i\ge2);  \\
\psi_{Ai}=\psi_{A2}\lambda_{A}^{-(i-2)},  (i\ge3).
\end{cases}
\end{equation}
$\psi_{\alpha i}$ is the amplitude of the wave function on the $i$th site of the $\alpha$th sublattice ($\alpha$=A, B, C; $i$=1, 2, 3, $\cdots$). The edge states demand $|\lambda_{\alpha}|>1$ ($\alpha$=A, B, C). $\psi_{\alpha i}$ depends implicitly on the wave vector $k_{y}$ through the following eigenfunctions that they obey
\begin{eqnarray}
&&t'\psi_{B1}=(E-\Delta)\psi_{A1},  \\
&&t'(\psi_{A1}+\psi_{C1})+t\psi_{A2}=E\psi_{B1},  \\
&&t'\psi_{B1}+t\psi_{B2}=(E+\Delta)\psi_{C1},  \\
&&\cdots\cdots   \notag  \\
&&t\psi_{B,i-1}+t'\psi_{Bi}=(E-\Delta)\psi_{Ai},  \\
&&t(\psi_{C,i-1}+\psi_{A,i+1})+t'(\psi_{Ai}+\psi_{Ci})=E\psi_{Bi},  \\
&&t'\psi_{Bi}+t\psi_{B,i+1}=(E+\Delta)\psi_{Ci},  \\
&&\cdots\cdots.  \notag
\end{eqnarray}
$t'=2t\cos(\frac{k_{y}a}{2})$. $i\ge2$ for Eqs.(29)-(31). $E$ is the eigenenergy of the edge state at $k_{y}$. Since we are seeking for solutions with $E\ne0$ and $E\ne\pm\Delta$, we get in terms of Eqs.(25)-(27)
\begin{equation}
\begin{cases}
\psi_{B1}=-\frac{E-\Delta}{t'}\psi_{C1}, \\
\psi_{A2}=-\frac{E(E-\Delta)}{tt'}\psi_{C1}. \\
\end{cases}
\end{equation}
Substituting Eqs.(25) and (32) into Eqs.(29)-(31), we get a set of equations for $E$ and $\lambda_{\alpha}$ ($\alpha$=A, B, C) that depend on the index $i$. In order to get a unique solution for $E$ and $\lambda_{\alpha}$ for all $i=2,3,\cdots$, we must have $\lambda_{A}=\lambda_{B}=\lambda_{C}=\lambda$. Then, taking the conditions $E\ne0$ and $E\ne\pm\Delta$ into account, we arrive at the following constraints on $E$ and $\lambda$
\begin{equation}
\lambda=\frac{tt'}{E^{2}-\Delta E-t^{2}}=\frac{t(\Delta-E)}{2t'E}.
\end{equation}
The second equality of Eq.(33) leads to the following cubic equation for the eigenenergy of the edge states
\begin{equation}
E^{3}-2\Delta E^{2}+(\Delta^{2}+2t'^{2}-t^{2})E+\Delta t^{2}=0.
\end{equation}
To solve the above cubic equation in terms of the Cardano's method, we introduce the following parameters
\begin{equation}
\begin{cases}
a=-2\Delta,  \\
b=\Delta^{2}+2t'^{2}-t^{2},  \\
c=\Delta t^{2},  \\
p=\frac{3b-a^{2}}{3}, \\
q=\frac{2a^{3}+27c-9ab}{27}. \\
\end{cases}
\end{equation}
The number of real roots is determined by the following discriminant
\begin{equation}
27q^{2}+4p^{3}=\frac{\Delta^{2}(2\Delta^{2}+36t'^{2}+9t^{2})^{2}}{27}+4(2t'^{2}-t^{2}-\frac{\Delta^{2}}{3})^{3}.
\end{equation}
In the range $k_{y}\in(-\frac{2\pi}{3a}, \frac{2\pi}{3a})$ of interest, $t'^{2}>t^{2}$. For all $\Delta^{2}\le 3t'^{2}$, the above discriminant is clearly positive definite, and Eq.(34) has a single real root. For these cases that we focus on, the single real root of Eq.(34) is
\begin{equation}
E=\sqrt[3]{-\frac{q}{2}+\sqrt{(\frac{q}{2})^{2}+(\frac{p}{3})^{3}}} +\sqrt[3]{-\frac{q}{2}-\sqrt{(\frac{q}{2})^{2}+(\frac{p}{3})^{3}}}-\frac{a}{3}.
\end{equation}
Substituting into Eq.(33) gives the corresponding decay parameter $\lambda$. Up to an arbitrary U(1) phase factor, the remaining parameter $\psi_{C1}$ of the wave function may be determined from the normalization of the wave function.

As shown in Fig.7(b) is a comparison of the analytical spectrum of Eq.(37) with the numerical results obtained for the spectrum of the AB-in ribbon for $N_{x}=50$. The parameters are taken as $t=1$ and $\Delta=0.2t$. In the whole range $k_{y}\in[-\frac{2\pi}{3a}, \frac{2\pi}{3a}]$, Eq.(37) agrees perfectly with the numerical results. The numerical wave functions for the states of interest also agree with the assumption of Eq.(25). The numerical value of the decay parameter agrees very well with $|\lambda|$ determined by Eq.(33). As shown in Fig.7(c) is a comparison between the decay parameter $|\lambda|$ determined by Eq.(33) for the present edge states and the decay parameter $|t'/t|$ for the edge states corresponding to an ideal bearded-type edge of a honeycomb lattice. The two only have a minor difference that is larger close to $k_{y}=\pm\frac{2\pi}{3a}$ and smallest at $k_{y}=0$. $|\lambda|$ is always slightly larger than $|t'/t|$ so that the edge state determined by Eq.(25) is more localized than the edge state of the bearded-type edge of the honeycomb lattice.

The above analysis is extended easily to the C-terminated right edge. In fact, the set of eigenequations are in exactly the same form. The only change to the above results is a sign reversal of $\Delta$. It is easy to see that Eq.(37) reverses its sign as $\Delta$ reverses sign. Therefore, the C-terminated right edge supports edge states of energy between $0$ and $\Delta$ in the range $k_{y}\in(-\frac{2\pi}{3a}, \frac{2\pi}{3a})$.

In summary, we have arrived at a complete understanding over the in-gap states at energies $\pm\Delta$ and smaller magnitudes. These features show both a beautiful connection and a stark distinction as compared to the in-gap states for the zigzag ribbons of the honeycomb lattice (i.e., graphene lattice) with zigzag-type or bearded-type edges \cite{fujita96,nakada96,klein94}.

\subsection{Dirac cone states of the BB ribbons and related}

The Dirac cones in the spectra of BB-in and BB-off ribbons are unique features of the dice lattice and are absent in the spectra of zigzag ribbons of the honeycomb lattice with either zigzag-type or bearded-type edges \cite{fujita96,nakada96,klein94,oriekhov18}. As the bulk bands of the unbiased dice model have a pair of Dirac cones which are gapped out in the symmetrically biased dice model with nonzero $\Delta$, the Dirac cones in the BB-in and BB-off ribbons might have fundamental differences for the unbiased and the symmetrically biased dice models.

We firstly consider the Dirac cone states in the BB-in and BB-off ribbons of the unbiased dice model.
Comparing the numerical results of Fig.4 for the various zigzag ribbons, the Dirac cones in Fig.4(e) for the BB ribbons are clearly not a property of a single edge and must depend sensitively on the simultaneous presence of the left and right B-sublattice edges. Although as $N_{x}$ increases the gaps of the various zigzag ribbons (other than BB) decrease continuously, the gap is exactly zero only for the BB ribbons. By solving the differential equations for the continuum approximation to the model close to $k_{y}=\pm\frac{2\pi}{3a}$, Oriekhov et al \cite{oriekhov18} found an analytical solution for the wave function of the Dirac cone states, which is valid for $k_{y}$ close to the Dirac points. The wave function of the Dirac cone states found by Oriekhov et al have $k_{y}$-independent constant amplitudes on the three sublattices in the ratio of $\psi_{A}:\psi_{B}:\psi_{C}=1:\pm\sqrt{2}:1$ \cite{oriekhov18}. The relative weights of this state on the three sublattices, $|\psi_{A}|^{2}:|\psi_{B}|^{2}:|\psi_{C}|^{2}=1:2:1$, are identical to those for the bulk dispersive bands \cite{betancur17}.

\begin{figure}[!htb]\label{fig8}
\centering
\hspace{-8.0cm} {\textbf{(a)}}\\
\includegraphics[width=8.2cm,height=1.42cm]{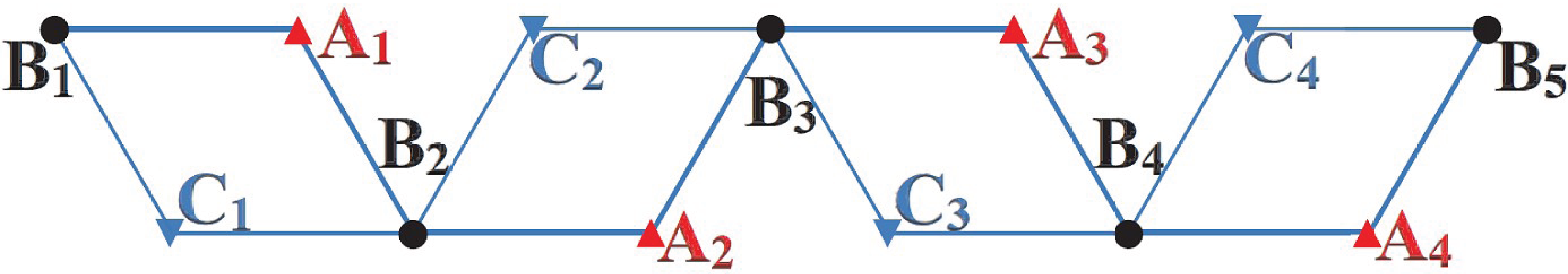} \\ \vspace{0.05cm}
\hspace{-8.0cm} {\textbf{(b)}}\\
\includegraphics[width=8.2cm,height=1.407cm]{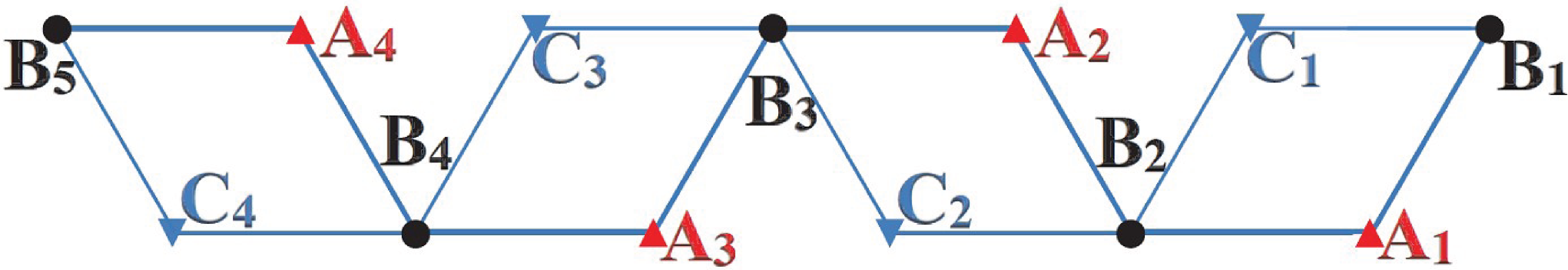}  \\
\caption{(a) and (b) are the unit cells of a BB-in ribbon of the dice lattice for $N_{x}=2$, the sites of which are numbered separately from left to right and from right to left.}
\end{figure}

Here, we study the Dirac cone states of the BB ribbons of the pure dice model in terms of the tight-binding model Eq.(13) defined on the ribbon lattice. Firstly, we want to know why only the BB ribbons support the zero-energy states constituting the Dirac points at $k_{y}=\pm2\pi/(3a)$. As shown in Figs.8(a) and 8(b) are the unit cell of a BB-in ribbon with $N_{x}=2$. The various sites are labeled from left to right in Fig.8(a) and from right to left in Fig.8(b). The unit cell for a BB-off ribbon may similarly be labeled in either of the two manners. The wave function amplitudes of any $E=0$ state of the BB ribbon at the Dirac point satisfy
\begin{equation}
h(k_{y}=\pm\frac{2\pi}{3a})\Psi=E\Psi=0.
\end{equation}
For $\Delta=0$ and $t'=t$ at the Dirac points, the above eigen-equation leads to the following constraints to the components of the zero-energy wave function
\begin{eqnarray}
&&\psi_{Ci}=-\psi_{Ai},  \\
&&\psi_{B,i+1}=-\psi_{Bi},
\end{eqnarray}
where $i=1,2,3,\cdots$. Eq.(39) and Eq.(40) define bulk states with uniform amplitudes across the full width of a BB ribbon. Eq.(39) is compatible with the flat band states determined by Eq.(24), for $\Delta=0$ and $t'=t$. Eq.(40), on the other hand, allows for a finite weight on the B sublattice sites, which is beyond the zero-energy flat bands defined by Eq.(24) for $\Delta=0$. This indicates the presence of additional zero-energy states at $k_{y}=\pm\frac{2\pi}{3a}$ for the BB ribbons. For all other zigzag ribbons of the unbiased dice model, all the $E=0$ states at $k_{y}=\pm\frac{2\pi}{3a}$ may be represented by the states of Eq.(24). Therefore, only the BB ribbons are compatible with additional zero-energy states at $k_{y}=\pm\frac{2\pi}{3a}$ that do not belong to the zero-energy flat bands.

Next, we try to solve for the analytical expressions for the Dirac cone states. While the Dirac cone states of Fig.4(e) at and close to the Dirac points are most probably bulk-like states according to the above analysis, the Dirac cone states of Fig.5(e) for the BB ribbons of the symmetrically biased dice model must be edge states because they lie within the gap region of the bulk bands. On the other hand, both the Dirac cone states of Fig.4(e) and Fig.5(e) connect continuously to the in-gap states at the BZ boundary corresponding to isolated BA or BC dimers at the left or right edge. Therefore, it seems reasonable to treat the biased (i.e., $\Delta\ne0$) and unbiased (i.e., $\Delta=0$) cases simultaneously in terms of a tunable $\Delta$. We firstly assume $\Delta\ne0$ and therefore consider the in-gap edge states for the BB ribbons of the symmetrically biased dice model. Then we study the properties of the solutions in the limit of $\Delta\rightarrow0$, for the unbiased dice model.

We thereby take the following ansatz for the wave function amplitudes on the various sites of the unit cell
\begin{equation}
\psi_{\alpha i}=\psi_{\alpha1}\lambda_{\alpha}^{-(i-1)},
\end{equation}
where $\alpha$=A, B, or C, $i=1,2,\cdots$. We firstly consider the edge states associated to the left edge terminated with a $y$-chain of B sublattice sites. The various sites within the unit cell of the ribbon are therefore labeled according to Fig.8(a). $i$ increases as the various lattice sites extend rightward away from the left edge. The components of the wave function for the edge state of energy $E$ now satisfy the following simultaneous equations
\begin{eqnarray}
&&t'\psi_{C1}+t\psi_{A1}=E\psi_{B1},  \\
&&t'\psi_{B1}+t\psi_{B2}=(E+\Delta)\psi_{C1},  \\
&&t\psi_{B1}+t'\psi_{B2}=(E-\Delta)\psi_{A1},  \\
&&\cdots \cdots  \notag \\
&&t(\psi_{C,i-1}+\psi_{Ai})+t'(\psi_{A,i-1}+\psi_{Ci})=E\psi_{Bi},  \\
&&t'\psi_{Bi}+t\psi_{B,i+1}=(E+\Delta)\psi_{Ci},  \\
&&t\psi_{Bi}+t'\psi_{B,i+1}=(E-\Delta)\psi_{Ai},  \\
&&\cdots \cdots,  \notag
\end{eqnarray}
where $t'=2t\cos(\frac{k_{y}a}{2})$, $i=2,3,\cdots$. In order to fulfill Eqs.(45)-(47) for all $i=2,3,\cdots$, we must have
\begin{equation}
\lambda_{A}=\lambda_{B}=\lambda_{C}=\lambda.
\end{equation}
From Eqs.(42)-(44), we get
\begin{equation}
\begin{cases}
\psi_{A1}=\frac{E(E+\Delta)+t^{2}-t'^{2}}{2tE}\psi_{B1},  \\
\psi_{C1}=\frac{E(E-\Delta)+t'^{2}-t^{2}}{2t'E}\psi_{B1}.  \\
\end{cases}
\end{equation}
We have assumed general $k_{y}$ for which $E\ne0$ and $E\ne\pm\Delta$. Substituting Eqs.(41), (48), and (49) into Eqs.(45)-(47), we get the following constraints to $\lambda$ and $E$
\begin{eqnarray}
&&(t^{2}+t'^{2})E^{2}-\Delta(t^{2}-t'^{2})E-(t^{2}-t'^{2})^{2}=0,  \\
&&\lambda=\frac{2tt'E}{E^{3}-(t^{2}+t'^{2}+\Delta^{2})E-(t^{2}-t'^{2})\Delta}.
\end{eqnarray}
Eq.(50) gives the following two branches of edge states
\begin{equation}
E_{\pm}(k_{y})=\frac{t^{2}-t'^{2}}{2(t^{2}+t'^{2})}[\Delta\pm\sqrt{\Delta^{2}+4(t^{2}+t'^{2})}].
\end{equation}
Substituting Eq.(52) into Eq.(51), the corresponding decay parameters are found to be
\begin{equation}
\lambda_{\pm}=-\frac{2(t^{2}+t'^{2})^{2}}{tt'[\sqrt{\Delta^{2}+4(t^{2}+t'^{2})}\pm\Delta]^{2}}.
\end{equation}

At the boundary of the 1D BZ, $k_{y}a=\pm\pi$, $t'=0$, the two eigenvalues $E_{\pm}(k_{y})$ determined by Eq.(52) reduce to the two eigenvalues of Eq.(20) for $\alpha=1$, and the decay parameters $\lambda_{\pm}$ diverge to negative infinity. The solutions reproduce perfectly the eigenstates for an isolated BA dimer. The two branches of the edge states are therefore associated with the isolated BA dimers close to the left edge terminated with a $y$-chain of B sublattice sites. As $k_{y}$ approaches $2\pi/(3a)$ or $-2\pi/(3a)$, $t'$ approaches $t$, the two branches of $E_{\pm}$ both approach zero. However, the numerical results [e.g., Fig.5(e)] show that the higher energy branch merges with the projection of the bulk band states, and only the lower energy branch remains to be in-gap states as we approach the Dirac points $\pm2\pi/(3a)$. The resolution to this contradiction between the analytical results of Eq.(52) and the numerical results such as Fig.5(e) lies in the properties of the decay parameters of Eq.(53). To be consistent with the numerical calculations of Fig.5, we assume $\Delta>0$ and $t>0$. In this case, the high-energy and low-energy bands of Eq.(52) are separately $E_{+}(k_{y})$ and $E_{-}(k_{y})$. For $\Delta>0$, we have $|\lambda_{-}|>1$ for all the wave vectors in the 1D BZ. $|\lambda_{+}|$, in contrast, is smaller than 1 for the wave vectors in the neighborhood of $k_{y}=2\pi/(3a)$ and $k_{y}=-2\pi/(3a)$. The critical wave vectors separating wave vectors for which $|\lambda_{+}|>1$ and wave vectors for which $|\lambda_{+}|<1$ are determined by $|\lambda_{+}|=1$, which gives
\begin{equation}
(t-t')^{4}=2\Delta^{2}tt',
\end{equation}
where $t'=2t\cos(\frac{k_{y}a}{2})$. Eq.(54) is an even function of $k_{y}$. Qualitatively, as $\Delta$ increases, there are three kinds of solutions to Eq.(54). For $\Delta\in(0,t/2)$, Eq.(54) gives four wave vectors, which we denote by $\pm k_{1}$ and $\pm k_{2}$, $k_{1}>\frac{2\pi}{3a}>k_{2}>0$. At $\Delta=t/2$, Eq.(54) gives three wave vectors, $\pm k_{1}$ and 0, with $k_{1}>\frac{2\pi}{3a}$. For $\Delta>t/2$, Eq.(54) gives only two wave vectors $\pm k_{1}$ with $k_{1}>\frac{2\pi}{3a}$. For $\Delta\in(0,t/2)$, $|\lambda_{+}|\le1$ for the wave vectors in two disconnected regions $k_{y}\in[-k_{1},-k_{2}]$ and $k_{y}\in[k_{2},k_{1}]$. The higher energy edge state solutions are unphysical for the wave vectors in these regions. There are only bulk states for these wave vectors. For $\Delta\ge t/2$, $|\lambda_{+}|\le1$ for the wave vectors in a single connected section $k_{y}\in[-k_{1},k_{1}]$. All the higher energy edge state solutions are unphysical for the wave vectors in this region.

In terms of the above analysis, Eqs.(52) and (53) completely determine the in-gap edge states associated with a left edge terminated with a $y$-chain of B sublattice sites. Together with Eq.(49), we may determine the corresponding wave function amplitude $\psi_{B1}$ up to an arbitrary U(1) phase factor, in terms of the normalization of the wave function.

For a right edge terminated with a $y$-chain of B sublattice sites, we may get the corresponding edge states by repeating the derivations of Eqs.(41)-(53). Alternatively, by comparing Figs.8(b) and 8(a), we see that by exchanging the labels A and C, and simultaneously replacing $\Delta$ with $-\Delta$, the above results directly map to the results for the B-terminated right edge. Therefore, Eq.(49) becomes
\begin{equation}
\begin{cases}
\psi_{C1}'=\frac{E(E-\Delta)+t^{2}-t'^{2}}{2tE}\psi_{B1}',  \\
\psi_{A1}'=\frac{E(E+\Delta)+t'^{2}-t^{2}}{2t'E}\psi_{B1}'.  \\
\end{cases}
\end{equation}
We have added a prime to the wave function components to indicate that they represent edge states associated with a B-terminated right edge. The two branches of in-gap bands have the following dispersions
\begin{equation}
E_{\pm}'(k_{y})=\frac{t^{2}-t'^{2}}{2(t^{2}+t'^{2})}[-\Delta\pm\sqrt{\Delta^{2}+4(t^{2}+t'^{2})}].
\end{equation}
The decay parameters are correspondingly
\begin{equation}
\lambda_{\pm}'=-\frac{2(t^{2}+t'^{2})^{2}}{tt'[\sqrt{\Delta^{2}+4(t^{2}+t'^{2})}\mp\Delta]^{2}}.
\end{equation}

\begin{figure}[!htb]\label{fig9}
\centering
\hspace{-8.0cm} {\textbf{(a)}}\\
\includegraphics[width=7.50cm,height=5.815cm]{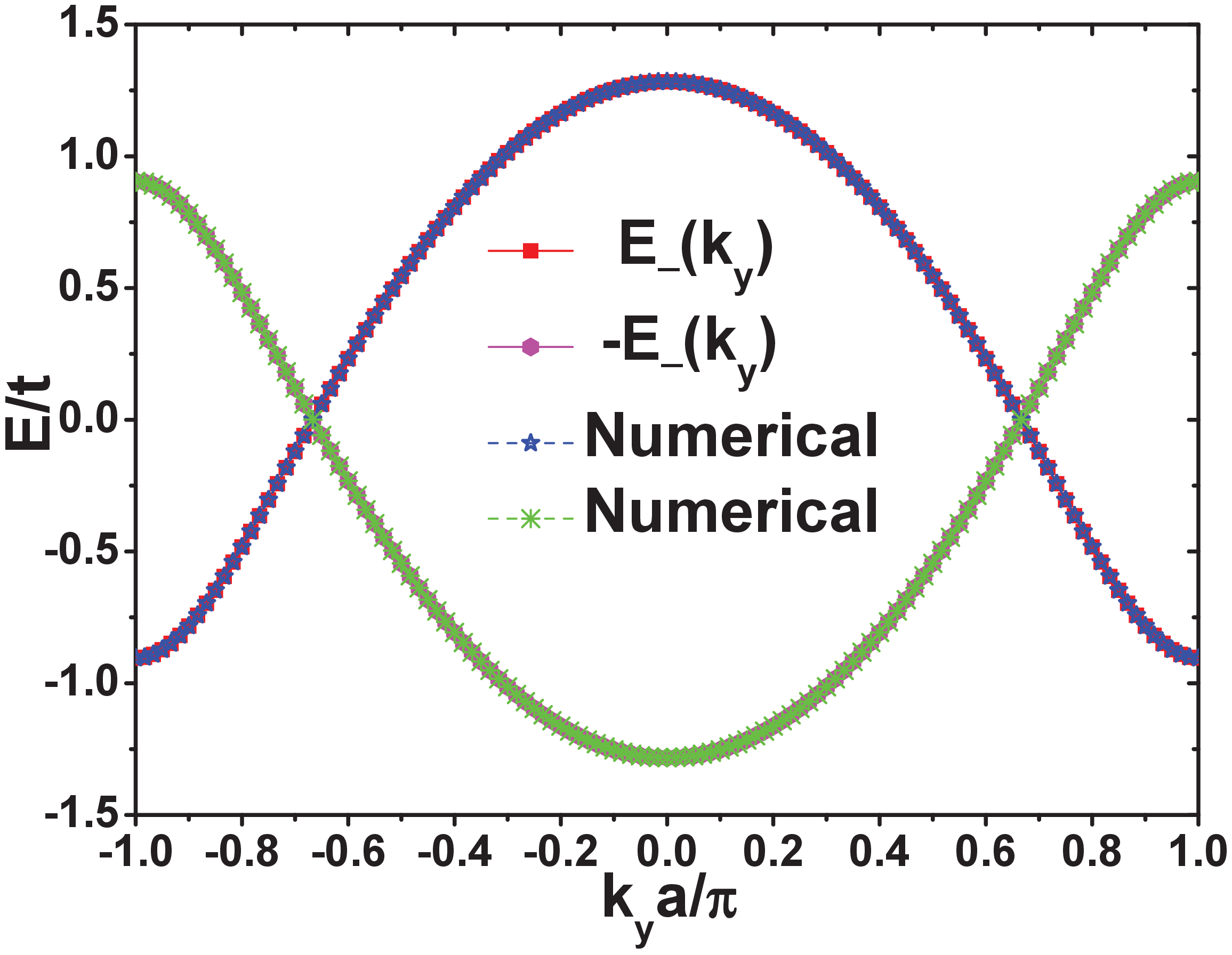}  \\\vspace{0.05cm}
\hspace{-8.0cm} {\textbf{(b)}}\\
\includegraphics[width=7.50cm,height=5.678cm]{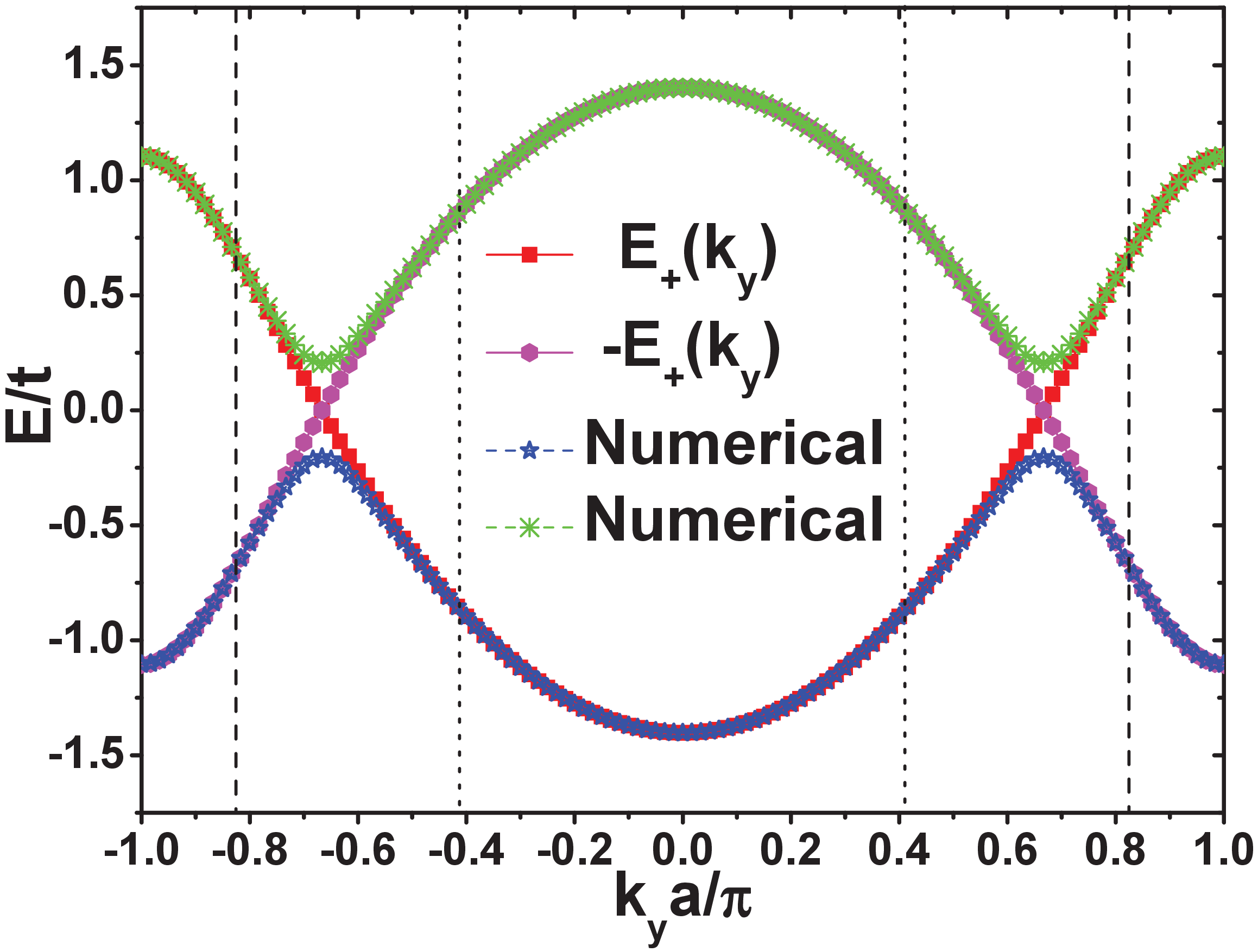}  \\\vspace{0.05cm}
\hspace{-8.0cm} {\textbf{(c)}}\\
\includegraphics[width=7.50cm,height=5.454cm]{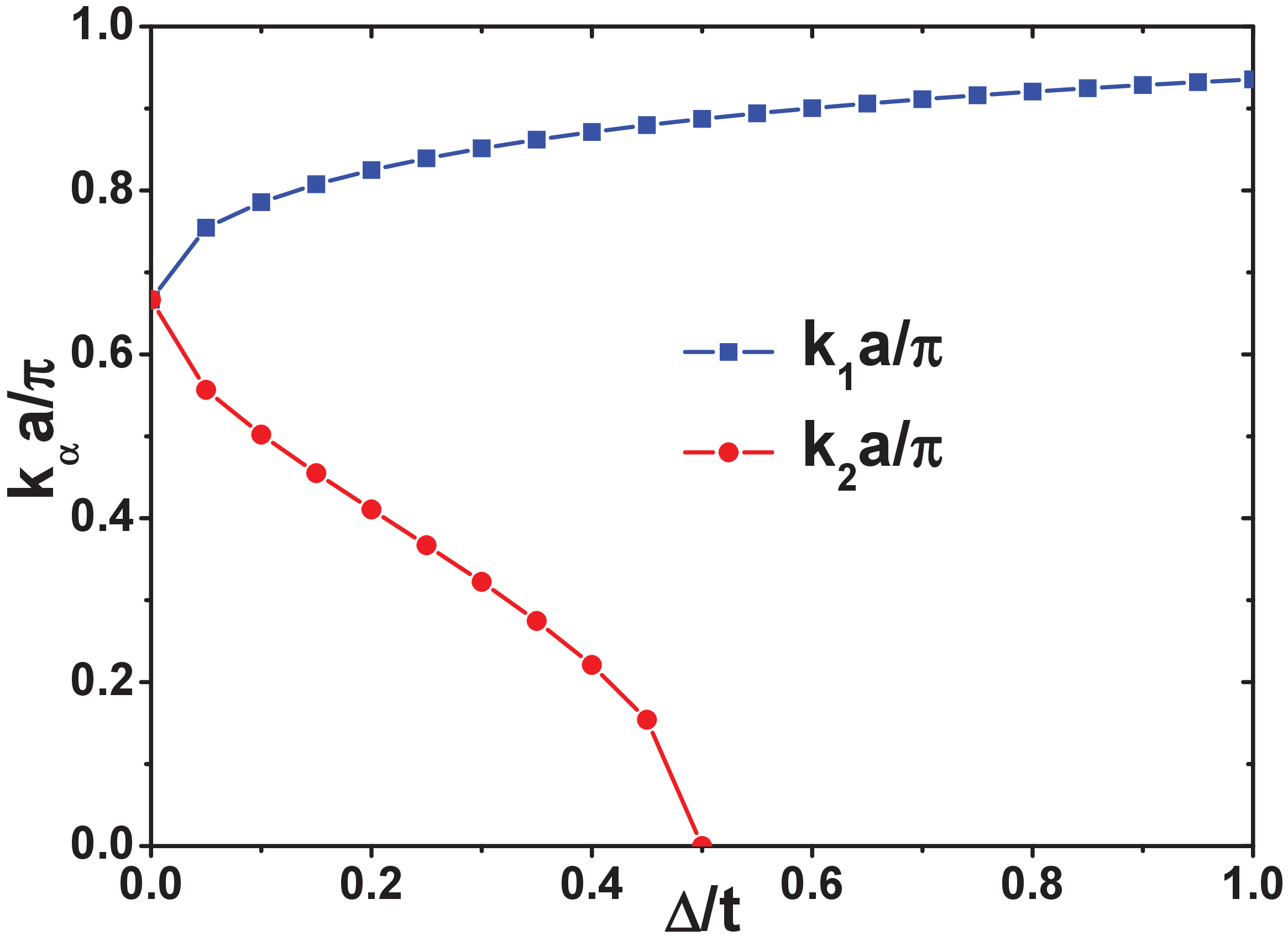}  \\
\caption{(a) and (b) are comparisons of the analytical spectrum of Eqs.(52) and (56) with the numerical spectrum for a BB-in ribbon of the symmetrically biased dice model. $-E_{-}(k_{y})=E_{+}'(k_{y})$ and $-E_{+}(k_{y})=E_{-}'(k_{y})$. We consider parameters $t=1$, $\Delta=0.2t$. The numerical results are for $N_{x}=50$. (c) shows the evolutions with $\Delta$ of the two positive critical wave vectors $k_{1}$ and $k_{2}$ that satisfy Eq.(54).}
\end{figure}

The eigenenergies and decay parameters of the left and right B-terminated edges are related through
\begin{equation}
\begin{cases}
E_{\pm}'(k_{y})=-E_{\mp}(k_{y}),  \\
\lambda_{\pm}'=\lambda_{\mp}.  \\
\end{cases}
\end{equation}
Therefore, for $\Delta>0$, the states in the two low-energy bands $E_{-}(k_{y})$ and $E_{+}'(k_{y})$ are all well-defined edge states. $E_{-}(k_{y})$ and $E_{+}'(k_{y})=-E_{-}(k_{y})$ are symmetrical with respect to $E=0$. They intersect linearly at the two Dirac points $k_{y}=\pm\frac{2\pi}{3a}$ and give the two 1D Dirac cones. For $\Delta>0$, the two high-energy bands $E_{+}(k_{y})$ and $E_{-}'(k_{y})=-E_{+}(k_{y})$ are only well-defined edge states for wave vectors sufficiently far from $k_{y}=\pm\frac{2\pi}{3a}$. Within the sections of wave vectors determined above, the high-energy bands $E_{+}(k_{y})$ and $E_{-}'(k_{y})$ are not well defined and replaced by bulk states. These features agree very well with the numerical results. As shown in Fig.9(a) is the comparison between the analytical and numerical results for the two low-energy edge states comprising the two 1D Dirac cones. The parameters are taken as $t=1$ and $\Delta=0.2t$. The numerical results are for a BB-in ribbon with $N_{x}=50$, corresponding exactly to Fig.5(e). The analytical and numerical results coincide perfectly. As shown in Fig.9(b) is a comparison between the two high-energy analytical bands $E_{+}(k_{y})$ and $E_{-}'(k_{y})$ and the numerical results for the corresponding bands. The numerical solutions to Eq.(54) for $t=1$ and varying $\Delta$ are as shown on Fig.9(c). For $\Delta=0.2t$, we have $k_{1}a\simeq0.825\pi$ and $k_{2}a\simeq0.411\pi$. The positions of $\pm k_{1}$ and $\pm k_{2}$ are marked separately by vertical dashed and vertical dotted lines in Fig.9(b). From Fig.9(b), it is clear that the analytical results agree perfectly with the numerical results for wave vectors where the edge states are well defined, including the wave vectors in the three regions $k_{y}\in[-\frac{\pi}{a},-k_{1})\cup(-k_{2},k_{2})\cup(k_{1},\frac{\pi}{a}]$.

We now turn to the unbiased system, which corresponds to the $\Delta=0$ limit of the above results. In this limit, the edge states for the left edge and the right edge are degenerate. The eigenenergies and the decay parameters are separately
\begin{eqnarray}
&&E_{\pm}(k_{y})=E_{\pm}'(k_{y})=\pm\frac{t^{2}-t'^{2}}{\sqrt{t^{2}+t'^{2}}}, \\
&&\lambda=\lambda_{+}=\lambda_{-}=-\frac{t^{2}+t'^{2}}{2tt'}.
\end{eqnarray}
For $\Delta=0$, Eq.(49) for the ratios between the wave function amplitudes of the edge states for the left B-terminated edge becomes
\begin{equation}
\begin{cases}
\psi_{A1}=\pm\frac{t}{\sqrt{t^{2}+t'^{2}}}\psi_{B1}, \\
\psi_{C1}=\mp\frac{t'}{\sqrt{t^{2}+t'^{2}}}\psi_{B1}. \\
\end{cases}
\end{equation}
Eq.(55) for the ratios between the wave function amplitudes of the edge states for the right B-terminated edge becomes
\begin{equation}
\begin{cases}
\psi_{C1}'=\pm\frac{t}{\sqrt{t^{2}+t'^{2}}}\psi_{B1}',  \\
\psi_{A1}'=\mp\frac{t'}{\sqrt{t^{2}+t'^{2}}}\psi_{B1}'.  \\
\end{cases}
\end{equation}

For wave vectors far away from the two Dirac points at $k_{y}a=\pm\frac{2\pi}{3a}$, and for a sufficiently wide ribbon, Eqs.(59)-(62) describe the two-fold degenerate in-gap states such as those in Fig.4(e). Close to the Dirac points, on the other hand, the Dirac cone states from the numerical calculations for a ribbon of finite width are only singly degenerate, as for example in Fig.4(e) for a BB-in ribbon at $N_{x}=50$. Where does the other branch of edge states goes?

Right at a Dirac point, we have $t'=t$, $E_{\pm}=0$, and $\lambda=-1$. The two states described by Eq.(61) are in fact identical to the two states described by Eq.(62). Consequently, there is indeed only a single Dirac cone, rather than a pair of degenerate Dirac cones, at the Dirac points. Eq.(61) and Eq.(62) agree with Eq.(39). $\lambda=-1$ agrees with Eq.(40). The weight of the wave function on the B sublattice is twice the weight on the A or C sublattice. These constraints and the normalization condition completely determine the wave functions for the two degenerate zero-energy states at the Dirac points.

For a wave vector $k_{y}$ very close to but deviating from the Dirac point $2\pi/(3a)$, we define
\begin{equation}
q=k_{y}-\frac{2\pi}{3a}
\end{equation}
to measure the tiny deviation of $k_{y}$ from $2\pi/(3a)$.
For such wave vectors, $t'\ne t$, the two states defined by Eq.(61) are different from the two states defined by Eq.(62). Expanding to the leading order of $q$, we have
\begin{eqnarray}
&&t'=2t\cos(\frac{k_{y}a}{2})\simeq t(1-\frac{\sqrt{3}}{2}qa),\\
&&E_{\pm}=E_{\pm}'\simeq \pm\frac{\sqrt{6}}{2}|t|qa, \\
&&\lambda\simeq-(1+\frac{3}{8}q^{2}a^{2}).
\end{eqnarray}
For wave vectors very close to the Dirac point, while $|\lambda|\simeq1+3q^{2}a^{2}/8>1$, the deviation of $|\lambda|$ from 1 is vanishingly small for $qa\ll1$. Although the four states defined by Eqs.(59)-(62) are in principle edge states, they extend through the whole breadth of a finite-width ribbon and are effectively bulk states. As a result, the two states defined by Eq.(61) associated to the left edge are cut off by the right edge, while the two states defined by Eq.(62) associated to the right edge are cut off by the left edge. These four states, different from the uniform state at a Dirac point, are not exact eigenstates of the finite-width ribbon. From the continuity of the wave functions in a band, and the similarity of the wave functions to the uniform wave function at the Dirac point, it is natural to conjecture that a linear combination of the edge states of Eqs.(59)-(62) in the form of an even function might be a good approximation to the Dirac cone states.

For the above purpose, we introduce another basis set out of those defined by Eqs.(59)-(62) as follows
\begin{equation}
\tilde{E}_{\pm}(k_{y})=\tilde{E}_{\pm}'(k_{y})=\pm\frac{|t^{2}-t'^{2}|}{\sqrt{t^{2}+t'^{2}}},
\end{equation}
\begin{equation}
\begin{cases}
\tilde{\psi}_{A1}=\pm s\frac{t}{\sqrt{t^{2}+t'^{2}}}\tilde{\psi}_{B1}, \\
\tilde{\psi}_{C1}=\mp s\frac{t'}{\sqrt{t^{2}+t'^{2}}}\tilde{\psi}_{B1}. \\
\end{cases}
\end{equation}
\begin{equation}
\begin{cases}
\tilde{\psi}_{C1}'=\pm s\frac{t}{\sqrt{t^{2}+t'^{2}}}\tilde{\psi}_{B1}',  \\
\tilde{\psi}_{A1}'=\mp s\frac{t'}{\sqrt{t^{2}+t'^{2}}}\tilde{\psi}_{B1}'.  \\
\end{cases}
\end{equation}
The parameter $s$ is a sign function defined as
\begin{equation}
s=\frac{t^{2}-t'^{2}}{|t^{2}-t'^{2}|}.
\end{equation}
The new basis set gathers all the positive (negative) energy states into the positive energy band (negative energy band) that contains the upper (lower) Dirac cones. The decay parameter $\lambda$ for the states of all four bands are still given by Eq.(60).
For clarity of discussions, we define the eigenvectors for $\tilde{E}_{+}(k_{y})$ and $\tilde{E}_{-}(k_{y})$ as $\tilde{\Psi}_{1}$ and $\tilde{\Psi}_{2}$, and define $\tilde{\Psi}_{3}$ and $\tilde{\Psi}_{4}$ as the eigenvectors for $\tilde{E}_{+}'(k_{y})$ and $\tilde{E}_{-}'(k_{y})$.

We consider the two degenerate positive energy branches, $\tilde{E}_{+}(k_{y})$ and $\tilde{E}_{+}'(k_{y})$. The analysis for the two negative energy branches is completely the same. To compare with numerical results, we consider the BB-in ribbons with $N_{x}=50$. According to the continuity of the wave functions, the states in the band containing the upper Dirac cones should correspond to an even combination of the states of the $\tilde{\Psi}_{1}(k_{y})$ and $\tilde{\Psi}_{3}(k_{y})$ bands. The orthogonal (complementary) state in this two-fold degenerate subspace corresponds to the odd combination of the $\tilde{\Psi}_{1}(k_{y})$ and $\tilde{\Psi}_{3}(k_{y})$ states. For simplicity and without losing generality, we take the wave function amplitudes on the outermost B sites to be real positive for the $\tilde{\Psi}_{1}(k_{y})$ and $\tilde{\Psi}_{3}(k_{y})$ states, namely $\tilde{\psi}^{(1)}_{B1}=\tilde{\psi}^{(3)}_{B1}>0$. This amounts to fixing the redundant $U(1)$ phase factors for the two bands. In this convention, the even and odd combinations of the two basis states are
\begin{equation}
\tilde{\Psi}_{\alpha}=N_{\alpha}(\tilde{\Psi}_{1}+\alpha\tilde{\Psi}_{3}),
\end{equation}
where $\alpha=\pm$ and $N_{\alpha}$ is a normalization factor. $\tilde{\Psi}_{+}$ and $\tilde{\Psi}_{-}$ are separately the even and odd combinations of the degenerate $\tilde{\Psi}_{1}(k_{y})$ and $\tilde{\Psi}_{3}(k_{y})$ bands.

It is straightforward to check that $\tilde{\Psi}_{+}$ and $\tilde{\Psi}_{-}$ are not exact eigenstates of $h(k_{y})$. They are however relevant from at least two points: Firstly, $\tilde{\Psi}_{+}$ evolves continuously to the exact uniform zero-energy eigenstate at the Dirac points, and $\tilde{\Psi}_{-}$ is orthogonal to $\tilde{\Psi}_{+}$. Secondly, far away from the Dirac point, where the decay parameter $|\lambda|$ becomes noticeably larger than 1 and so the edge states become true edge states for a finite-width ribbon, $\tilde{\Psi}_{+}$ and $\tilde{\Psi}_{-}$ are simply the even and odd combinations of two decoupled edge states. In this case, $\tilde{\Psi}_{+}$ and $\tilde{\Psi}_{-}$ are exact eigenstates of $h(k_{y})$ with degenerate eigenenergies equal to $\tilde{E}_{+}(k_{y})=\tilde{E}_{+}'(k_{y})$. To test the relevance of these two states to the states close to the Dirac points, we calculate the expectation values of the model under these two states
\begin{equation}
\bar{E}_{\alpha}=\langle\tilde{\Psi}_{\alpha}|h(k_{y})|\tilde{\Psi}_{\alpha}\rangle.
\end{equation}
We focus on the states close to the Dirac point $k_{y}=\frac{2\pi}{3a}$ and compare $\bar{E}_{\alpha}$ ($\alpha=\pm$) with the numerical results for the upper Dirac cone ($E_{1}$ hereafter) and the lowest positive band ($E_{2}$ hereafter) above the upper Dirac cone. We consider parameters the same as Fig.4(e), with $t=1$, $\Delta=0$, and $N_{x}=50$.

\begin{figure}[!htb]\label{fig10}
\centering
\hspace{-8.0cm} {\textbf{(a)}}\\
\includegraphics[width=7.0cm,height=5.572cm]{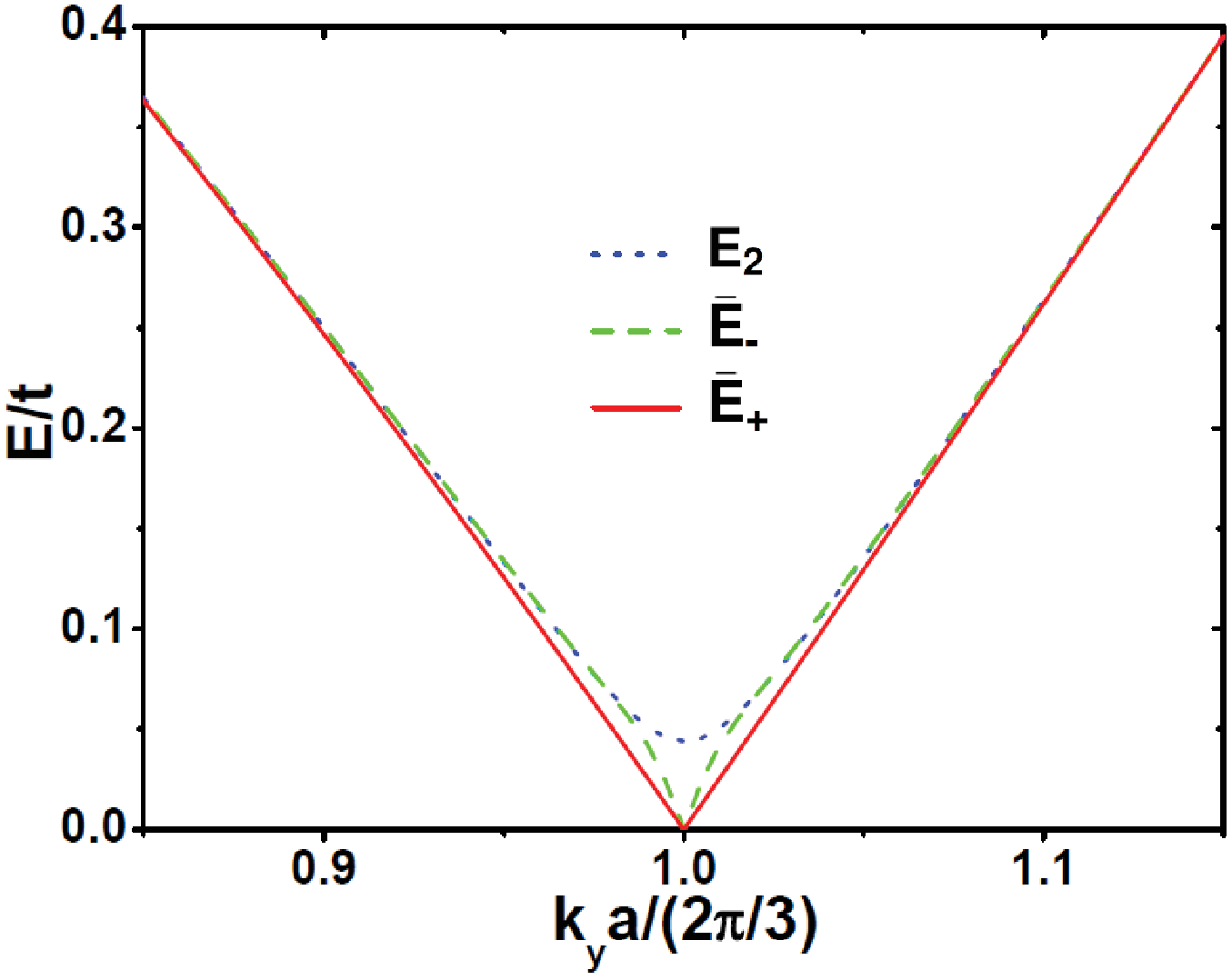}  \\\vspace{0.05cm}
\hspace{-8.0cm} {\textbf{(b)}}\\
\includegraphics[width=7.0cm,height=5.296cm]{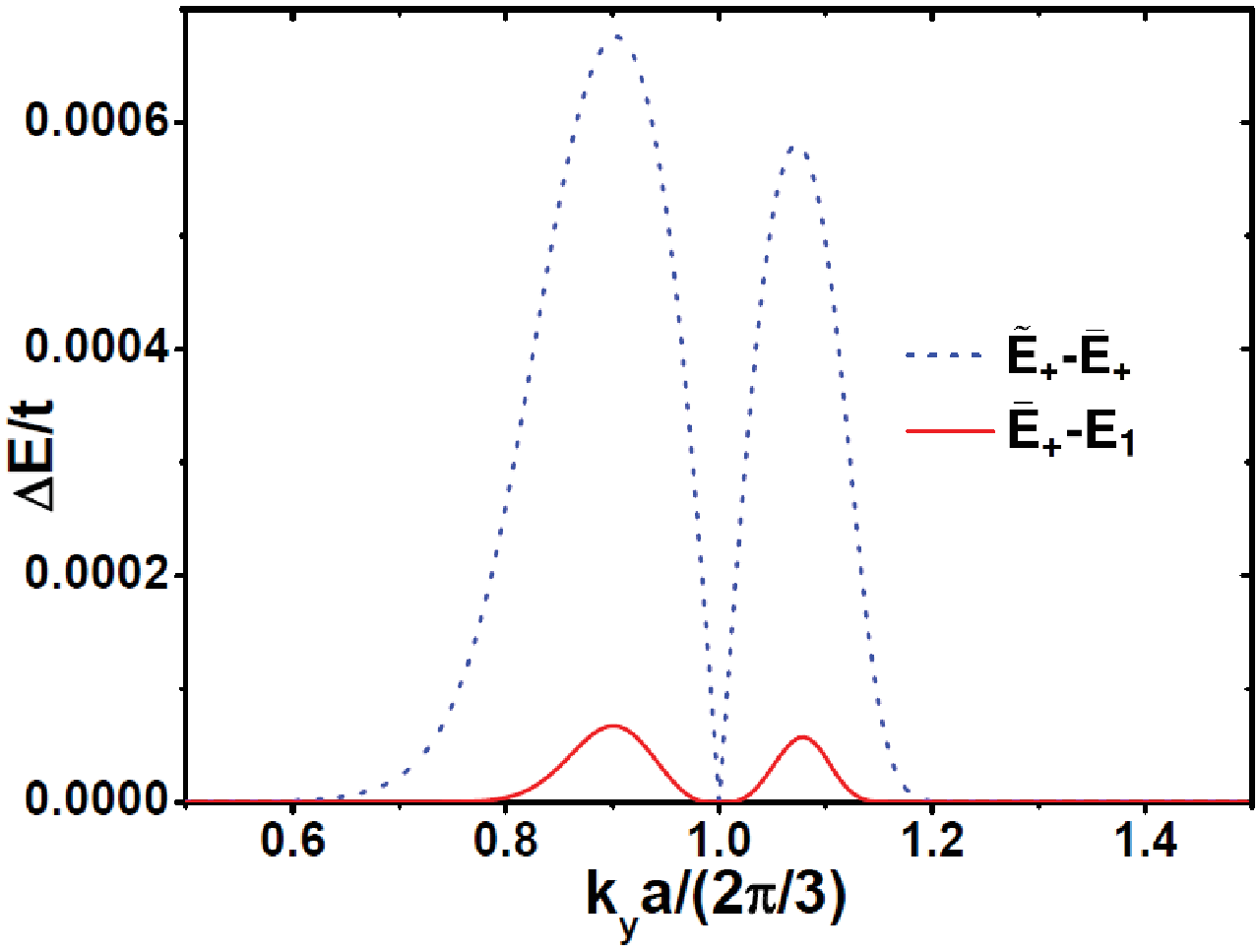}  \\\vspace{0.05cm}
\hspace{-8.0cm} {\textbf{(c)}}\\
\includegraphics[width=7.0cm,height=5.402cm]{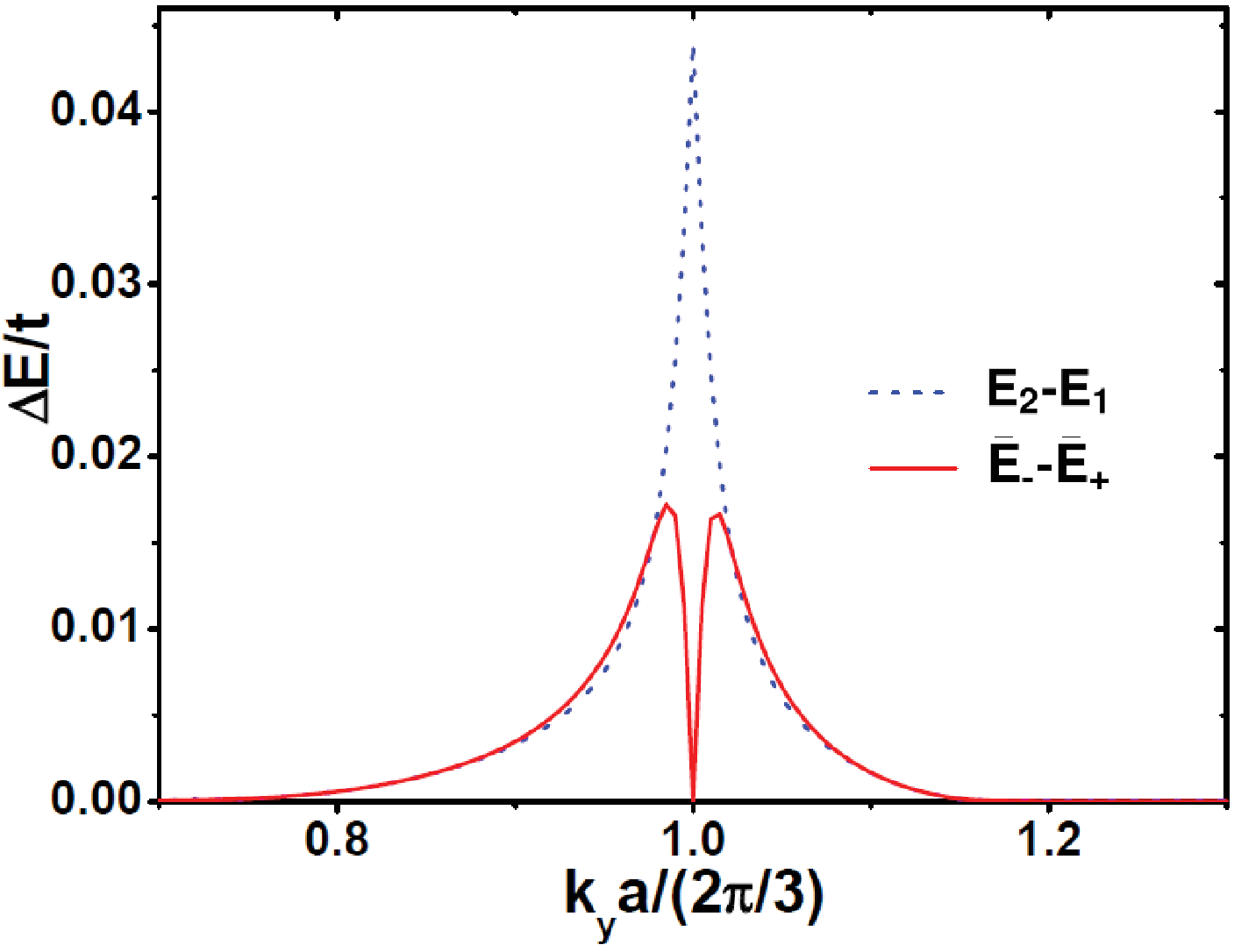}  \\
\caption{Plots of the low-energy bands or energy differences between bands of the pure dice model, in the neighborhood of the Dirac point $k_{y}=\frac{2\pi}{3a}$. (a) Plots of $\bar{E}_{+}$ (solid red line), $\bar{E}_{-}$ (dashed green line), and $E_{2}$ (dotted blue line). $E_{2}$ is defined as the lowest positive energy band above the upper Dirac cone, in the numerical results. (b) Plots of two energy differences, $\bar{E}_{+}-E_{1}$ (solid red line) and $\tilde{E}_{+}-\bar{E}_{+}$ (dotted blue line). $E_{1}$ is the numerical upper Dirac cone band. (b) Plots of two energy differences, $\bar{E}_{-}-\bar{E}_{+}$ (solid red line) and $E_{2}-E_{1}$ (dotted blue line). All calculations consider a BB-in ribbon with $t=1$, $\Delta=0$, and $N_{x}=50$.}
\end{figure}

As shown in Fig.10(a) is the result of $\bar{E}_{+}$ (the solid red line). $\tilde{E}_{+}$ and $E_{1}$ are not drawn directly, because they are indistinguishable from $\bar{E}_{+}$ in figure. We instead draw the energy differences between two pairs of bands in Fig.10(b), including $\bar{E}_{+}-E_{1}$ (solid red line) and $\tilde{E}_{+}-\bar{E}_{+}$ (dotted blue line). Another energy difference, $\tilde{E}_{+}-E_{1}$, is the sum of the two energy differences shown in Fig.10(b). It is clear that $\bar{E}_{+}$ is a much better approximation to $E_{1}$ than $\tilde{E}_{+}$ is. This is particularly true in the close neighborhood of the Dirac point.

For wave vectors close to the Dirac point, the weights of $\tilde{\Psi}_{+}$ on various sites of the unit cell of the ribbon also agree very well with the numerical results, such as those for Fig.4(e). For wave vectors very close to but different from the Dirac point, Eqs.(68) and (69) tell us that the weights of the wave function on the A and C sublattices are unequal. On the other hand, $|\lambda|>1$ and so the wave function amplitudes of $\tilde{\Psi}_{1}$ decay as the sites go from left to right and the wave function amplitudes of $\tilde{\Psi}_{3}$ decay as the sites go from right to left. As a result, $\tilde{\Psi}_{+}$ is dominated by $\tilde{\Psi}_{1}$ for sites closer to the left edge and dominated by $\tilde{\Psi}_{3}$ for sites closer to the right edge. For $|k_{y}|<2\pi/(3a)$, $t'>t$, we have $|\tilde{\psi}_{Ci}/\tilde{\psi}_{Ai}|=|\tilde{\psi}_{Ai}'/\tilde{\psi}_{Ci}'|=|t'/t|>1$. Therefore, the wave function amplitudes of $\tilde{\Psi}_{+}$ are larger on the C sublattice sites than those on the nearby A sublattice sites in the neighborhood of the left edge, which is reversed in the neighborhood of the right edge. These trends agree with the numerical results for the wave functions. For $|k_{y}|>2\pi/(3a)$, $t'<t$, we have $|\tilde{\psi}_{Ci}/\tilde{\psi}_{Ai}|=|\tilde{\psi}_{Ai}'/\tilde{\psi}_{Ci}'|=|t'/t|<1$. Correspondingly, the wave function amplitudes of the Dirac cone states should be larger on the A sublattice sites than those on the nearby C sublattice sites in the neighborhood of the left edge, which is reversed in the neighborhood of the right edge. These trends also agree very well with the numerical results for the wave functions of the Dirac cone states. In comparison, the analytical results of the wave function for the Dirac cone states obtained by Oriekhov et al \cite{oriekhov18} predict an eigenvector that do not depend on the value of $k_{y}$ for all $k_{y}$ close to a Dirac point. The results of Oriekhov et al \cite{oriekhov18} agree with the bulk bands but disagree with the numerical results for the BB ribbons.

We finally consider the $\tilde{\Psi}_{-}$ state and the corresponding expectation value $\bar{E}_{-}$. Note that since $\tilde{\Psi}_{1}=\tilde{\Psi}_{3}$ at a Dirac point, $\tilde{\Psi}_{-}$ is not well defined at the Dirac points. We however may define $\tilde{\Psi}_{-}$ at a Dirac point by a limiting process of letting the wave vector approach the Dirac point. In this manner we may get a full band of $\bar{E}_{-}$ across the whole 1D BZ, as shown in Fig.10(a) close to the Dirac point $k_{y}=\frac{2\pi}{3a}$ (the dashed green line). As expected, the resulting $\bar{E}_{-}$ is degenerate with $\bar{E}_{+}$ for wave vectors far away from the Dirac points. Close to a Dirac point, such as $k_{y}=2\pi/(3a)$, $\bar{E}_{-}$ splits off and is higher than $\bar{E}_{+}$. $\bar{E}_{-}$ is therefore similar to $E_{2}$, which is the lowest positive energy band above the upper Dirac cone $E_{1}$. However, the splitting between $\bar{E}_{-}$ and $\bar{E}_{+}$ is nonmonotonic, which firstly increases and then decreases to zero as the wave vector approaches the Dirac point [solid red line of Fig.10(c)]. This is in contrast to the monotonous increase of the splitting between $E_{2}$ and $E_{1}$ from the numerical results as $k_{y}$ approaches a Dirac point [dotted blue line of Fig.10(c)]. The turning point in the gap between $\bar{E}_{-}$ and $\bar{E}_{+}$ corresponds roughly to the wave vector where $\bar{E}_{-}$ starts to become clearly different from the numerical spectrum of $E_{2}$ [dotted blue line of Fig.10(a)].

Comparison of $\tilde{\Psi}_{-}$ and the numerical wave functions for the $E_{2}$ band also shows a partial agreement. On one hand, they are both odd functions across the unit cell and vanish at the central B sublattice site. The weights of the wave functions on the various lattice sites of the unit cell are also very close to each other, for wave vectors not too close to the Dirac points. On the other hand, right at the Dirac point, $\tilde{\Psi}_{-}$ defined by the limit process has weights only on the A and C sublattices across the whole unit cell, whereas the numerical wave function has weights on all the three sublattices. Overall, in comparison to $\tilde{\Psi}_{+}$ which makes a very good description for the upper Dirac cone band $E_{1}$, $\tilde{\Psi}_{-}$ makes a good description for the lowest gapped band $E_{2}$ only for wave vectors not too close to the Dirac points. The inaccuracy of $\tilde{\Psi}_{-}$ for $E_{2}$ in the neighborhood of a Dirac point seems to be related to the pathology of the definition Eq.(71) for $\tilde{\Psi}_{-}$ around a Dirac point, which takes subtraction between two wave functions that are identical or nearly identical. In other words, $\tilde{\Psi}_{1}=\tilde{\Psi}_{3}=\tilde{\Psi}_{+}$ is the same state at a Dirac point, and it is impossible to define a meaningful different state out of this single state. In this respect, the results of Oriekhov et al \cite{oriekhov18} based on solving a continuum model provide a better description for the $E_{2}$ band in the neighborhood of a Dirac point. $\tilde{\Psi}_{-}$, on the other hand, provides a good description for the $E_{2}$ band for all the wave vectors of the 1D BZ except for the neighborhood of the two Dirac points.

\section{Summary}

In summary, we have investigated the structure-spectrum correspondences for the distinct edge termination morphologies of the zigzag dice lattice ribbons. We introduce a general formula to predict the number of all distinct edge termination configurations for the regular ribbons of a 2D lattice with a specific edge orientation, which is illustrated for ribbons of the Lieb lattice, the honeycomb lattice, and the dice lattice. Then, focusing on the zigzag ribbons of the dice lattice, which fall into 18 distinct categories according to the formula, we find a one-to-one correspondence between the electronic spectra and the edge termination morphologies, for the symmetrically biased dice model. On the other hand, there are qualitative degeneracies among the 18 spectra for the pure dice model. We clarify the nature of several interesting features in the electronic spectra of the 18 types of zigzag ribbons of the pure dice model and the symmetrically biased dice model. These features include the number and wave functions of the zero-energy flat bands, and various novel in-gap states including segmental flat bands and 1D Dirac cones. We find both interesting similarities and also salient distinctions compared to the spectra for the zigzag ribbons of the honeycomb lattice. The present study sheds new light on the interesting electronic states of the dice lattice. Further studies on the electronic spectrum of other dice lattice ribbons, of different edge orientations or model parameters, are highly desirable. Properties related to the 1D flat bands and novel in-gap states of the various zigzag dice lattice ribbons are also intriguing new research directions.



\end{document}